\documentclass[a4paper,11pt]{article}

\usepackage{jcappub} 

\usepackage[T1]{fontenc} 
\usepackage[utf8]{inputenc}

\usepackage{comment}
\usepackage{graphicx}	
\usepackage{amsmath}	
\usepackage{amssymb}	
\usepackage{lmodern}
\usepackage{bm}
\usepackage{cases}
\usepackage{float}
\usepackage{colortbl}
\usepackage{multirow}
\usepackage{booktabs}
\usepackage{tabularx}
\usepackage{makecell}
\usepackage{hyperref}
\usepackage{MnSymbol}
\usepackage{mathtools}
\usepackage{subcaption} 

\newcommand{\bk}{\mathbf{k}}

\newcommand{\bx}{\mathbf{x}}
\newcommand{\Mpc}{{\rm Mpc}}

\title{Towards cosmological constraints from the compressed modal bispectrum: \\a robust comparison of real-space bispectrum estimators}

\author[a,1]{Joyce Byun,\note{Corresponding author.}}
\author[b,c,d]{Andrea Oddo,}
\author[e]{Cristiano Porciani,}
\author[f,c,d]{and Emiliano Sefusatti}

\affiliation[a]{Depart\'ement de Physique Th\'eorique and Center for Astroparticle Physics (CAP), University of Geneva, 24 quai Ernest Ansermet, CH-1211 Geneva, Switzerland}
\affiliation[b]{SISSA - International School for Advanced Studies, Via Bonomea 265, 34136 Trieste, Italy}
\affiliation[c]{Institute for Fundamental Physics of the Universe, Via Beirut 2, 34151 Trieste, Italy}
\affiliation[d]{Istituto Nazionale di Fisica Nucleare, Sezione di Trieste, via Valerio 2, 34127 Trieste, Italy}
\affiliation[e]{Argelander Institut f{\"u}r Astronomie der Universit{\"a}t Bonn, Auf dem H{\"u}gel 71, 53121 Bonn, Germany}
\affiliation[f]{Istituto Nazionale di Astrofisica, Osservatorio Astronomico di Trieste, via Tiepolo 11, 34143 Trieste, Italy}

\emailAdd{joyce.byun@unige.ch}
\emailAdd{andrea.oddo@sissa.it}
\emailAdd{porciani@astro.uni-bonn.de}
\emailAdd{emiliano.sefusatti@inaf.it}

\abstract{
Higher-order clustering statistics, like the galaxy bispectrum, can add complementary cosmological information to what is accessible with two-point statistics, like the power spectrum. 
While the standard way of measuring the bispectrum involves estimating a bispectrum value in a large number of Fourier triangle bins, the compressed modal bispectrum approximates the bispectrum as a linear combination of basis functions and estimates the expansion coefficients on the chosen basis.
In this work, we compare the two estimators by using parallel pipelines to analyze the real-space halo bispectrum measured in a suite of $N$-body simulations corresponding to a total volume of $\sim 1{,}000 \, h^{-3}\,{\rm Gpc}^3$, with covariance matrices estimated from 10,000 mock halo catalogs.
We find that the modal bispectrum yields constraints that are consistent and competitive with the standard bispectrum analysis: for the halo bias and shot noise parameters within the tree-level halo bispectrum model up to $k_{\rm max} \approx 0.06 \, (0.10) \, h\,{\rm Mpc}^{-1}$, only 6 (10) modal expansion coefficients are necessary to obtain constraints equivalent to the standard bispectrum estimator using $\sim 20$ to 1,600 triangle bins, depending on the bin width.
For this work, we have implemented a modal estimator pipeline using Markov Chain Monte Carlo simulations for the first time, and we discuss in detail how the parameter posteriors and modal expansion are robust to, or sensitive to, several user settings within the modal bispectrum pipeline.
The combination of the highly efficient compression that is achieved and the large number of mock catalogs available allows us to quantify how our modal bispectrum constraints depend on the number of mocks that are used to estimate covariance matrices and the functional form of the likelihood.
}

\begin{document}
\maketitle
\flushbottom


\section{Introduction}
\label{sec:intro}

Gravitational clustering and nonlinear bias induce a non-Gaussianity in the large-scale structure (LSS) of the Universe that can be quantified by higher-order correlation functions, like the 3-point correlation function (3PCF) and its Fourier counterpart, the bispectrum.
The measurement of the galaxy bispectrum in redshift surveys can contribute additional constraining power towards a wide range of science goals, improving on what is achievable using only 2-point correlations, like the power spectrum.

To date, the most precise measurements of the galaxy bispectrum and 3PCF are from the SDSS Baryon Oscillation Spectroscopic Survey \cite{Gil-Marin:2014sta,Gil-Marin:2014baa,Gil-Marin:2016wya,Slepian:2016kfz,Pearson:2017wtw,PearsonSamushia2018errata,Sugiyama:2018yzo}.
In the near future, spectroscopic galaxy surveys like 
DESI,\footnote{\url{https://www.desi.lbl.gov}}
Euclid,\footnote{\url{https://www.euclid-ec.org}}
SPHEREx,\footnote{\url{https://spherex.caltech.edu}}
and the Roman Space Telescope\footnote{\url{https://wfirst.ipac.caltech.edu}}
\cite{Levi:2013gra,Laureijs:2011gra,Dore:2014cca,Spergel:2015sza}
will map galaxy distributions over larger areas of the sky, to higher redshifts, and with more precision than before, opening up new opportunities for higher-order galaxy clustering statistics to be used as stronger probes of $\Lambda$CDM (the standard cosmological model dominated by a cosmological constant called $\Lambda$ and cold dark matter), dark energy, 
modified gravity theories, primordial non-Gaussianity, and massive neutrinos (e.g.~\cite{
Chan:2016ehg,Byun:2017fkz, 
Song:2015gca,Gagrani:2016rfy,Yankelevich:2018uaz,Gualdi:2020ymf,Agarwal:2020lov, 
Yamauchi:2017ibz,Bose:2018zpk,Bose:2019wuz, 
Tellarini:2016sgp,Karagiannis:2018jdt, 
Ruggeri:2017dda,Hahn:2019zob}). 

Bispectrum data sets are naturally much larger than for power spectra, because they typically capture the correlation amplitudes for a large number of triangle bins, $B(k_1,k_2,k_3)$, rather than being a function of only one wavenumber, like $P(k)$.
This presents practical challenges that can potentially limit the full exploitation of the data that will be available from upcoming LSS surveys.
One particularly acute challenge for bispectrum analyses is the large number of mock catalogs that are typically necessary to accurately estimate the large data covariance matrices for galaxy clustering analyses.
One remedy to this problem has been to explore the accuracy of fast, approximate simulation codes that reduce the computational resources needed for generating mocks \cite{Monaco:2016pys,Colavincenzo:2018cgf}. 
Alternatively, it may be possible to obtain equivalent covariance matrices using fewer or smaller volume mocks (e.g.~\cite{
Joachimi:2016xhk, 
FriedrichEifler2018,Hall:2018umb, 
Pearson:2015gca, 
Howlett:2017vwp}) 
or even no mocks, if an accurate theoretical model of the covariance matrix is available (e.g.~\cite{Mohammed:2016sre,Sugiyama:2019ike,Wadekar:2019rdu,Taruya:2020qoy}). 
A third strategy that has been pursued in tandem is to develop methods that compress the information contained in the bispectrum into smaller, more manageable data sets.

Much work to date falls into this last category of bispectrum compression methods.
Using the standard bispectrum estimator, choosing to use wider wavenumber bins reduces the total number of triangle bins, but at the same time erases some of the triangle-dependence that encodes cosmological information.
Other compression methods and compressed bispectrum observables include: 
Karhunen-Lo\`{e}ve compression of the standard bispectrum estimator \cite{Gualdi:2017iey,Gualdi:2018pyw} 
(which is similar to the MOPED algorithm \cite{Heavens:1999am,Heavens:2020spq}),
subspace projection of the standard bispectrum estimator \cite{Philcox:2020zyp},
binning triangles based on their geometrical properties \cite{Gualdi:2019ybt,Gualdi:2019sfc},
skew-spectra \cite{Pratten:2011kh,Schmittfull:2014tca,MoradinezhadDizgah:2019xun},
position-dependent power spectra (also called integrated bispectra) \cite{Chiang:2014oga,Chiang:2015eza,Chiang:2015pwa},
line correlation functions \cite{Obreschkow:2012yb,Wolstenhulme:2014cla,Eggemeier:2015ifa,Eggemeier:2016asq,Franco:2018yag,Ali:2018sdk,Byun:2020hun},
and the modal bispectrum.
The focus of this work is to implement and explore the last of these, and to compare it with a standard bispectrum analysis.

The modal bispectrum describes the bispectrum as a linear combination of smooth 3-dimensional basis functions, such that the observable data are the expansion coefficients over a chosen basis.
If the bispectrum is relatively smooth, and the chosen basis is suitable for describing changes in the bispectrum induced by the model parameters we wish to constrain, we expect that the modal expansion coefficients will provide an efficient compression of the cosmological information that is typically distributed over a large number of triangle bins.
The modal expansion method was originally developed in the context of primordial non-Gaussianity in the cosmic microwave background \cite{Fergusson:2009nv,Fergusson:2010dm,Ade:2013ydc} before it was adapted for the LSS bispectrum \cite{Fergusson:2010ia,Regan:2011zq,Schmittfull:2012hq}.
It has been used to test and develop theoretical models of the matter bispectrum \cite{Schmittfull:2012hq,Lazanu:2015rta,Lazanu:2015bqo} and compare the matter bispectra measured from different dark matter simulation codes \cite{Hung:2019ygc} and mock-making prescriptions \cite{Hung:2019nma}.
These previous works have been in real space, and an extension of the modal expansion method to redshift space was outlined in \cite{Regan:2017vgi}.

The first direct comparison between the modal bispectrum and the standard bispectrum estimators was in \cite{Byun:2017fkz}, where Fisher forecasted constraints on $\Lambda$CDM cosmological parameters and galaxy bias using the real-space matter modal bispectrum and standard bispectrum estimators were equivalent.
For this work, we have implemented a new modal bispectrum analysis pipeline that uses the real-space halo bispectrum to constrain galaxy bias and shot noise parameters.
This builds on previous work by implementing the modal bispectrum method in a Markov chain Monte Carlo (MCMC) analysis pipeline for the first time.
Where possible, we have adhered closely to the standard bispectrum analysis in \cite{Oddo:2019run}, so the two estimators can be rigorously compared.
In the process of implementing the new modal bispectrum pipeline, we have explored several technical details of the modal method's implementation that have not previously been presented in the literature, and we discuss how the modal estimator pipeline is sensitive (or not) to these details.

The main message of this work is that the modal bispectrum estimator provides an extremely efficient compression of the information contained within the halo bispectrum, resulting in parameter constraints that are at least as strong as the standard bispectrum estimator.
Depending on the specific settings within the modal bispectrum pipeline, we find that as few as 10 modal expansion coefficients are necessary for galaxy bias and shot noise parameter constraints to converge when the smallest scale is given by $k_{\rm max} \approx 0.10 \, h\,{\rm Mpc}^{-1}$.
While the modal decomposition method does require some additional calculations and machinery, we show that this overhead is small, especially when compared to the benefits of having a highly compressed data set.
For example, in this work we use the modal bispectrum pipeline to show how bispectrum constraints can depend on the number of mock catalogs that are used to estimate the covariance matrix---a result that has not been attempted using the standard bispectrum estimator due to the extremely large number of mocks that would be required.

The outline of the paper is as follows. 
In Section \ref{sec:method} we review the modal decomposition method. 
In Section \ref{sec:data_and_analysis} we describe the data that we use, our likelihood modeling, and MCMC simulation details. 
We discuss our new results in Section \ref{sec:results} and summarize our main conclusions in Section \ref{sec:conclusions}.


\section{Modal decomposition method}
\label{sec:method}

The basic premise of the modal decomposition method is that the bispectrum is a relatively smooth function of Fourier-space triangles, so the bispectrum  that is normally measured in a large number of triangle bins, $N_{\rm triangles}$, is very well approximated by a linear combination of a smaller number, $N_{\rm modes}$, of basis mode functions. 
We write this as
\begin{equation}
w(k_1,k_2,k_3)B(k_1,k_2,k_3) \approx \sum_{n=0}^{N_{\rm modes}-1} \beta^Q_n \, Q_n(k_1,k_2,k_3),
	\label{eq:wB_expansion}
\end{equation}
where $w(k_1,k_2,k_3)$ is a weighting function, $B(k_1,k_2,k_3)$ is the bispectrum, $\beta^Q_n$ are a set of modal expansion coefficients, and $Q_n(k_1,k_2,k_3)$ are the modal basis functions.
The modal coefficients therefore correspond to the amplitudes of template bispectra in the data.

The optimal estimator for the amplitude of a single bispectrum template for an isotropic and statistically homogeneous density field in the limit of weak non-Gaussianity is \cite{Babich:2005en}
\begin{equation}
	\hat{\varepsilon} = \frac{1}{N_\varepsilon} 
	\int_{\bk_1} 
	\int_{\bk_2} 
	\int_{\bk_3} 
	(2\pi)^3 \delta_D(\bk_{123}) 
	B^{\rm template}(k_1,k_2,k_3)
	\frac{\delta_{\bk_1} \delta_{\bk_2} \delta_{\bk_3}}
	{P(k_1)P(k_2)P(k_3)},
\end{equation}
where we have introduced the shorthand notations $\int_{\bk} \equiv \int \frac{{\rm d}^3k}{(2\pi)^3}$ and $\bk_{123} \equiv \bk_1+\bk_2+\bk_3$ to make our expressions more compact. 
$\delta_\bk$ is the observed density field on a discretized Fourier-space grid, $P(k_i)$ is the total power spectrum (including the shot noise), and $N_\varepsilon$ is a normalization constant.\footnote{
We choose our Fourier transform convention so that the forward and backward transforms are
$\delta(\bk) = \int {\rm d}^3x \, \delta(\bx) e^{i \bk \cdot \bx}$ and 
$\delta(\bx) = (2\pi)^{-3} \int {\rm d}^3k \, \delta(\bk) e^{-i \bk \cdot \bx}$.
}
After defining the quantity
\begin{equation}
	\hat{\mathcal{B}}(\bk_1,\bk_2,\bk_3) \equiv \frac{\delta_{\bk_1} \delta_{\bk_2} \delta_{\bk_3}}{V} \bf{1}_{\bk_{123}},
\end{equation}
where $V$ is the survey volume and 
$\bf{1}_{\bk_
{123}}$ is a Kronecker symbol that is unity if $\bk_{123}=0$ and zero otherwise,
we see that the amplitude estimator is effectively a normalized weighted inner product between $B^{\rm template}$ and $\hat{\mathcal{B}}$,
\begin{equation}
	\hat{\varepsilon} = \frac{V}{N_\varepsilon} \llangle wB^{\rm template} | w\hat{\mathcal{B}} \rrangle,
\end{equation}
where the definition of the inner product is
\begin{equation}
	\llangle wB^{\rm template} | w\hat{\mathcal{B}} \rrangle \equiv 
	\int_{\bk_1}
	\int_{\bk_2}
	\int_{\bk_3}
	(2\pi)^3 \delta_D(\bk_{123})  
	\frac{wB^{\rm template}(k_1,k_2,k_3) w\hat{\mathcal{B}}(\bk_1,\bk_2,\bk_3)}
	{k_1k_2k_3}
	\label{eq:innerprod}
\end{equation}
and the weighting function is 
\begin{equation}
	w(k_1,k_2,k_3) \equiv \frac{\sqrt{k_1k_2k_3}}{\sqrt{P(k_1)P(k_2)P(k_3)}}.
	\label{eq:w}
\end{equation}
Then the normalization constant must be $N_\varepsilon \equiv V \llangle wB^{\rm template} | wB^{\rm template} \rrangle$ so that the ensemble average of the estimated amplitude $\langle\hat{\varepsilon}\rangle \rightarrow 1$ if $\langle\hat{\mathcal{B}}\rangle \rightarrow B^{\rm template}$.

If the ensemble average $\langle \hat{\mathcal{B}}(\bk_1,\bk_2,\bk_3) \rangle = B^{\rm obs}(k_1,k_2,k_3)$, we can perform the angular integrals in the inner product analytically, following steps detailed in \cite{Fergusson:2010ia}, which we also review here.
We first use 
\begin{equation}
	\delta_D(\bk_{123}) = \frac{1}{(2\pi)^3} \int {\rm d}^3x \, e^{i\bk_{123}\cdot\bx}
	\label{eq:delta}
\end{equation}
and rewrite the exponential part as
\begin{equation}
	e^{i\bk\cdot\bx} = 
	4\pi \sum_{\ell m} i^\ell j_\ell(kx) 
	Y_{\ell m}(\hat{\bk}) Y_{\ell m}^*(\hat{\bx})
\end{equation}
to get 
\begin{align}
	\llangle wB^{\rm template} | wB^{\rm obs} \rrangle = 
	\int {\rm d}^3x (4\pi)^3
	&\left[ 
	\int_{\bk_1}
	\sum_{\ell_1 m_1} i^{\ell_1} j_{\ell_1}(k_1x) 
	Y_{\ell_1 m_1}(\hat{\bk}_1) Y_{\ell_1 m_1}^*(\hat{\bx})
	\right] \nonumber \\
	\times &\left[ 
	\int_{\bk_2}
	\sum_{\ell_2 m_2} i^{\ell_2} j_{\ell_2}(k_2x) 
	Y_{\ell_2 m_2}(\hat{\bk}_2) Y_{\ell_2 m_2}^*(\hat{\bx})
	\right] \nonumber \\
	\times &\left[ 
	\int_{\bk_3}
	\sum_{\ell_3 m_3} i^{\ell_3} j_{\ell_3}(k_3x) 
	Y_{\ell_3 m_3}(\hat{\bk}_3) Y_{\ell_3 m_3}^*(\hat{\bx})
	\right] \nonumber \\
	\times &
	\frac{wB^{\rm template}(k_1,k_2,k_3) wB^{\rm obs}(k_1,k_2,k_3)}
	{k_1k_2k_3}.
	\label{eq:inner product with jYY}
\end{align}
The integral over $\hat{\bk}_i$ inside each pair of square brackets is\footnote{Our spherical harmonics are normalized such that $\int {\rm d}\Omega_{\bk} Y_{\ell m}(\hat{\bk})^2 = 1$ and $Y_{00} = 1/\sqrt{4\pi}$.}
\begin{equation}
	\int {\rm d}\Omega_{\bk_i} Y_{\ell_i m_i}(\hat{\bk}_i) = \sqrt{4\pi}\delta_{\ell_i0}\delta_{m_i0},
\end{equation}
which forces all $\ell_i$ and $m_i$ in eq.~\eqref{eq:inner product with jYY} to be zero, giving 
\begin{align}
	\llangle wB^{\rm template} | wB^{\rm obs} \rrangle = \int {\rm d}^3x \frac{(4\pi)^{9/2}}
	{(2\pi)^9} & \left[ \int {\rm d}k_1 \, k_1^2 \, j_0(k_1x) Y_{00}(\hat{\bx}) \right] \nonumber \\
	\times & \left[ \int {\rm d}k_2 \, k_2^2 \, j_0(k_2x) Y_{00}(\hat{\bx}) \right] \nonumber \\
	\times & \left[ \int {\rm d}k_3 \, k_3^2 \, j_0(k_3x) Y_{00}(\hat{\bx}) \right] \nonumber \\
	\times & \frac{wB^{\rm template}(k_1,k_2,k_3) wB^{\rm obs}(k_1,k_2,k_3)}{k_1k_2k_3}.
	\label{eq: inner product with jY}
\end{align} 
In the final step, integration over $\bx$ using 
\begin{align}
	\int {\rm d}x \,x^2 j_0(k_1x)j_0(k_2x)j_0(k_3x) &= \frac{\pi}{8k_1k_2k_3} \\
	\int {\rm d}\Omega_{\bx} Y_{00}(\hat{\bx})^3 &= \frac{1}{\sqrt{2\pi}}
\end{align}
shows that the inner product is
\begin{eqnarray}
	\llangle wB^{\rm template} | wB^{\rm obs} \rrangle = \frac{1}{8\pi^4}
	\int_{\mathcal{V}_T}  {\rm d}k_1 \, {\rm d}k_2 \, {\rm d}k_3
	\, wB^{\rm template}(k_1,k_2,k_3) wB^{\rm obs}(k_1,k_2,k_3),
	\label{eq:inner product tetrapyd}
\end{eqnarray}
where the subscript $\mathcal{V}_T$ signifies that the 3-dimensional integral must only cover the volume, sometimes called a \textit{tetrapyd}, where $(k_1,k_2,k_3)$ can form a closed triangle.
Therefore, estimating the amplitude of a given template bispectrum is closely related to calculating a weighted inner product between the template and observed bispectrum over the tetrapyd space.

In previous literature on the modal bispectrum, the inner product in eq.~\eqref{eq:inner product tetrapyd} is sometimes written in terms of $(x_1,x_2,x_3)$ instead of $(k_1,k_2,k_3)$, where $x_i \equiv (k_i-k_{\rm min})/(k_{\rm max}-k_{\rm min})$, such that the allowed $(x_1,x_2,x_3)$ form a tetrapyd that fits inside of a unit cube.
We will sometimes use a different notation to define the inner product over this unit tetrapyd,
\begin{equation}
	\langle f | g \rangle \equiv \int_{\mathcal{V}_T}  {\rm d}x_1 \, {\rm d}x_2 \, {\rm d}x_3 \, f(x_1,x_2,x_3) g(x_1,x_2,x_3).
\end{equation}

\subsection{Modal estimator}
\label{subsec:estimator}

In this work, it is not the amplitude of one template that we are interested in, but the expansion coefficients of a general bispectrum on a chosen set of basis functions.
In this case, we make the replacement $wB \rightarrow w\hat{\mathcal{B}}$ in eq.~\eqref{eq:wB_expansion} and take the inner product of both sides with $Q_m$ to obtain
\begin{equation}
	\llangle Q_m | w\hat{\mathcal{B}} \rrangle = \sum_{n=0}^{N_{\rm modes}-1} \hat{\beta}^Q_n \, \gamma_{nm},
	\label{eq:modal_linear_eq}
\end{equation}
where we have defined the positive-definite symmetric matrix $\gamma_{nm} \equiv \llangle Q_n | Q_m \rrangle$.\footnote{In \cite{Byun:2017fkz}, the inner product over the unit tetrapyd, $\overline{\gamma} \equiv \langle Q|Q \rangle$, was also defined and used to convert between $\gamma$ matrices calculated over different $k$-ranges, given by $k_{\rm min}$ and $k_{\rm max}$. 
This was motivated by the assumption that $\overline{\gamma}$ could be computed once, and thereafter $\gamma$ over a general $k$-range could be computed as $\gamma = (k_{\rm max}-k_{\rm min})^3/(8\pi^4) \overline{\gamma}$.
However, we note that this was not quite correct; this rescaling can only be done when $k_{\rm min}=0$, because in general $\gamma$ depends on $k_{\rm min}$ and $k_{\rm max}$ in a way that cannot be factored out. 
We thank Dionysios Karagiannis for noticing this.}
To estimate the modal coefficients, $\hat{\beta}^Q_n$, we first measure the $N_{\rm modes}$ inner products, $\llangle Q_n | w\hat{\mathcal{B}} \rrangle$ on the left-hand side, and solve the linear matrix equation in eq.~\eqref{eq:modal_linear_eq}.

To make the measurement of
\begin{eqnarray} 
	\llangle Q_n | w\hat{\mathcal{B}} \rrangle 
	&=& \frac{1}{V}
	\int_{\bk_1}
	\int_{\bk_2}
	\int_{\bk_3} 
	(2\pi)^3 \delta_D(\bk_{123}) 
	\frac{Q_n(k_1,k_2,k_3) \, \delta_{\bk_1} \delta_{\bk_2} \delta_{\bk_3}}
	{\sqrt{k_1k_2k_3}\sqrt{P(k_1)P(k_2)P(k_3)}}
	\label{eq:QwB estimator}
\end{eqnarray}
computationally tractable, we require that the $Q_n$ basis functions can be written in separable form as a product of three 1-dimensional functions,
\begin{equation}
	Q_n(k_1,k_2,k_3) = q_{\{p}(k_1) q_r(k_2)q_{s\}}(k_3).
\end{equation}
The $p$, $r$, and $s$ subscripts on the right side index the different 1-dimensional functions that we have chosen, and the curly brackets require that the $Q_n$ functions are invariant to permutations of $k_1$, $k_2$, and $k_3$.
In Appendix \ref{app:1d qn}, we give additional details on how we compute the $q_n(k)$ from either normal or Legendre polynomials and how we have chosen the mapping between $\{ prs \} \leftrightarrow n$.

Taking advantage of the separability of $Q_n$ and using eq.~\eqref{eq:delta} to rewrite the delta function in its exponential form, 
the final expression for $\llangle Q_n | w\hat{\mathcal{B}} \rrangle$ simplifies into a computationally tractable expression written concisely as
\begin{equation}
	\llangle Q_n | w\hat{\mathcal{B}} \rrangle =  \frac{1}{V} \int {\rm d}^3x \, M_{\{p}(\bx)M_r(\bx)M_{s\}}(\bx),
	\label{eq:QwB estimator with Mrx}
\end{equation}
where 
\begin{eqnarray}
	M_r(\bx) \equiv \int \frac{{\rm d}^3k}{(2\pi)^3} \frac{e^{i\bk\cdot\bx}}{\sqrt{kP(k)}} \, q_r(k) \, \delta_{\bk}.
	\label{eq:Mrx}
\end{eqnarray}
Therefore the inner product can be computed very efficiently using fast Fourier transform (FFT) routines (such as FFTW\footnote{\url{http://www.fftw.org}} or Intel MKL\footnote{\url{https://software.intel.com/content/www/us/en/develop/tools/math-kernel-library.html}} libraries), if the basis of $Q_n$ functions are multiplicative separable.

We note that this estimator requires minimal modifications to the standard bispectrum estimator, which takes the form
\begin{align}
	\hat{B}(k_1,k_2,k_3) =& \frac{V}{N_\triangle(k_1,k_2,k_3)}
	\int_{\mathbf{q}_1}
	\int_{\mathbf{q}_2} 
	\int_{\mathbf{q}_3}
	(2\pi)^3 \delta_D(\mathbf{q}_{123})
	\tilde{\Pi}_{k_1}(\mathbf{q}_1) 
	\tilde{\Pi}_{k_2}(\mathbf{q}_2) 
	\tilde{\Pi}_{k_3}(\mathbf{q}_3) 
	\delta_{\mathbf{q}_1} \delta_{\mathbf{q}_2} \delta_{\mathbf{q}_3} \nonumber \\
	=& \frac{V}{N_\triangle(k_1,k_2,k_3)} \int {\rm d}^3x 
	\, \mathcal{D}_{k_1}(\bx) \mathcal{D}_{k_2}(\bx) \mathcal{D}_{k_3}(\bx),
	\label{eq:B estimator}
\end{align}
where 
\begin{equation}
	\mathcal{D}_k(\bx) \equiv \int \frac{{\rm d}^3q}{(2\pi)^3} \, e^{i \mathbf{q} \cdot \bx} \, \tilde{\Pi}_k(\mathbf{q}) \, \delta_{\mathbf{q}},
\end{equation}
and $\tilde{\Pi}_k(\mathbf{q})$ is a binning function that is 1 if $|\mathbf{q}| \in [ k - \Delta k/2, k + \Delta k/2 ]$ and zero otherwise.
$N_\triangle$ is the number of $(\mathbf{q}_1,\mathbf{q}_2,\mathbf{q}_3)$ triangles that are averaged inside a $(k_1,k_2,k_3)$ triangle bin, 
\begin{align}
	N_\triangle(k_1,k_2,k_3) &\equiv V^2
	\int_{\mathbf{q}_1}
	\int_{\mathbf{q}_2}
	\int_{\mathbf{q}_3}
	\delta_D(\mathbf{q}_{123}) \,
	\tilde{\Pi}_{k_1}(\mathbf{q}_1) 
	\tilde{\Pi}_{k_2}(\mathbf{q}_2) 
	\tilde{\Pi}_{k_3}(\mathbf{q}_3) \nonumber \\
	&= \frac{V^2}{(2\pi)^3} \int {\rm d}^3x \, \Pi_{k_1}(\bx) \Pi_{k_2}(\bx) \Pi_{k_3}(\bx),
\end{align}
where $\Pi_k(\bx)$ is the inverse Fourier transform of $\tilde{\Pi}_k(\mathbf{q})$.
Comparing the standard bispectrum estimator with the estimator for $\llangle Q_n | w\hat{\mathcal{B}} \rrangle$, we see that they both require very similar computational steps, and the modal estimator recovers the bispectrum estimator by making the replacement $q_r(k) / \sqrt{kP(k)} \rightarrow  \tilde{\Pi}_k (\mathbf{q})$.
The critical difference, however, is that for a single realization, while the bispectrum estimator is computed once per $(k_1,k_2,k_3)$ triangle bin, the modal estimator is computed once for each $Q_n$.
Also, the memory requirement for the bispectrum estimator is such that each $k_i$ bin requires a full Fourier grid to store the corresponding $\mathcal{D}_{k_i}(\bx)$, while for the modal estimator each 1-dimensional basis function $q_r$ requires its own grid to store $M_r(\bx)$.
Therefore the modal estimator is typically more computationally efficient compared to the standard bispectrum estimator. 

Once the $\llangle Q_n | w\hat{\mathcal{B}} \rrangle$ have been measured, we estimate $\hat{\beta}^Q_n$ by numerically solving the linear equation in eq.~\eqref{eq:modal_linear_eq}.
Using the $\hat{\beta}^Q$ coefficients, we can also calculate a reconstructed bispectrum as
\begin{equation}
	B_{\rm rec} = \frac{1}{w} \sum_n \hat{\beta}^Q_n \, Q_n.
\label{eq:Brec}
\end{equation}
Later, in Section \ref{subsec:Brec}, we compare this bispectrum, $B_{\rm rec}$, with the bispectrum measured using the standard estimator, $\hat{B}$.

We note that the $\gamma$ matrix only needs to be computed once for a desired wavenumber range $(k_{\rm min},k_{\rm max})$ and choice of $Q_n$ basis functions.
Different methods for calculating the inner products in $\gamma$ have been discussed to date in the literature, and one of the goals of this work is to compare these methods.
In the next section, we summarize the inner product methods that we implement and compare in this work.


\subsection{Inner product methods}
\label{subsec:gamma_methods}

Here, we briefly describe the four different methods we have implemented in this work for calculating the inner product matrix 
\begin{eqnarray}
	\gamma \equiv \llangle Q_n | Q_m \rrangle = \frac{1}{8\pi^4}
	\int_{\mathcal{V}_T}  {\rm d}k_1 \, {\rm d}k_2 \, {\rm d}k_3
	\, Q_n(k_1,k_2,k_3) \, Q_m(k_1,k_2,k_3).
	\label{eq:QnQm_tetrapyd}
\end{eqnarray}

\subsubsection*{Monte Carlo integration} 
This method uses the Monte Carlo algorithm called Vegas included in the public Cuba library for multidimensional numerical integration \cite{Hahn:2004fe,Hahn:2014fua} to calculate the inner product via random sampling of the 3-dimensional tetrapyd space.
The free parameters for this method are the convergence tolerance and the maximum number of samples. 
In addition to being very slow to converge, we find it quite challenging, despite different choices in the integration parameters, to avoid a non-positive definite $\gamma$, which cannot be used for the modal analysis pipeline.

\subsubsection*{Voxels} 
This method divides the cubic volume of $(k_1,k_2,k_3)$ from $k_{\rm min}$ up to $k_{\rm max}$ into a grid of smaller cubes, called \textit{voxels}, and calculates eq.~\eqref{eq:QnQm_tetrapyd} by integrating over each voxel using tri-linear interpolation of the integrand within each voxel (as described in Appendix A2 of \cite{Byun:2017fkz}). 
Because of the shape of the tetrapyd volume, some care must be taken to properly integrate over voxels that intersect with the tetrapyd boundary.
The only free parameter of this method is the grid resolution set by the number of individual voxels, $N_v$, spanning the chosen $k$-range in each of the three dimensions.

\subsubsection*{3D FFT} 
This method calculates the inner product in the same way that the modal estimator does: we take the expression for $\llangle Q_n | w\hat{\mathcal{B}} \rrangle $ in eq.~\eqref{eq:QwB estimator} and make the replacement $w\hat{\mathcal{B}} \rightarrow Q_m$ to find
\begin{eqnarray} 
	\llangle Q_n | Q_m \rrangle 
	&=& 
	\int_{\bk_1}
	\int_{\bk_2}
	\int_{\bk_3} 
	(2\pi)^3 \delta_D(\bk_{123}) 
	\frac{Q_n(k_1,k_2,k_3) \, Q_m(k_1,k_2,k_3)}{k_1 k_2 k_3} \\
	&=& \int {\rm d}^3x \int_{\bk_1} \int_{\bk_2} \int_{\bk_3} 
	e^{i \bk_{123} \cdot \bx}
	\,
	\frac{q_{\{p}(k_1) q_r(k_2)q_{s\}}(k_3)
	\, q_{\{a}(k_1) q_b(k_2)q_{c\}}(k_3)}{k_1 k_2 k_3}.
\end{eqnarray}
Then, similarly to eq.~\eqref{eq:QwB estimator with Mrx}, we write this in a compact form as \cite{Hung:2019ygc}
\begin{eqnarray}
\llangle Q_n | Q_m \rrangle = \frac{1}{6} \int {\rm d}^3x
	 && \{M_{pa}(\bx)	\left[ M_{rb}(\bx) M_{sc}(\bx) + M_{rc}(\bx) M_{sb}(\bx) \right] \nonumber \\
	&&+ M_{pb}(\bx)	\left[ M_{rc}(\bx) M_{sa}(\bx) + M_{ra}(\bx) M_{sb}(\bx) \right] \nonumber \\
	&&+ M_{pc}(\bx)	\left[ M_{ra}(\bx) M_{sb}(\bx) + M_{rb}(\bx) M_{sa}(\bx) \right] \},
\end{eqnarray}
where we have defined
\begin{equation}
	M_{pa}(\bx) \equiv \int_\bk e^{i\bk\cdot\bx} \, \frac{q_p(k) \, q_a(k)}{k}.
\end{equation}
Like the modal estimator that we have already discussed, this expression for $\llangle Q_n | Q_m \rrangle$ can be computed quickly using existing FFT software.
The free parameters of this method are, as with any discrete Fourier transform, the real-space volume and the FFT grid resolution.

\subsubsection*{1D FFT\footnote{We thank Dionysios Karagiannis for suggesting the 1D FFT method.}} This method computes the inner product using 1-dimensional FFTs by evaluating the expression in eq.~\eqref{eq: inner product with jY} after the replacements $wB^{\rm template} \rightarrow Q_n$ and $wB^{\rm obs} \rightarrow Q_m$.
In this case, using $j_0(k_ix) = \sin(k_ix)/k_ix$ and $Y_{00} = 1/\sqrt{4\pi}$, the inner product becomes
\begin{eqnarray}
	\llangle Q_n|Q_m \rrangle = \frac{1}{2\pi^5} \int {\rm d}x \, \frac{1}{x} 
	&& \{ F_{pa}(x) [ F_{rb}(x) F_{sc}(x) + F_{rc}(x) F_{sb}(x)] \nonumber \\
	&&+ F_{pb}(x) \left[ F_{rc}(x) F_{sa}(x) + F_{ra}(x) F_{sb}(x) \right] \nonumber \\
	&&+ F_{pc}(x) \left[ F_{ra}(x) F_{sb}(x) + F_{rb}(x) F_{sa}(x) \right] \},
	\label{eq:1dfft QQ}
\end{eqnarray}
where
\begin{equation}
	F_{pa}(x) \equiv \int {\rm d}k \,q_p(k)\,q_a(k)\,\sin(kx).
	\label{eq:1dfft Fx}
\end{equation}
The integral over $k$ in $F_{pa}(x)$ can be performed using 1-dimensional FFTs, as described in Chapter 13.9 of \textit{Numerical Recipes} \cite{Press1996}, while the outermost 1-dimensional integral over $x$ can be done using standard numerical integration methods (in our case, the Cuhre routine included in the Cuba library).
We have deferred the numerical details of this calculation to Appendix \ref{app:1dfft}.
This method has two free parameters corresponding to the grid resolutions in $k$ and $x$, and we find that the resulting $\gamma$ is positive-definite only when these resolutions are sufficiently high.

\subsubsection*{Summary} 
We have described four different methods for computing the inner product matrix, $\gamma \equiv \llangle Q|Q \rrangle$.
The Monte Carlo routine, called Vegas in the Cuba library, fails to converge to positive-definite $\gamma$, so we do not use it subsequently in this work.
The remaining three methods we have implemented are calculated independently and give numerically different results for $\gamma$, stemming from the fact that each method makes different assumptions and approximations about the inner product.
The voxel and 1D FFT methods assume that the $\bk_i$ wavevectors are sampled very finely, such that the inner product is effectively a continuous integral.
These two methods are still completely different in their numerical implementation.
In contrast, the 3D FFT method assumes each $\bk_i$ is discretely sampled in three dimensions, and so it is the only method that accounts explicitly for the discrete sampling of Fourier space, treating this sampling in the same way that the modal estimator is applied to the data through $\llangle Q_n | w\mathcal{B} \rrangle$ in eq.~\eqref{eq:QwB estimator with Mrx}.
Our benchmark constraints use the 3D FFT method, and part of Section \ref{subsec:checks} investigates whether the methods described here have an impact on the resulting parameter constraints.

\subsection{Bispectrum model and custom modes}
\label{subsec:model}

We use a tree-level standard perturbation theory (SPT) model for the real-space halo bispectrum that matches the modeling in \cite{Oddo:2019run}:
\begin{eqnarray}
	B_h(\bk_1,\bk_2,\bk_3) = && b_1^3 B_m(\bk_1,\bk_2,\bk_3) + b_2 b_1^2 \Sigma(\bk_1,\bk_2,\bk_3) + 2 \gamma_2 b_1^2 K(\bk_1,\bk_2,\bk_3) \nonumber \\
	&& + \frac{1+\alpha_1}{\overline{n}} b_1^2 [P_L(k_1)+P_L(k_2)+P_L(k_3)]
	+ \frac{1+\alpha_2}{\overline{n}^2},
	\label{eq:Bh}
\end{eqnarray}
where the tree-level matter bispectrum is
\begin{align}
	B_m(\bk_1,\bk_2,\bk_3) = 2[&F_2(\bk_1,\bk_2) P_L(k_1) P_L(k_2) \nonumber \\
	+ &F_2(\bk_1,\bk_3) P_L(k_1) P_L(k_3) \nonumber \\
	+ &F_2(\bk_2,\bk_3) P_L(k_2) P_L(k_3)]
\end{align}
and $P_L(k)$ is the linear matter power spectrum.
$b_1$ and $b_2$ are the linear and quadratic bias parameters, while $\gamma_2$ is the tidal bias.
The shot noise terms in the second row of eq.~\eqref{eq:Bh} are parametrized by $\alpha_1$ and $\alpha_2$, such that $\alpha_1=\alpha_2=0$ correspond to Poissonian shot noise.
The kernel definitions are
\begin{eqnarray}
	F_2(\bk_1,\bk_2) \equiv && \frac{5}{7} + \frac{1}{2} \frac{\bk_1 \cdot \bk_2}{k_1k_2} \left( \frac{k_1}{k_2} + \frac{k_2}{k_1} \right) + \frac{2}{7} \left( \frac{\bk_1 \cdot \bk_2}{k_1k_2} \right)^2 \label{eq:F2} \\
	\Sigma(\bk_1,\bk_2,\bk_3) \equiv && P_L(k_1)P_L(k_2) + P_L(k_1)P_L(k_3) + P_L(k_2)P_L(k_3) \\
	K(\bk_1,\bk_2,\bk_3) \equiv && \left[ (\hat{\bk}_1 \cdot \hat{\bk}_2)^2 - 1\right] P_L(k_1)P_L(k_2) \nonumber \\
	&+&\left[ (\hat{\bk}_1 \cdot \hat{\bk}_3)^2 - 1\right] P_L(k_1)P_L(k_3) \nonumber \\
	&+&\left[ (\hat{\bk}_2 \cdot \hat{\bk}_3)^2 - 1\right] P_L(k_2)P_L(k_3).
\end{eqnarray}

In this work, we consider the M5 model in \cite{Oddo:2019run}, where the cosmological parameters are fixed and we only vary the bias and shot noise parameters, $(b_1,b_2,\gamma_2,\alpha_1,\alpha_2)$. 
This allows the theory predictions for the halo bispectrum to be computed very quickly, as a linear combination of precomputed terms corresponding to the $B_m$, $\Sigma$, $K$, and $P_L$ terms in eq.~\eqref{eq:Bh} after accounting for binning effects.
Similarly, in this work we require fast theoretical predictions for the modal coefficients, $\beta^Q_n$, and we achieve this by taking advantage of \textit{custom modes}.

First proposed in \cite{Hung:2019ygc}, custom modes are a set of four separable basis functions that by design reproduce exactly the tree-level matter bispectrum.
We can see that this should be possible by rewriting the $F_2$ perturbation theory kernel in eq.~\eqref{eq:F2} as
\begin{equation}
	F_2(k_1,k_2,k_3) = \frac{5}{7} + \frac{1}{2} \left(\frac{k_3^2-k_1^2 - k_2^2}{2k_1k_2}\right)\left(\frac{k_1}{k_2} + \frac{k_2}{k_1}\right) + \frac{2}{7}\left(\frac{k_3^2-k_1^2 - k_2^2}{2k_1k_2}\right)^2,
\end{equation}
where, after expanding this expression, we see that each term will be separable in $k_1$, $k_2$, and $k_3$.
More specifically, the weighted bispectrum, $wB_m$, can be written as a linear combination of four modes,
\begin{eqnarray}
	Q_0^{\rm tree}(k_1,k_2,k_3) &=& q_{\{0}^{\rm tree}(k_1) q_1^{\rm tree}(k_2) q_{1\}}^{\rm tree}(k_3) \\
	Q_1^{\rm tree}(k_1,k_2,k_3) &=& q_{\{0}^{\rm tree}(k_1) q_2^{\rm tree}(k_2) q_{3\}}^{\rm tree}(k_3) \\
	Q_2^{\rm tree}(k_1,k_2,k_3) &=& q_{\{1}^{\rm tree}(k_1) q_3^{\rm tree}(k_2) q_{4\}}^{\rm tree}(k_3) \\
	Q_3^{\rm tree}(k_1,k_2,k_3) &=& q_{\{3}^{\rm tree}(k_1) q_3^{\rm tree}(k_2) q_{5\}}^{\rm tree}(k_3),
\end{eqnarray}
where the custom 1-dimensional basis functions are
\begin{eqnarray}
	q_0^{\rm tree}(k) &=& \sqrt{\frac{k}{P(k)}} \frac{5}{14} \label{eq:q0tree} \\
	q_1^{\rm tree}(k) &=& \sqrt{\frac{k}{P(k)}} P_L(k) \\
	q_2^{\rm tree}(k) &=& -\,\sqrt{\frac{k}{P(k)}} P_L(k) k^2 \\
	q_3^{\rm tree}(k) &=& \sqrt{\frac{k}{P(k)}} \frac{P_L(k)}{k^2} \\
	q_4^{\rm tree}(k) &=& \sqrt{\frac{k}{P(k)}} \frac{3}{14} k^2 \\
	q_5^{\rm tree}(k) &=& \sqrt{\frac{k}{P(k)}} \frac{1}{14} k^4. \label{eq:q5tree} 
\end{eqnarray}
We note that it is important to distinguish between $P_L(k)$, which is the linear power spectrum appearing in the tree-level matter bispectrum model, from $P(k)$ (without a subscript) which is the power spectrum appearing in the definition of the weighting function, $w(k_1,k_2,k_3)$.
In addition to the four custom modes above, in this work we add two more, 
\begin{eqnarray}
	Q_4^{\rm tree}(k_1,k_2,k_3) &=& q_{\{0}^{\rm tree}(k_1) q_0^{\rm tree}(k_2) q_{1\}}^{\rm tree}(k_3) \\
	Q_5^{\rm tree}(k_1,k_2,k_3) &=& q_{\{0}^{\rm tree}(k_1) q_0^{\rm tree}(k_2) q_{0\}}^{\rm tree}(k_3),
\end{eqnarray}
to model the shot noise terms.

With these definitions for six custom modes in total, $Q_n^{\rm tree}$ for $n=0,...,5$, we can reproduce the tree-level halo bispectrum model in eq.~\eqref{eq:Bh} corresponding to values of the parameters $(b_1,b_2,\gamma_2,\alpha_1,\alpha_2)$ by choosing the modal coefficients, $\beta_n^{\rm tree}$, as shown in Table \ref{tab:betaQcustom}, i.e.
\begin{equation}
	wB_h(k_1,k_2,k_3) = \sum_{n=0}^{5} \beta_n^{\rm tree} Q_n^{\rm tree}(k_1,k_2,k_3).
\end{equation}
The fact that we can write  the model predictions for $\beta_n^{\rm tree}$ as trivial functions of $(b_1,b_2,$ $\gamma_2,\alpha_1,\alpha_2)$ means that, as in the standard bispectrum analysis of \cite{Oddo:2019run}, the calculation of the model predictions is very fast, allowing for MCMC simulations to run quickly.

We state for emphasis that this expansion of the halo bispectrum model is \textit{exact}---it is not an approximation.
In our subsequent analysis, unless otherwise mentioned explicitly, we always use a basis where the first six basis functions are these $Q_n^{\rm tree}$, and starting with the seventh basis function, we use the $Q_n$ that we have introduced earlier, which are either constructed from 1-dimensional normal or Legendre polynomials, which we call $q_n(k)$.

For a general model of the halo bispectrum or a general cosmological parameter set, it may not be possible to predict theoretical values of $\beta^Q_n$ as quickly as what we use here.
Finding a general strategy to manage this problem is outside the scope of this work, but it is an interesting challenge for future work.
We note that this computational bottleneck has a counterpart in the standard bispectrum pipeline, where it is necessary to quickly calculate predictions for $B(k_1,k_2,k_3)$ in all triangle bins, accounting for the bin size, for a general bispectrum model and parameter set.

\begin{center}
\begin{table}
\begin{center}
{\renewcommand{\arraystretch}{1.5}%
  \begin{tabular}{lccccc}
  \hline
  $\beta_0^{\rm tree}=$ & $6b_1^3$ & $+\frac{42}{5}b_1^2b_2$ & $-\frac{42}{5}b_1^2\gamma_2$ & & \\
  $\beta_1^{\rm tree}=$ & $6b_1^3$ & & $-\frac{42}{5}b_1^2\gamma_2$ &&\\
  $\beta_2^{\rm tree}=$ & $6b_1^3$ & & $-28b_1^2\gamma_2$ &&\\
  $\beta_3^{\rm tree}=$ & $6b_1^3$ & & $+21b_1^2\gamma_2$ &&\\
  $\beta_4^{\rm tree}=$ & & & & $\frac{588}{25} b_1^2 \left( \frac{1+\alpha_1}{\overline{n}} \right)$&\\
  $\beta_5^{\rm tree}=$ & & & & & $\frac{2744}{125} \left( \frac{1+\alpha_2}{\overline{n}^2} \right)$ \\
  \hline
  \end{tabular}}
  \end{center}
  \caption[Theoretical predictions for modal coefficients]{For an input model with free parameters $(b_1,b_2,\gamma_2,\alpha_1,\alpha_2)$, we list the modal coefficients for the six custom modes, $Q_n^{\rm tree}$ for $n=0,...,5$, that reproduce the tree-level halo bispectrum model in eq.~\eqref{eq:Bh}.}
  \label{tab:betaQcustom}
\end{table}
\end{center}

\subsection{Orthonormal basis}
\label{subsec:orthonormal}

Finally, in this section we introduce one more set of basis functions that are rotations of any general set of separable $Q_n$ (which may include custom modes).
We label this new basis $R_n$ and call it the orthonormal basis because it satisfies
\begin{equation}
	\llangle R_n | R_m \rrangle = \frac{(k_{\rm max}-k_{\rm min})^3}{8\pi^4} \delta_{nm}.
	\label{eq:RR equals delta}
\end{equation}
We comment that the factor of $(k_{\rm max}-k_{\rm min})^3/(8\pi^4)$ on the right hand side only changes the overall amplitude of all $R_n$ by a constant factor, and it is only present here because we made the arbitrary choice to require that the $R_n$ basis is orthonormal in the unit tetrapyd space, i.e.~$\langle R_n | R_m \rangle = \delta_{nm}$.

Then we require that $R_n$ is a linear combination of the $Q_n$ basis as 
\begin{equation}
	R_n \equiv \sum_m \lambda^{-1}_{nm} \, Q_m,
	\label{eq:R equals lambda Q}
\end{equation}
and by deriving $\llangle R|R \rrangle$ and setting it equal to the identity matrix, we see that $\gamma$ and $\lambda$ are related by
\begin{equation}
\gamma = \frac{(k_{\rm max}-k_{\rm min})^3}{8\pi^4} \lambda \cdot \lambda^T,	
\end{equation}
where $\lambda$ is the lower triangular matrix resulting from the Cholesky decomposition.

\begin{figure}[th]
  \begin{subfigure}[b]{\textwidth}
    \includegraphics[width=0.33\textwidth]{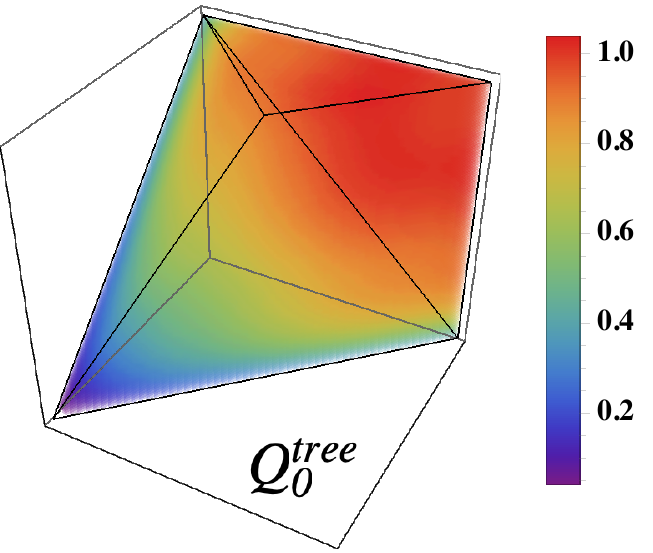}%
    \includegraphics[width=0.33\textwidth]{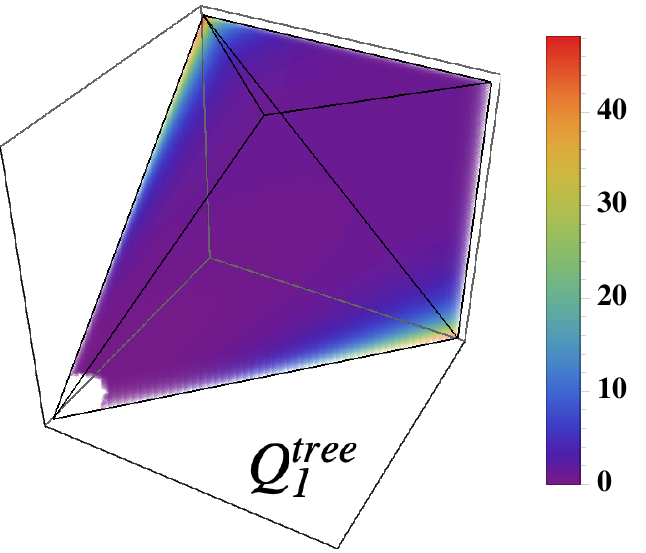}%
    \includegraphics[width=0.33\textwidth]{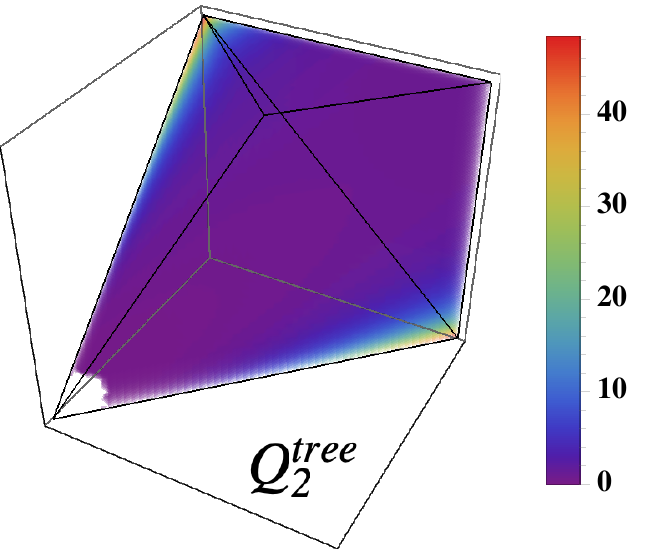}%
  \end{subfigure}
  \begin{subfigure}[b]{\textwidth}
    \includegraphics[width=0.33\textwidth]{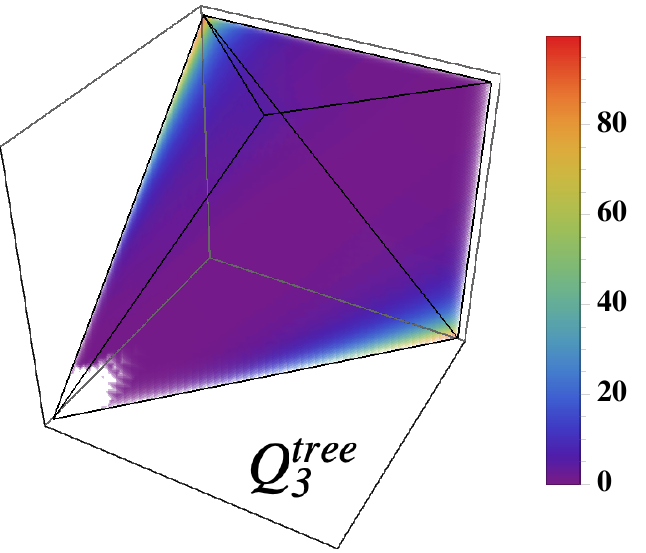}%
    \includegraphics[width=0.33\textwidth]{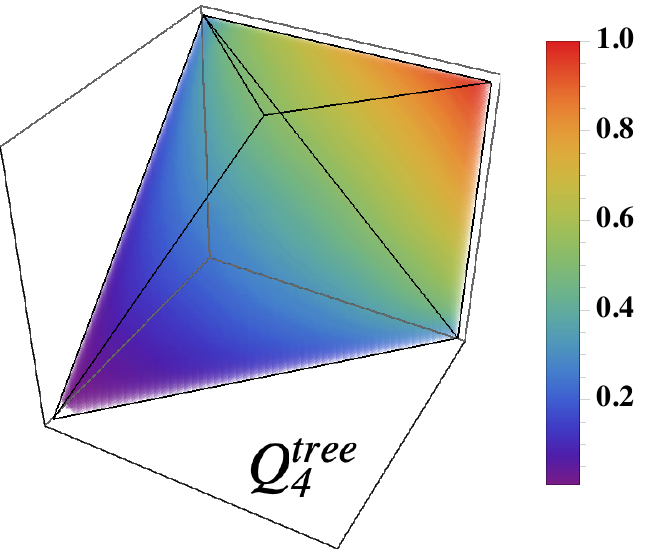}%
    \includegraphics[width=0.33\textwidth]{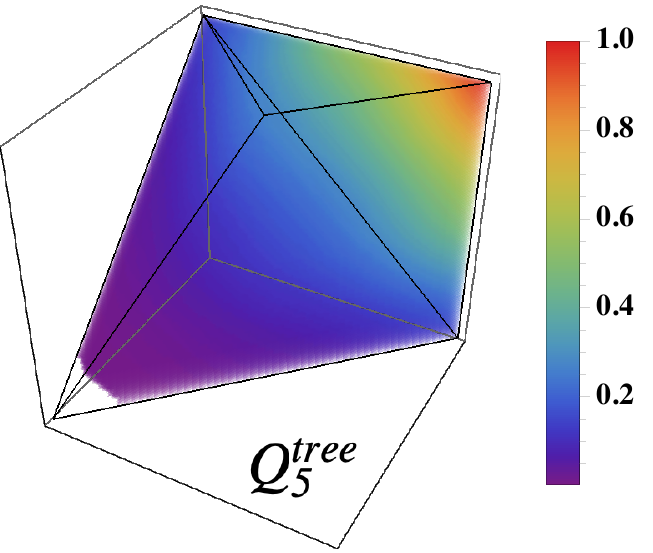}%
  \end{subfigure}
  \begin{subfigure}[b]{\textwidth}
    \includegraphics[width=0.33\textwidth]{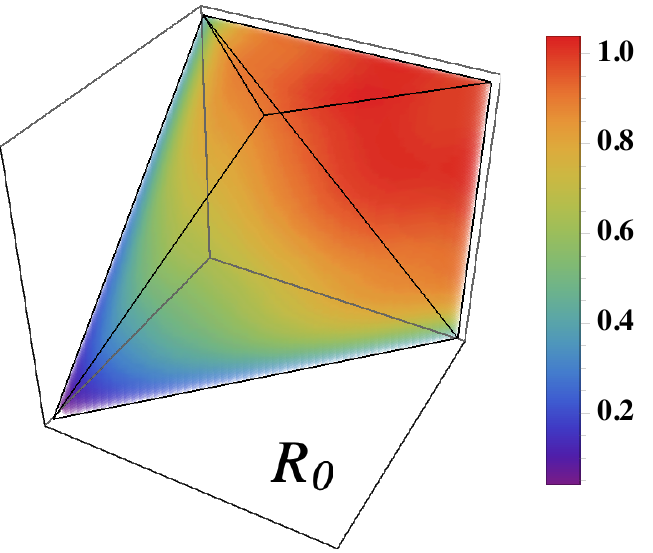}%
    \includegraphics[width=0.33\textwidth]{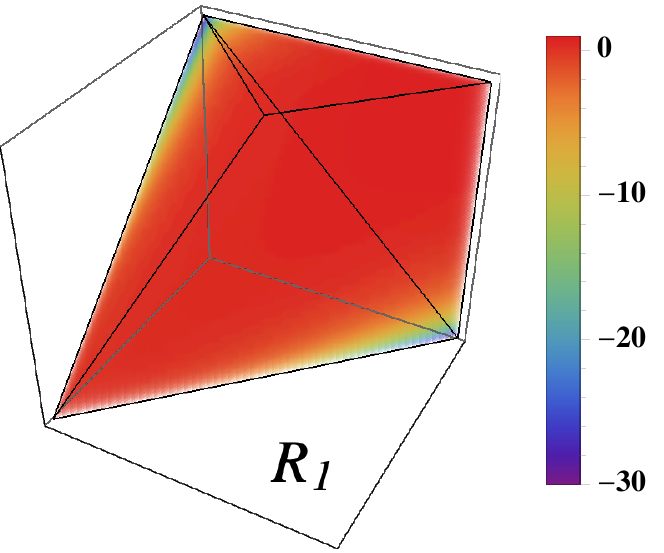}%
    \includegraphics[width=0.33\textwidth]{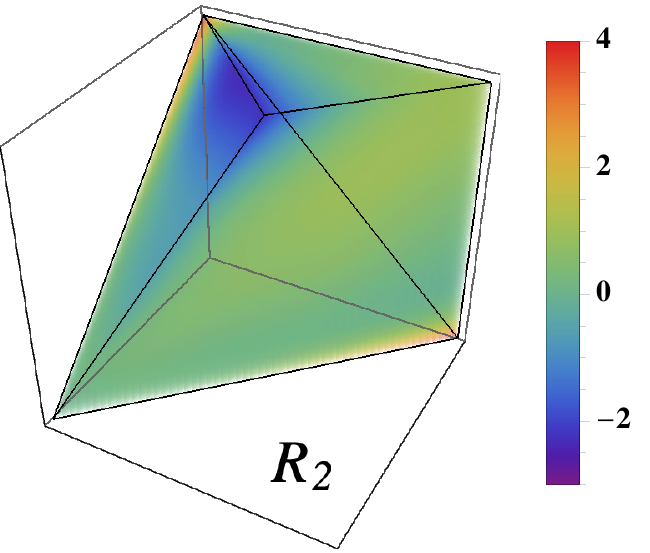}%
  \end{subfigure}
  \begin{subfigure}[b]{\textwidth}
    \includegraphics[width=0.33\textwidth]{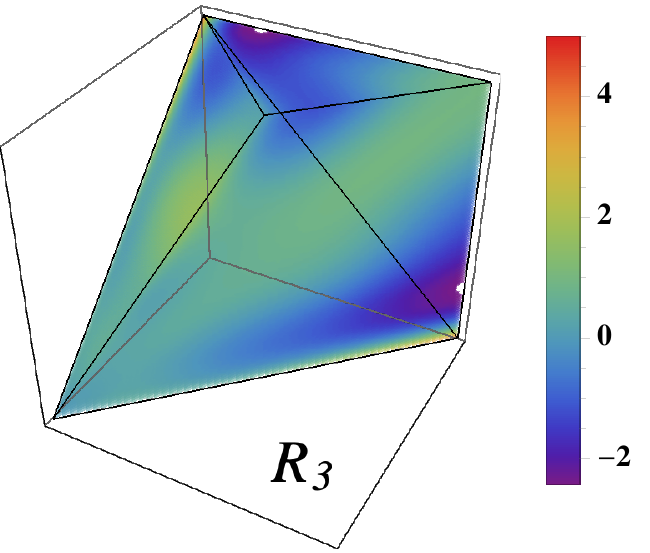}%
    \includegraphics[width=0.33\textwidth]{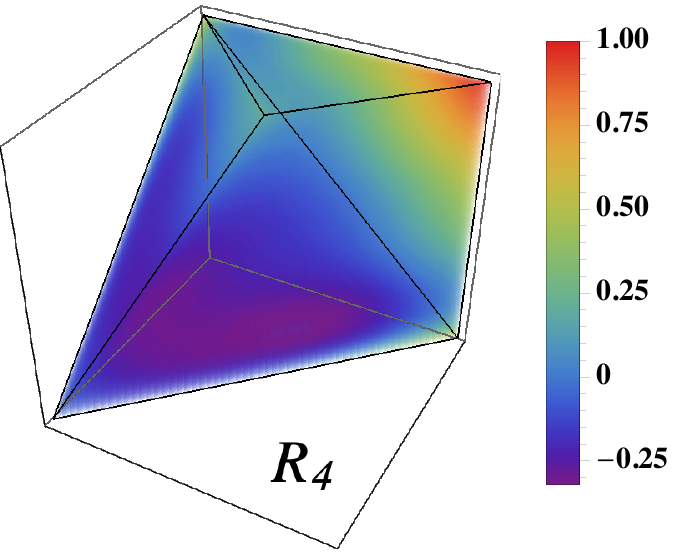}%
    \includegraphics[width=0.33\textwidth]{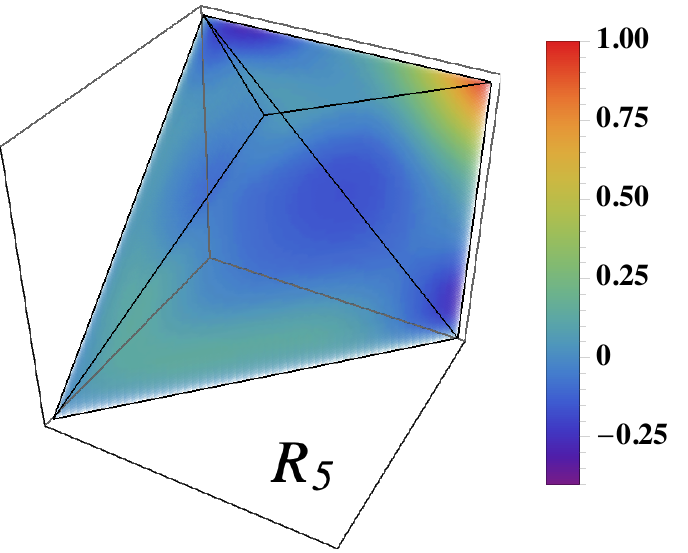}%
  \end{subfigure}
  \caption[$Q_n$ and $R_n$ basis functions]{Plots of the six custom $Q_n$ basis functions (first two rows) and the first six $R_n$ basis functions (bottom two rows) for $k_{\rm max} \approx 0.10 \, h\,{\rm Mpc}^{-1}$.
  Each function is plotted over the unit tetrapyd of allowed $(x_1,x_2,x_3)$ combinations, with the origin in the lower left corner, and for easier readability is normalized to unity at $x_1=x_2=x_3=1$ in the upper right corner.
  These plots show only half of the tetrapyd to show more of the interior region.}
  \label{fig:QnRn}
\end{figure}

Fig.~\ref{fig:QnRn} shows the six custom modes $Q_n^{\rm tree}$ and the first six $R_n$ for $k_{\rm max} \approx 0.10 \, h \,{\rm Mpc}^{-1}$, with the 4-dimensional data represented in a plot similar in style to \cite{Fergusson:2009nv}.\footnote{The $R_n$ shown in the figure were calculated the default modal settings described in detail at the beginning of Section \ref{sec:results}.}
The three axes in each plot are the $x_i$ defined by $(k_i - k_{\rm min})/(k_{\rm max} - k_{\rm min})$, where the origin is in the lower left corner.
To focus on the overall triangle-dependence of each basis function, we have normalized each one to equal unity when $x_1=x_2=x_3=1$ in the upper right corner.
We have also removed the half of the tetrapyd with $x_1 > x_2$, if the vertical axis is $x_3$, to show the interior of the tetrapyd region.
We notice that the $Q^{\rm tree}_1$, $Q^{\rm tree}_2$, and $Q^{\rm tree}_3$ modes are similar and peak (in red) at the edges of the tetrapyd corresponding to squeezed triangles, while the other three custom modes are largest at equilateral triangles with $k_1=k_2=k_3=k_{\rm max}$.
The plots showing $Q^{\rm tree}_0$ and $R_0$ are identical because $R_0 \propto Q^{\rm tree}_0$ by construction.
Unlike the $Q_n^{\rm tree}$, however, all of the $R_n$ look dissimilar because they have been defined to be orthogonal to each other, i.e. $\langle R_n | R_m \rangle = \delta_{nm}$.

Using the fact that the weighted bispectrum must be the same regardless of the basis,
\begin{equation}
	\sum_n \beta^Q_n \, Q_n = \sum_m \beta^R_m \, R_m,
\end{equation}
we see that the two sets of coefficients are related by
\begin{equation}
	\beta^R = \lambda^T \cdot \beta^Q.
\end{equation}

In the modal pipeline of this work, it is never necessary to calculate anything with $R_n$ itself directly. We just use its definition to work with the $\beta^R$ coefficients as our data, rather than the $\beta^Q$.
Whether we do our MCMC analysis in terms of $\beta^Q$ or $\beta^R$ does not matter, but the $\beta^R$ have the advantage that the numerical values of the $\beta^R_n$ coefficients for $n < N_{\rm modes}$ will not change if the size of the basis, given by $N_{\rm modes}$, is increased.
This is because $\lambda^{-1}$ in eq.~\eqref{eq:R equals lambda Q} is lower triangular, so $R_n$ only depends on $Q_m$ with $m \leq n$, and $\beta^R$ can also be expressed as $\llangle R_n | wB \rrangle = (k_{\rm max}-k_{\rm min})^3/(8\pi^4) \beta^R_n$.
On the other hand, we can see from $\llangle Q | wB \rrangle = \gamma \cdot \beta^Q$ that all numerical values of $\beta^Q$ will change as the basis set is increased.

The definition of $R_n$ in eq.~\eqref{eq:RR equals delta} corresponds to defining $\beta^R$ that are orthogonal in the limit of Gaussian covariance.
We note that the $\beta^R_n$ can be written as
\begin{eqnarray}
	\beta^R_n &=& \frac{8\pi^4}{(k_{\rm max}-k_{\rm min})^3} \llangle R_n | w\mathcal{B} \rrangle \\
	&=& \frac{8\pi^4}{(k_{\rm max}-k_{\rm min})^3} \frac{1}{V} 
		\int_{\bk_1}
	\int_{\bk_2}
	\int_{\bk_3} 
	(2\pi)^3 \delta_D(\bk_{123}) 
	\frac{R_n(k_1,k_2,k_3) \, \delta_{\bk_1} \delta_{\bk_2} \delta_{\bk_3}}
	{\sqrt{k_1k_2k_3}\sqrt{P(k_1)P(k_2)P(k_3)}},
\end{eqnarray}
such that the covariance $\left< \beta^R_n \beta^R_m \right>$ requires evaluating $\langle \delta_{\bk_1} \delta_{\bk_2} \delta_{\bk_3} \delta_{\bk_1'} \delta_{\bk_2'} \delta_{\bk_3'} \rangle$. The leading-order Gaussian contribution to this is
\begin{equation}
	\langle \delta_{\bk_1} \delta_{\bk_2} \delta_{\bk_3} \delta_{\bk_1'} \delta_{\bk_2'} \delta_{\bk_3'} \rangle_G = 6 (2\pi)^9 \delta_D(\bk_1+\bk_1') \delta_D(\bk_2+\bk_2') \delta_D(\bk_3+\bk_3') P(k_1)P(k_2)P(k_3),
	\label{eq:gaussian 6pt correlator}
\end{equation}
such that the Gaussian covariance for the modal coefficients is
\begin{align}
	\left< \beta^R_n \beta^R_m \right>_G &= \frac{6}{V} \left[ \frac{8\pi^4}{(k_{\rm max}-k_{\rm min})^3} \right]^2 \llangle R_n | R_m \rrangle \\
	&= \frac{6}{V}\frac{8\pi^4}{(k_{\rm max}-k_{\rm min})^3} \delta_{nm}.
	\label{eq:betaR gaussian covariance}
\end{align}
Non-Gaussian contributions to the covariance will generally couple orthonormal modal coefficients with different $n$ and $m$.
Later, in Section \ref{subsec:cov}, we evaluate the impact of assuming the Gaussian covariance on the parameter constraints.

\subsection{Summary}
\label{subsec:method summary}

Here we put together the practical steps of the modal method necessary to implement it, and summarize the expressions necessary to estimate and model modal coefficients, $\beta^Q$ and $\beta^R$, with or without custom modes included.

In the first step, we choose a $k$-range, bounded by $k_{\rm min}$ and $k_{\rm max}$, and we choose the 1-dimensional basis functions $q_n(k)$ that are combined to get the separable basis of $Q_n$, which may or may not include custom modes, depending on the choice of the user and the bispectrum model.
After the $Q_n$ have been defined, $\gamma$ is computed for this basis, using a chosen method---we have discussed four options in Section \ref{subsec:gamma_methods}.
Then we use the Cholesky decomposition to numerically calculate $\lambda$, which defines the orthonormal $R_n$ basis.

In the second step, we obtain our measurements from simulations. This is done by measuring $\llangle Q | w\mathcal{B} \rrangle$, and then solving 
\begin{align}
	\llangle Q | w\mathcal{B} \rrangle &= \gamma \cdot \hat{\beta}^Q \\
	\llangle Q | w\mathcal{B} \rrangle &= \frac{(k_{\rm max}-k_{\rm min})^3}{8\pi^4} \lambda \cdot \hat{\beta}^R
\end{align}
to get the modal coefficients. 

In the last step, we need a function for predicting $\beta^Q(\theta)$ and $\beta^R(\theta)$.
When custom modes are used, we set the modal coefficients to be
\begin{align}
  \beta^Q_n(\theta)&=\left\{
  \begin{array}{@{}ll@{}}
    \beta^{\rm tree}_n(\theta) & \text{if}\ 0\leq n \leq 5 \\
    0 & \text{otherwise}
  \end{array}\right. \\
  \beta^R(\theta) &= \lambda^T \cdot \beta^Q(\theta).
\end{align} 
When custom modes are \textit{not} used, we have to estimate the $\beta^Q$ in
\begin{align}
	\sum_{n=0}^5 \beta^{\rm tree}_n(\theta) \, Q_n^{\rm tree} \approx \sum_m \beta^Q_m \, Q_m,
\end{align}
where the sum over $m$ on the right side does not include any custom modes.
Taking the inner product of this with $Q$ on both sides (where $Q$ again does not include any custom modes), we find that we need to solve
\begin{align}
	\llangle Q | Q^{\rm tree} \rrangle \cdot \beta^{\rm tree}(\theta) &= \gamma \cdot \beta^Q(\theta) \\
	\llangle Q | Q^{\rm tree} \rrangle \cdot \beta^{\rm tree}(\theta) &= \frac{(k_{\rm max}-k_{\rm min})^3}{8\pi^4} \lambda \cdot \beta^R(\theta)
\end{align}
for $\beta^Q$ and/or $\beta^R$.
Therefore, when custom modes are not included in the modal basis, we need to precompute the rectangular matrix $\llangle Q|Q^{\rm tree} \rrangle$ as a means to obtaining theory predictions quickly.

\section{Data and analysis}
\label{sec:data_and_analysis}

\subsection{Simulations and mock halo catalogs}
\label{subsec:data}

We use the same simulation and halo catalog data as in \cite{Oddo:2019run}, since the aim of our work is to compare the modal bispectrum constraints with the results from the standard bispectrum in that work.
The data consist of two sets of simulations.
The first is a suite of 298 $N$-body simulations, called Minerva, created using the Gadget-2 code and first presented in \cite{Grieb:2015bia}.
Each realization is a $L=1500 \,h^{-1}\,{\rm Mpc}$ box evolved to $z=1$ based on the same fiducial flat $\Lambda$CDM cosmology.
Halos are identified using a friends-of-friends algorithm such that the minimum halo mass is $1.12 \times 10^{13} \, h^{-1}\,M_\odot$, and the mean number density is $\overline{n} = 2.13 \times 10^{-4} \, h^3\,{\rm Mpc}^{-3}$.
The measurements from these Minerva simulations are the data that we fit in our analysis.

We also use a set of 10,000 mock halo catalogs generated using the approximate $N$-body code Pinocchio \cite{Monaco:2001jg,Monaco:2013qta,Munari:2016aut}, which in \cite{Oddo:2019run} were used to obtain bispectrum covariance matrices.
298 of these realizations have initial conditions that match those of the Minerva realizations.
The halos in the Pinocchio simulations were chosen with a different mass threshold (compared to the Minerva data) such that the large-scale amplitude of the total halo power spectrum matches what is measured in the Minerva $N$-body simulations, because in the Gaussian limit the bispectrum covariance depends directly on the total power spectrum \cite{Oddo:2019run}.

For the modal estimator, as with the bispectrum measurements in \cite{Oddo:2019run}, we map the halo positions to the grid of halo density values using the 4th-order interlacing method in \cite{Sefusatti:2015aex} obtained with the public PowerI4\footnote{\url{https://github.com/sefusatti/PowerI4}} code and run the estimator using a FFT grid size of $N_g = 256$.

We note that by simultaneously fitting to 298 Minerva realizations, the results we present in Section \ref{sec:results} correspond to a total volume of $\approx 1{,}000 \, h^{-3}\,{\rm Gpc}^3$, which is much larger than any real galaxy survey planned for the near future. 
Therefore the results we present should be interpreted as a proof of principle of the modal bispectrum method, while the exact numerical values of the parameter constraints will change for more realistic survey scenarios in smaller volumes.

\subsection{Likelihood and MCMC}
\label{subsec:likelihood}

We implement two likelihood functions: a Gaussian likelihood, which we take to be our benchmark case, and the likelihood proposed by Sellentin and Heavens in  \cite{Sellentin:2015waz} (SH in what follows).
The two likelihoods differ in how they account for errors in the estimated covariance matrix, due to the fact that it is estimated from a finite number of mocks, but they both assume that the observable data is Gaussian distributed. 
We have checked that the probability distribution functions of the $\beta^R_n$ modal coefficients measured in the $N$-body simulations and mock catalogs do not show any strong indications of non-Gaussianity.
In Section \ref{subsec:shlike}, we compare results from the two likelihoods.

The Gaussian likelihood for our analysis is
\begin{equation}
	\ln \mathcal{L} = - \frac{1}{2} \sum_n \sum_m \delta\beta^R_n \; \hat{C}^{-1}_{nm} \; \delta\beta^R_m
	\equiv - \frac{1}{2} (\delta\beta^R)^T \cdot \hat{C}^{-1} \cdot \delta\beta^R,
	\label{eq:lnL gaussian}
\end{equation}
where $\delta\beta^R_n \equiv \beta^R_n(\theta) - \beta^{R, \,{\rm sims}}_n$. 
The covariance matrix estimated from $N_s$ mock catalogs is
\begin{equation}
	\widetilde{C}_{nm} \equiv \frac{1}{N_s-1} \sum_i^{N_s} (\beta^{R(i)}_n - \overline{\beta}^R_n)
	(\beta^{R(i)}_m - \overline{\beta}^R_m).
\end{equation}
Though this is an unbiased estimator of the covariance matrix, taking the inverse of this $\widetilde{C}_{nm}$ will result in a biased estimate of the precision matrix, which can be statistically debiased by a multiplicative factor \cite{Kaufmann1967,Anderson2003,Hartlap2007},
\begin{equation}
	\hat{C}^{-1} = \Gamma \, \widetilde{C}^{-1},
\end{equation}
where
\begin{equation}
\Gamma \equiv \frac{N_s - N_{\rm bins} -2}{N_s-1}
\end{equation}
and $N_{\rm bins}$ is the number of data bins.
We note however that any one particular $\widetilde{C}$ will have statistical noise such that applying the $\Gamma$ factor may actually bring the final parameter constraints closer to, or further away from, what we would have obtained using the true precision matrix.
In other words, the $\Gamma$ factor does not take into account the statistical nature of the estimated precision matrix---the fact that it is still a noisy estimate of an unknown quantity.

Instead, SH derives a likelihood that is the Gaussian likelihood marginalized over the unknown covariance matrix, conditioned on our estimate of it, to arrive at
\begin{equation}
	\ln \mathcal{L} = -\frac{N_s}{2} \ln \left[ 1 + \frac{(\delta\beta^R)^T \cdot \widetilde{C}^{-1} \cdot \delta\beta^R}{N_s-1} \right] + \ln\left(\frac{\overline{c}_p}{\sqrt{\det \widetilde{C}}}\right).
	\label{eq:lnL ng}
\end{equation}
$\overline{c}_p$ is a constant that depends only on $N_s$ and $N_{\rm bins}$, and since we assume the covariance matrix does not depend on the parameters, we drop the second $\ln(...)$ term.
For one particular estimate of the covariance matrix, this SH likelihood will also have errors that are too large or too small, and be biased.
However, on average the SH likelihood should yield parameter constraints that are closer to the true one.

The two likelihoods should equal each other, and approach the true answer, in the limit that the covariance matrix is well-estimated by a sufficiently large $N_s$.
Conversely, they will show differences when $N_s$ is small, e.g. $\Gamma \ll 1$.

As in \cite{Oddo:2019run}, we simultaneously fit all 298 Minerva realizations, which means that our total likelihood is
\begin{equation}
	\ln \mathcal{L}_{\rm total} = \sum_i^{N_R} \ln \mathcal{L}_i,
\end{equation}
where $\ln \mathcal{L}_i$ is the likelihood for one realization, and equal to eq.~\eqref{eq:lnL gaussian} or eq.~\eqref{eq:lnL ng}.

We use wide, uniform priors for all parameters, and we compare our results with those from \cite{Oddo:2019run} using their `broad' priors, which were $b_1 \in [0.5,5]$, $b_2,\gamma \in [-5,5]$, $\alpha_1 \in [-10,10]$, and $\alpha_2 \in [-100,100]$.
The analysis in \cite{Oddo:2019run} also explored the effect of narrower priors on $\alpha_1$ and $\alpha_2$, as well as fitting the simulation data with models with fewer than five parameters. 
However, in this work we only consider the five parameter model (called M5 in \cite{Oddo:2019run}) with broad priors, because we expect that a modal estimator pipeline fitting for more parameters with less informative priors will present a stronger test of the modal method.

Our MCMC simulations are run using the code emcee\footnote{\url{https://emcee.readthedocs.io}} \cite{ForemanMackey:2012ig}. 
Each chain uses 100 walkers that are started within a small sphere around an approximate maximum likelihood point. 
Our convergence criteria are that the integrated autocorrelation time $\tau$ is stable to within 1\% and that the chain is at least $50\tau$ long. 
After the chains have converged, we use the getdist\footnote{\url{https://getdist.readthedocs.io}} package to analyze the chains and produce contour plots of the posteriors \cite{Lewis:2019xzd}.

\section{Results}
\label{sec:results}

Unless otherwise stated, the results we present use following default settings for the modal estimator pipeline.
The first six $Q_n$ basis functions are the six custom modes defined in Section \ref{subsec:model}, and the rest of the $Q_n$ basis functions are constructed from 1-dimensional $q_n$ functions that are normal polynomials.
The basis functions are defined over the wavenumber range with $k_{\rm min}=1.5 \, k_f \approx 0.006 \, h\,\Mpc^{-1}$ and one of two choices of $k_{\rm max}$, $13.5\,k_f \approx 0.06 \, h\,{\rm Mpc}^{-1}$ or $24.5\,k_f \approx 0.10 \, h\,{\rm Mpc}^{-1}$, where $k_f = 2\pi/L$ is the fundamental wavenumber and $L=1500 \,h^{-1}\,{\rm Mpc}$ is the size of a simulation box.
We use the 3D FFT method to compute the inner product matrix $\gamma$, where the grid resolution is $N_g=256$ and the size of the real-space Fourier volume matches that of the simulation boxes.
The power spectrum that appears in the weighting which defines the basis is the average total halo power spectrum $P_h$ measured in the Minerva simulations.

We present our data consisting of the measured modal coefficients in Section \ref{subsec:mean and cov}.
Sections \ref{subsec:Brec} and \ref{subsec:benchmark} compare the modal bispectrum and standard bispectrum estimators, first by considering the similarities and differences in the bispectra that are measured, and then by comparing their resulting parameter constraints.
The subsequent sections then focus exclusively on the modal bispectrum.
In Section \ref{subsec:checks}, we explore how the parameter constraints from the modal bispectrum can depend on a variety of settings within the modal bispectrum pipeline, while in Section \ref{subsec:correlators} we discuss other ways that the convergence of the modal expansion could be estimated.
Sections \ref{subsec:cov} and \ref{subsec:shlike} quantify how our results depend on the estimated covariance matrices and likelihood modeling that are used in our analysis.

\subsection{Mean and covariance of modal coefficients}
\label{subsec:mean and cov}

\begin{figure}[t]
\centering 
\includegraphics[width=\textwidth]{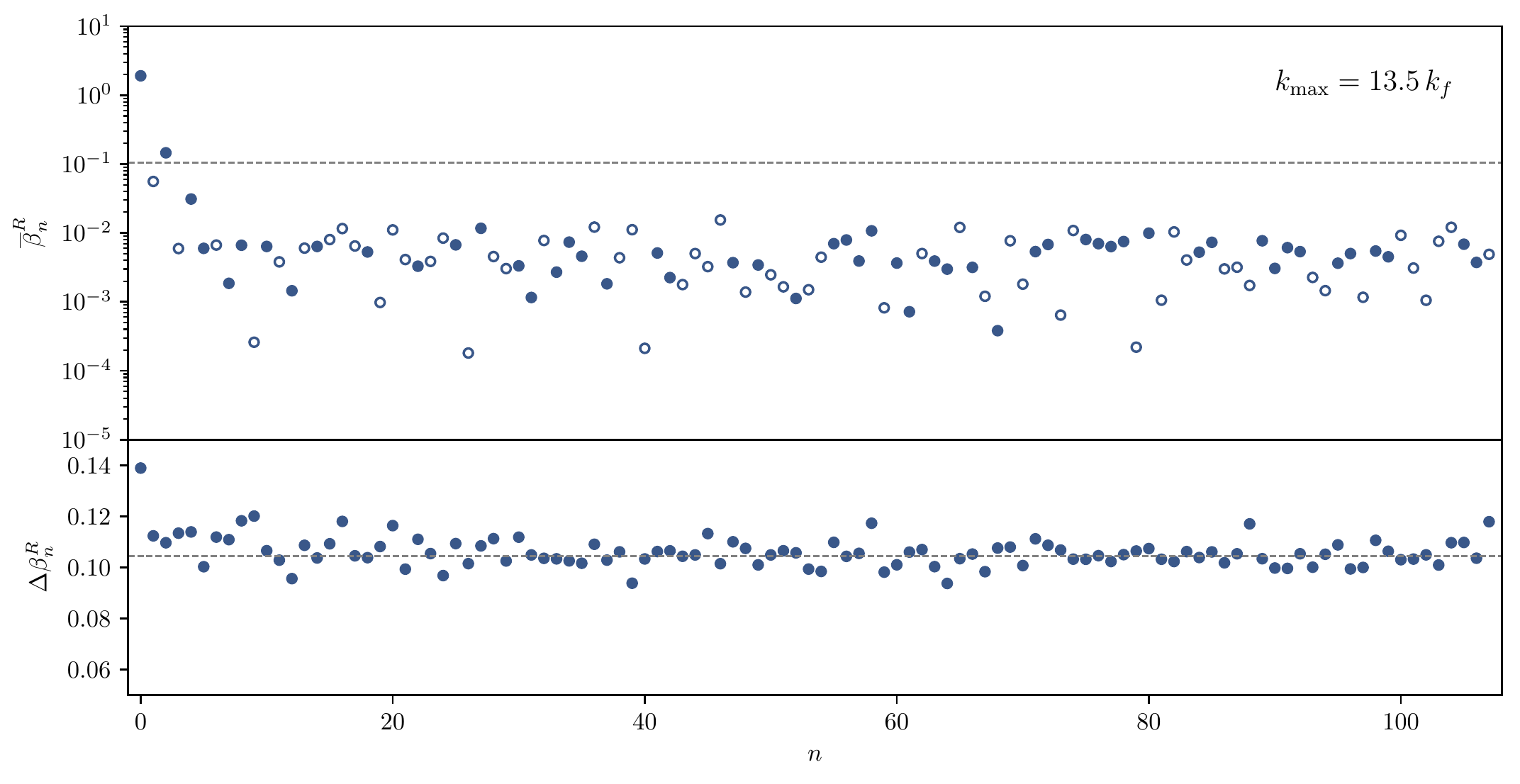}
\includegraphics[width=\textwidth]{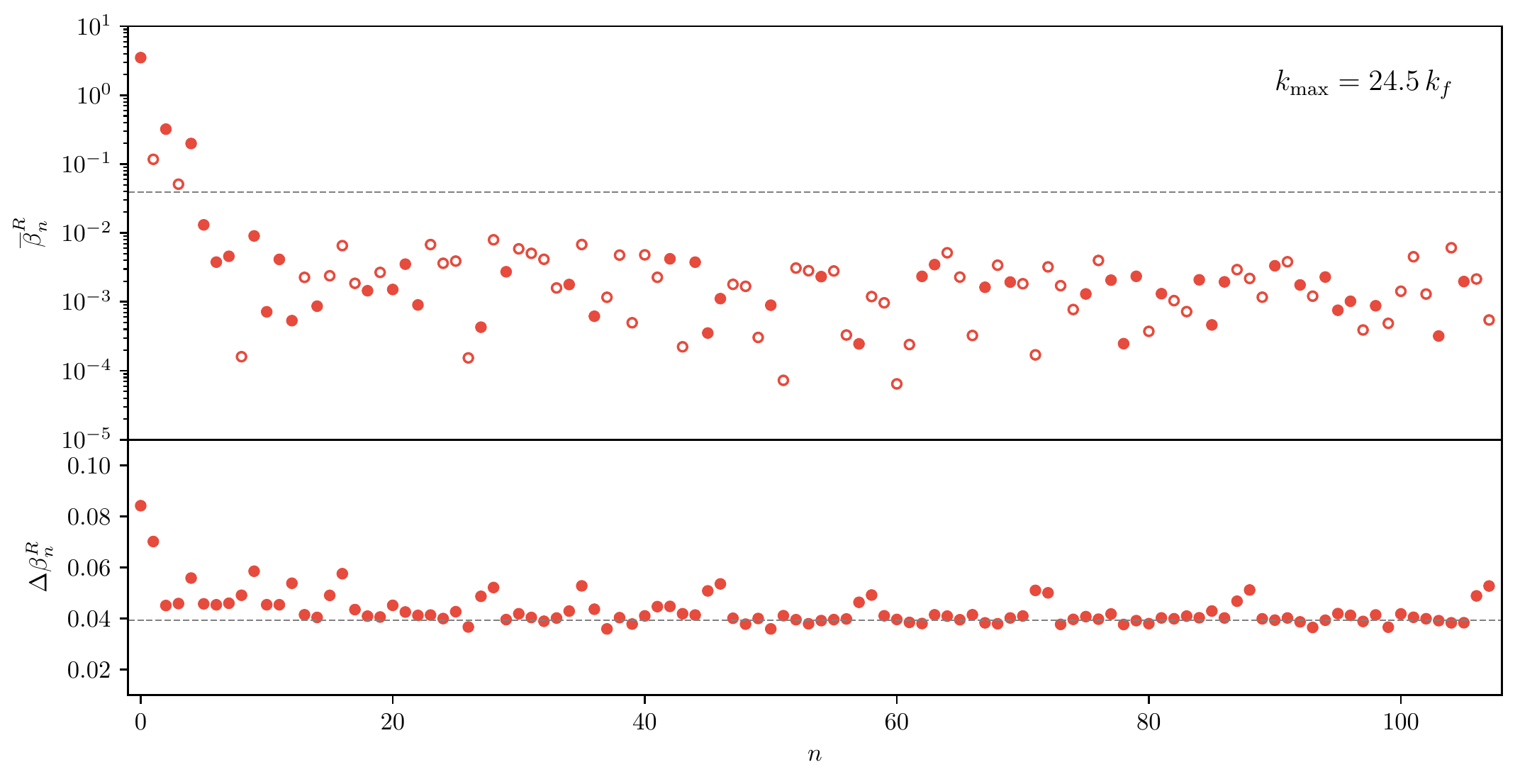}
\hfill
\caption[Modal expansion coefficients measured from Minerva $N$-body simulations]{Means $\overline{\beta}^R_n$ and errors $\Delta\beta^R_n$ of modal expansion coefficients measured from 298 Minerva $N$-body simulations. The two panels show the measurements for $k_{\rm max} = 13.5 \, k_f$ (top, blue circles) and $24.5 \, k_f$ (bottom, red circles). Filled (empty) circles correspond to positive (negative) values. The gray dashed lines correspond to the Gaussian predictions for the error given in eq.~\eqref{eq:betaR gaussian covariance}.}
\label{fig:minerva} 
\end{figure}

In Fig.~\ref{fig:minerva}, we show the means, $\overline{\beta}^R_n$, and errors, $\Delta\beta^R_n$, of modal coefficients that are measured in the 298 realizations of Minerva $N$-body simulations for $k_{\rm max} = 13.5 \, k_f$ and $24.5 \, k_f$, where $k_{\rm min}=1.5 \, k_f$ in both cases.
The dashed gray horizontal lines show the Gaussian predictions for the error in eq.~\eqref{eq:betaR gaussian covariance}, which depends on $k_{\rm max}$.
 
$\overline{\beta}^R_n$ for the first few $n$ are more easily detected, and if $k_{\rm max}$ is higher, more of the first few $n$ have a higher signal-to-noise ratio, $\overline{\beta}^R_n / \Delta\beta^R_n$: for $k_{\rm max} = 13.5 \, k_f$ ($24.5 \, k_f$), two (five) modes have a signal-to-noise ratio greater than 1.
The errors on the coefficients measured from simulations are in good agreement with the Gaussian predictions, but there are some deviations, particularly for higher $k_{\rm max}$ and small $n$.
These observations can be explained with the reasoning that, since clustering is more non-linear on smaller scales, the case with higher $k_{\rm max}$ will have larger $\overline{\beta}^R_n$ and more non-Gaussian $\Delta\beta^R_n$.
The fact that these effects are concentrated at the low $n$ modes can be explained by how the orthonormal basis of $R_n$ functions have been defined.
Since each $R_n$ is by definition a linear combination of $Q_0$, ..., $Q_n$ that is orthogonal to (i.e. in the Gaussian limit, has no covariance with) all previous $R_m$ for $m \leq n$, we generally expect higher $R_n$ modes to have smaller amplitudes in the data, making the Gaussian error approximation more accurate.

\begin{figure}[t]
\centering 
\includegraphics[width=\textwidth]{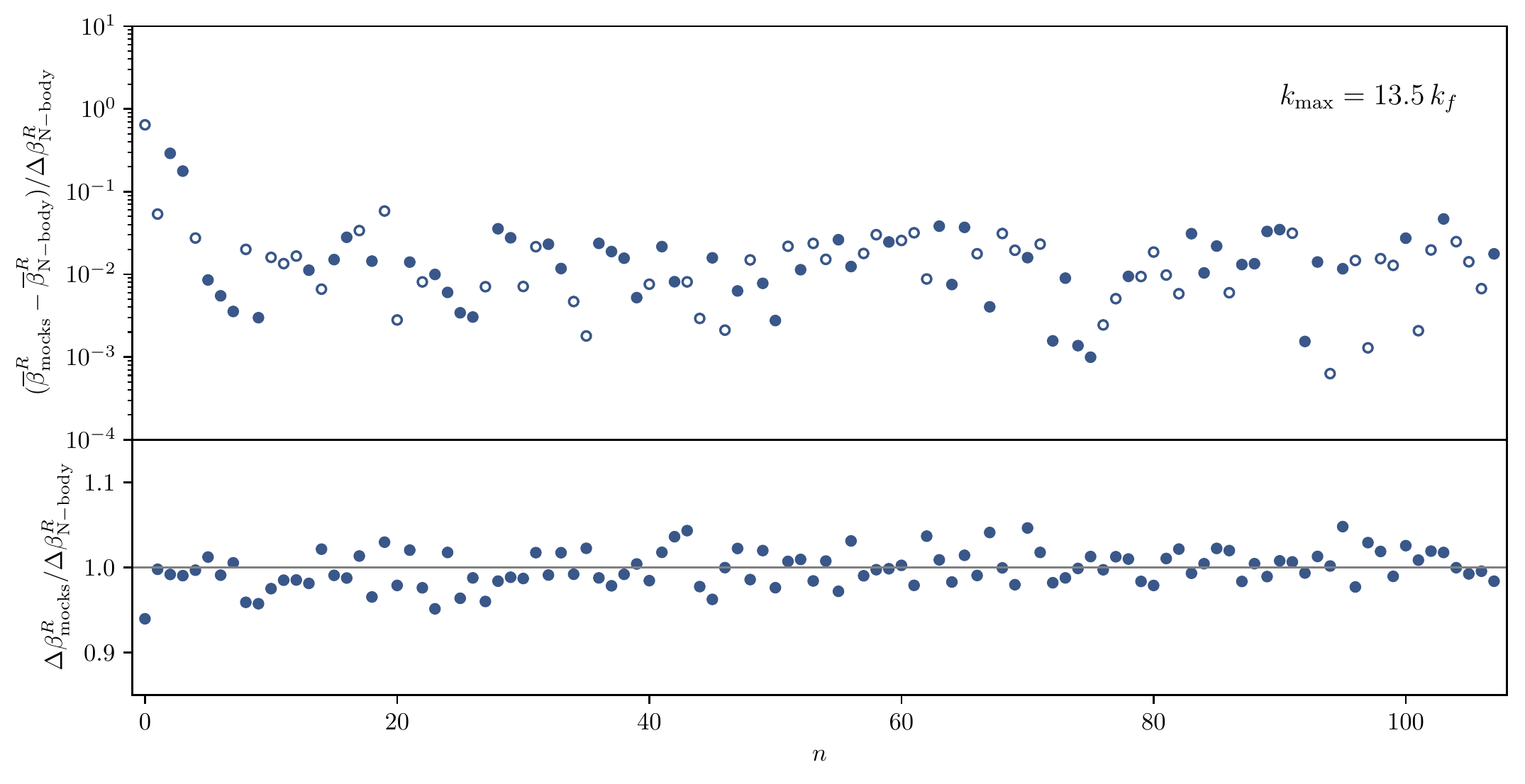}
\includegraphics[width=\textwidth]{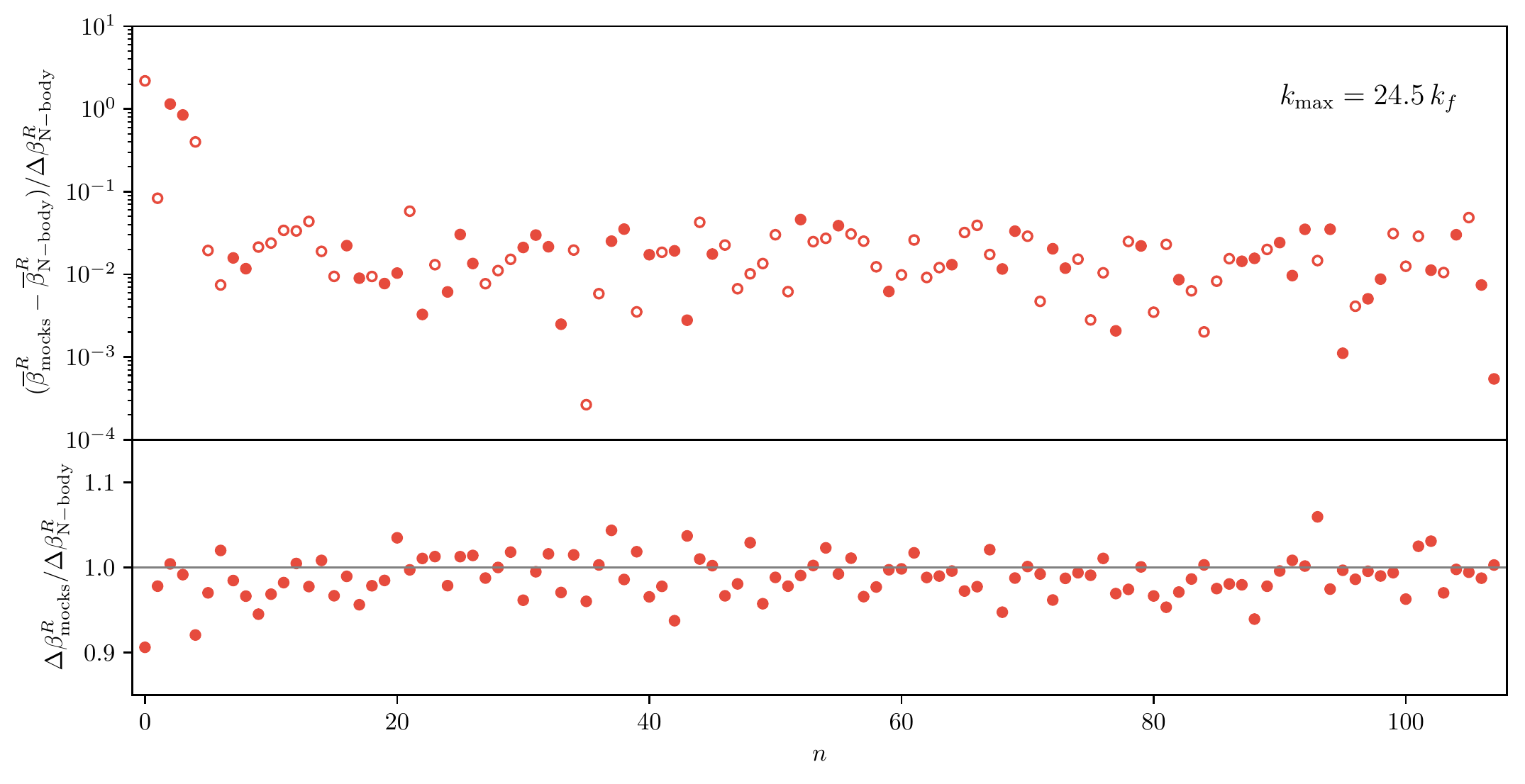}
\hfill
\caption[Modal expansion coefficients measured from Pinocchio mock halo catalogs]{Means and errors of modal expansion coefficients measured from 298 realizations of Pinocchio mock halo catalogs are compared with the same measurements from the Minerva $N$-body simulations with matching initial conditions. The two panels show the measurements for $k_{\rm max} = 13.5 \, k_f$ (top, blue circles) and $24.5 \, k_f$ (bottom, red circles). Filled (empty) circles correspond to positive (negative) values. The bottom subplots show that the errors from the mocks are always within 10\% of the $N$-body simulations.}
\label{fig:pinocchio} 
\end{figure}

In Fig.~\ref{fig:pinocchio}, we compare the means and errors of the modal expansion coefficients from the 298 matched Pinocchio realizations to the $N$-body measurements. 
The errors measured from Pinocchio are always within 10\% of the errors from Minerva.
For the first few modes, we note that the mean $\beta^R$ measured from Pinocchio can differ from the $N$-body result by more than $\sim 0.1 \, \sigma$, but this is not unexpected.
We recall that the halos in the mock catalogs were selected such that the total power spectrum, or the bispectrum variance, matched that of Minerva.
Thus, by construction, the bispectrum variance from the mocks agrees with that of the $N$-body simulations, while the mean bispectrum from the mocks is suppressed relative to the $N$-body mean \cite{Oddo:2019run}. 
This overall suppression is what causes the $\beta^R_0$ from Pinocchio to be less than what is measured in Minerva.

\begin{figure}[ht]
\centering 
\includegraphics[width=\textwidth]{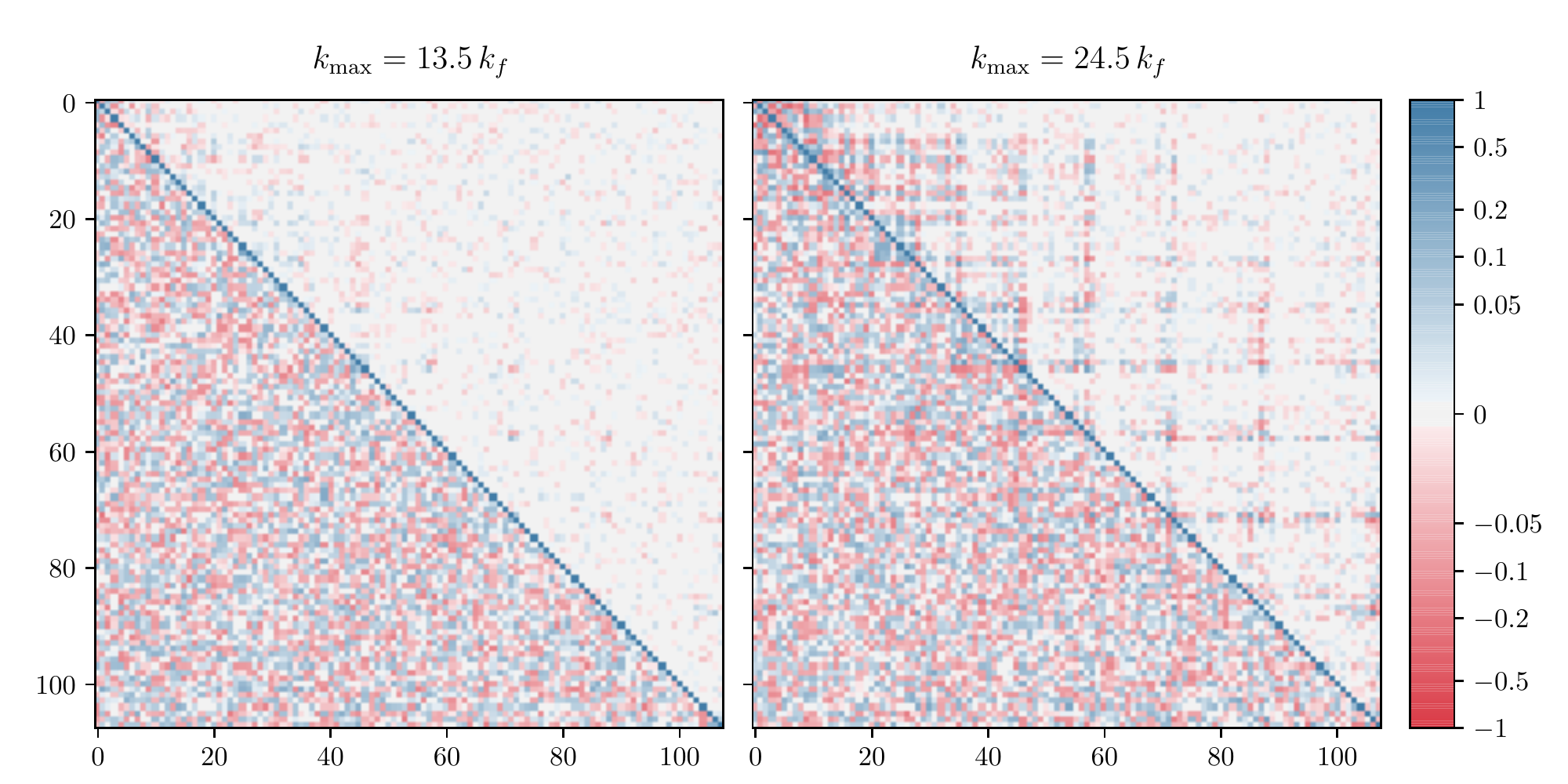}
\hfill
\caption[Correlation matrices for $\beta^R$]{Correlation matrices for 108 $\beta^R$ modal coefficients at $k_{\rm max} = 13.5 \, k_f$ (left) and $24.5 \, k_f$ (right). Within each matrix, the lower triangular elements are from the 298 Minerva simulations while the upper triangular elements are from the full set of 10,000 Pinocchio realizations. Correlation coefficients tend to be closer to zero for lower $k_{\rm max}$ and when more realizations are used.}
\label{fig:covariance} 
\end{figure}

In Fig.~\ref{fig:covariance}, we compare the correlation matrices from Minerva vs the 10,000 Pinocchio mocks. 
As in the bispectrum case \cite{Oddo:2019run}, we find that the correlation coefficients are generally closer to zero when 10,000 mocks are used, compared to 298 Minerva simulations.
When only the 298 Pinocchio mocks with matched initial conditions are compared, the correlation coefficients agree very well (though we do not show this in a plot).
Modal coefficients are more correlated when $k_{\rm max}$ is higher.
We compare how the different estimates of the covariance matrix (using 298 Minerva simulations, 298 matched Pinocchio mocks, or all 10,000 Pinocchio mocks) impact the parameter constraints in Section \ref{subsec:cov}.

\subsection{Standard bispectrum vs reconstructed bispectrum}
\label{subsec:Brec}

In this section, we look at the relationship between the standard bispectrum estimator in eq.~\eqref{eq:B estimator} and the reconstructed bispectrum in eq.~\eqref{eq:Brec}
at the level of an individual realization, the mean of many simulations, and the resulting covariance.
Both estimators capture information about the bispectrum, but they are not equivalent, and in this section we examine which properties of the two estimators are the same or not.

For the comparisons in this section, we include scales up to $k_{\rm max}=13.5 \, k_f$, and adopt triangle bins of width $\Delta k = s \, k_f$ with $s=1$ for the standard bispectrum estimator.
In this case, the standard bispectrum estimator measured the bispectrum in 294 triangle bins.
Within each bin, we compare the standard estimator measurement with the bin-averaged value of $B_{\rm rec}$ in eq.~\eqref{eq:Brec}. 
We have also done the comparison using $B_{\rm rec}$ evaluated on effective triangles with sorted side lengths, which was shown in \cite{Oddo:2019run} to provide a good approximation to the true bin average, and the results that follow are not changed.

\begin{figure}[t]
\centering 
\includegraphics[width=\textwidth]{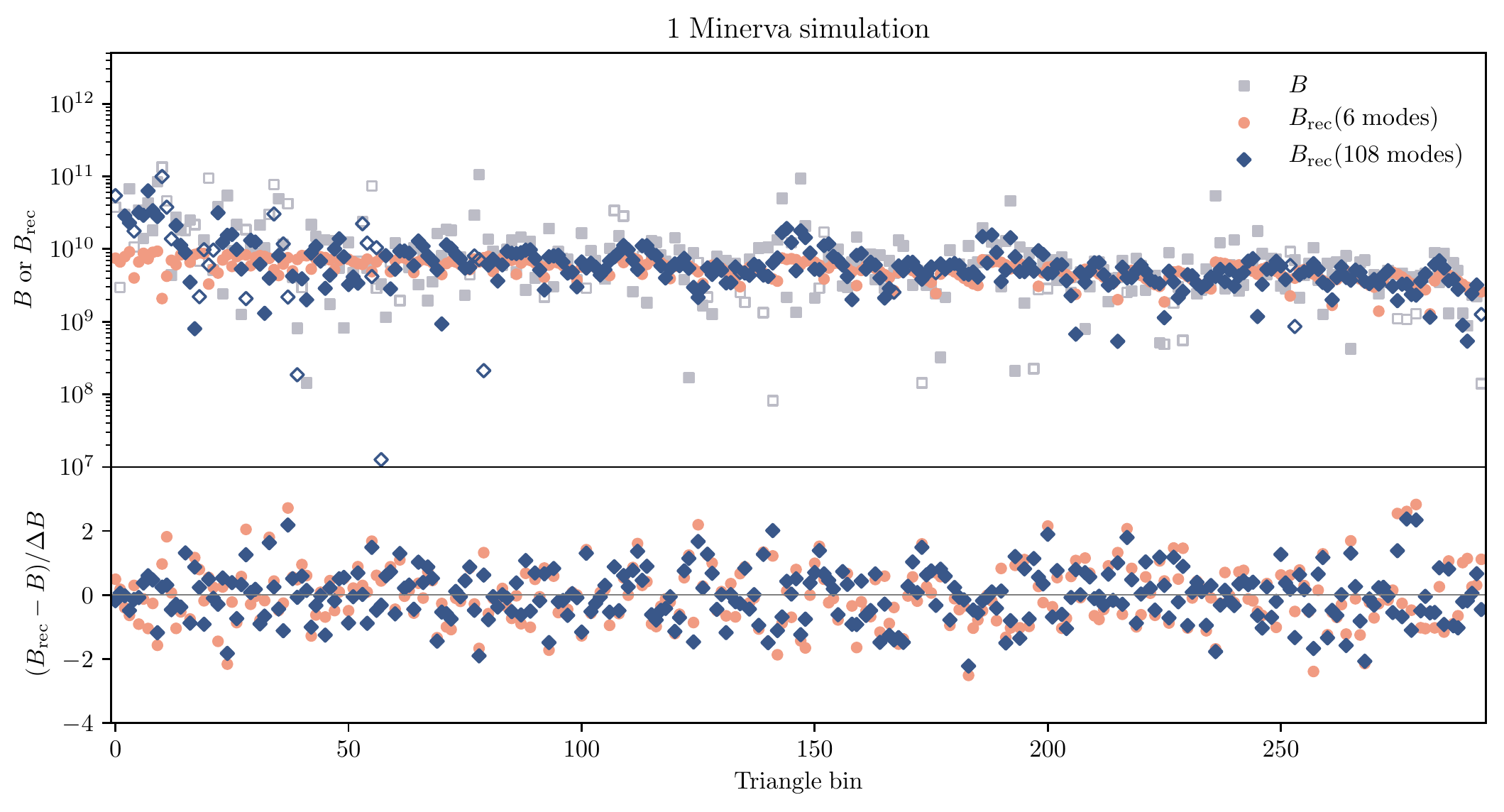}
\hfill
\caption[Standard bispectrum vs reconstructed bispectrum from 1 realization]{Comparison of the standard bispectrum vs reconstructed bispectrum on one Minerva simulation for $k_{\rm max}=13.5 \, k_f$. 
In the top panel, filled (empty) markers signify positive (negative) values.
The standard bispectrum measurement and the reconstructed bispectra do not obtain the same values in each bin, though the scatter between them is typically within two times the $1\,\sigma$ error of the standard estimator, $\Delta B$.
}
\label{fig:B vs Brec 1sim} 
\end{figure}

Fig.~\ref{fig:B vs Brec 1sim} compares the standard bispectrum estimate $B$ with $B_{\rm rec}$ for a single realization, showing that the two do not measure the same value in each bin. 
This is not unexpected, as the two estimators perform rather different operations on the same density grid in Fourier space in order to produce these bispectrum estimates.

\begin{figure}[ht]
\centering 
\includegraphics[width=\textwidth]{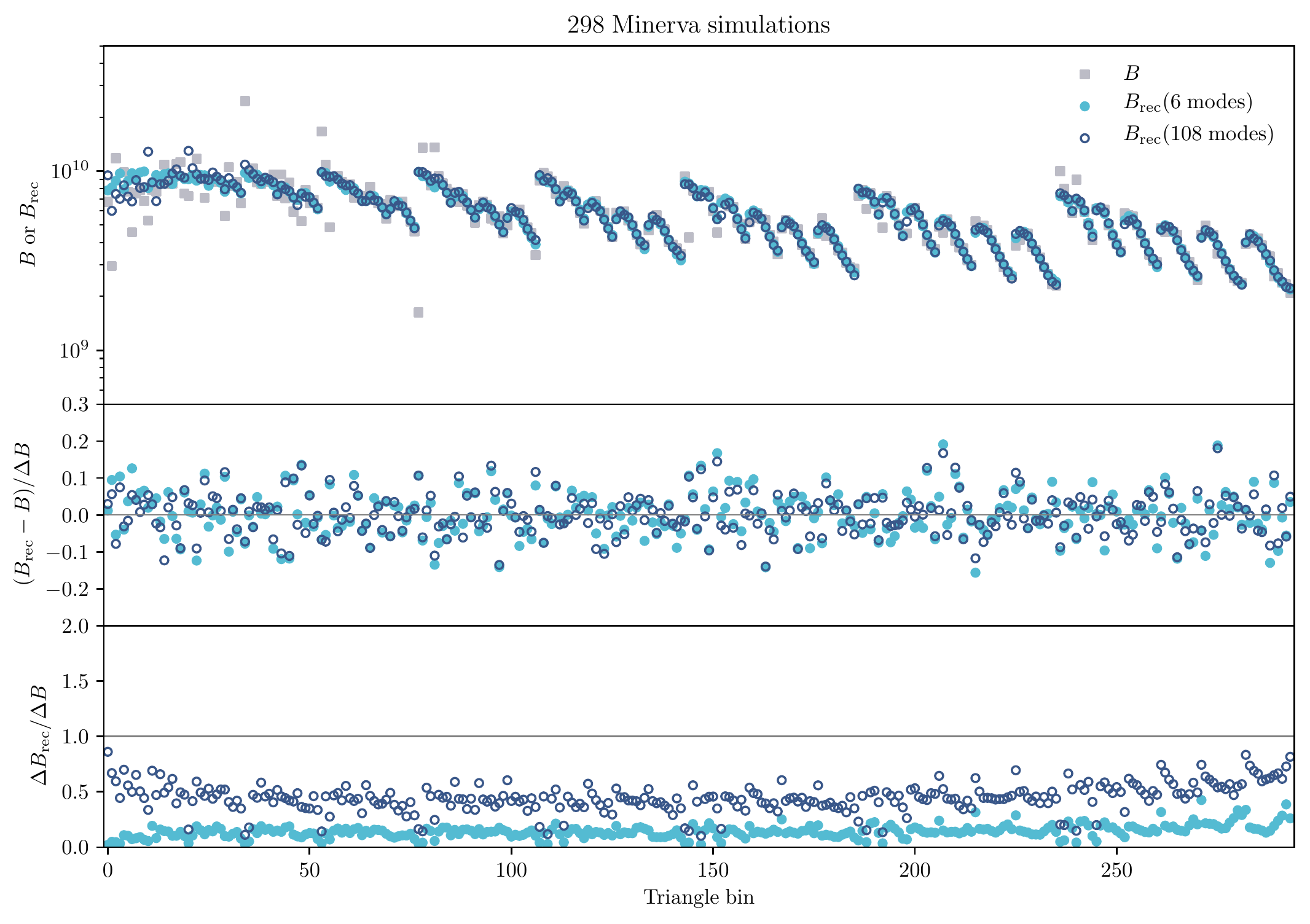}
\hfill
\caption[Standard bispectrum vs reconstructed bispectrum averaged over all Minerva simulations]{Comparison of the standard bispectrum vs reconstructed bispectrum means and errors over all 298 Minerva simulations for $k_{\rm max}=13.5\,k_f$.
While the means agree to within $\sim 10\%$ of the error of the standard bispectrum estimator (middle panel), the error in the reconstructed bispectrum is generally suppressed relative to the error in the standard bispectrum measurement (bottom panel). 
This suppression is larger when fewer modes are used in the reconstruction.
}
\label{fig:B vs Brec allsims} 
\end{figure}

\begin{figure}[ht]
\centering 
\includegraphics[width=\textwidth]{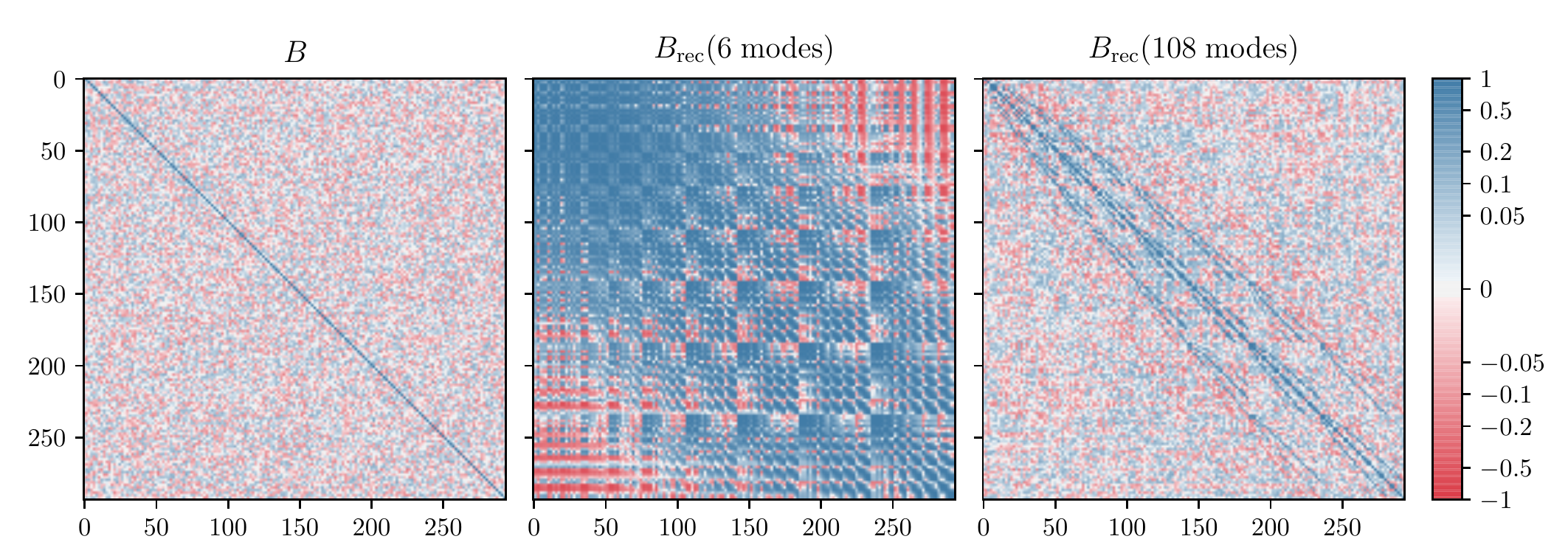}
\hfill
\caption[Standard bispectrum vs reconstructed bispectrum correlation matrices]{Comparison of correlation matrices from the standard bispectrum and reconstructed bispectra for 294 triangle bins and $k_{\rm max}=13.5 \, k_f$.
The noticeable differences in the correlation matrices for $B$ and $B_{\rm rec}$ indicate that the covariance matrix for the standard bispectrum estimator cannot be recovered by the modal bispectrum.
This discrepancy is especially pronounced when fewer modes are used in the reconstruction.
}
\label{fig:B vs Brec corr} 
\end{figure}

Then, in Fig.~\ref{fig:B vs Brec allsims}, we extend the comparison to the suite of all 298 Minerva simulations.
In this case, the mean $B$ and $B_{\rm rec}$ are in much better agreement, on a bin-by-bin basis, where the differences are typically $\lesssim 10\%$ of the error on the standard bispectrum estimate (middle panel).
The difference is that the standard bispectrum estimates have more scatter, while the reconstructed bispectrum with fewer modes has less scatter.
In the bottom panel of Fig.~\ref{fig:B vs Brec allsims}, we show one key difference between the standard bispectrum and modal bispectrum estimates: the error on the reconstructed bispectrum is typically suppressed relative to the error on the standard bispectrum, and it depends on the number of modes used in the reconstruction.
This is shown also for the correlation matrices in Fig.~\ref{fig:B vs Brec corr}, where we see that the triangle bins are much more correlated (and anti-correlated) when the reconstructed bispectrum is used, especially with fewer modes.

The underlying reason for this discrepancy is that the full $N_{\rm tri}$-dimensional Gaussian distribution that is captured by the standard bispectrum covariance cannot be compressed into a $N_{\rm modes}$-dimensional one, unless $N_{\rm modes} = N_{\rm tri}$, in which case there is no practical compression of the data.
This can be illustrated by calculating the covariance of the reconstructed bispectrum as 
\begin{align}
	{\rm Cov} \left[ B_{\rm rec}(\Delta_i), B_{\rm rec}(\Delta_j) \right] &= 
	\frac{1}{w(\Delta_i)w(\Delta_j)} \sum_n \sum_m {\rm Cov} \left[ \beta^R_n, \beta^R_m \right] R_n(\Delta_i)R_m(\Delta_j).
\end{align}
In the limit of Gaussian covariance for $\Delta_i = \Delta_j$,
\begin{align}
	{\rm Var} \left[ B_{\rm rec} (\Delta_i) \right] &= \frac{1}{w(\Delta_i)^2} \sum_n {\rm Var} \left[ \beta^R_n \right] R_n(\Delta_i)^2,
\end{align}
such that adding more modes to the modal expansion will always increase the variance on $B_{\rm rec}$, and conversely, using fewer modes will suppress the variance on $B_{\rm rec}$ (as seen in the bottom panel of Fig.~\ref{fig:B vs Brec allsims}).

The modal method produces errors on $B_{\rm rec}$ that are suppressed relative to errors from the standard bispectrum estimator. 
This implies that measurements of the modal coefficients cannot give estimates of bispectrum errors that are directly relevant for standard bispectrum pipelines. 
For example, errors on $B_{\rm rec}$ should not be used to judge how well theoretical model predictions for the bispectrum would work in a pipeline using the standard bispectrum estimator. 


\subsection{Benchmark comparison: modal bispectrum vs standard bispectrum}
\label{subsec:benchmark}

Here we compare parameter constraints from the modal bispectrum and the standard bispectrum for the default modal pipeline settings detailed at the beginning of Section \ref{sec:results}.
In subsequent sections, we demonstrate how the modal bispectrum constraints can depend on these settings.

To put the bispectrum constraints in context, we also compare them with an independent constraint of $b_1 = 2.7081 \pm 0.0012$ from using chi-squared minimization to fit the ratio of cross-power spectra $P_{hm}/P_{mm}$ from the Minerva simulations \cite{Oddo:inprep}.
The fitting function is $P_{hm}(k)/P_{mm}(k) = b_1 + {\rm coefficient} \times k^2$ up to $k_{\rm max}=0.044 \,h\,{\rm Mpc}^{-1}$, where the $k^2$ term is used to account for scale-dependent loop corrections, which for the small error bars corresponding to the total volume of the Minerva simulations can be important even at scales as large as $k \approx 0.02 \, h\,{\rm Mpc}^{-1}$.

\begin{figure}[t]
\centering 
\includegraphics[width=\textwidth]{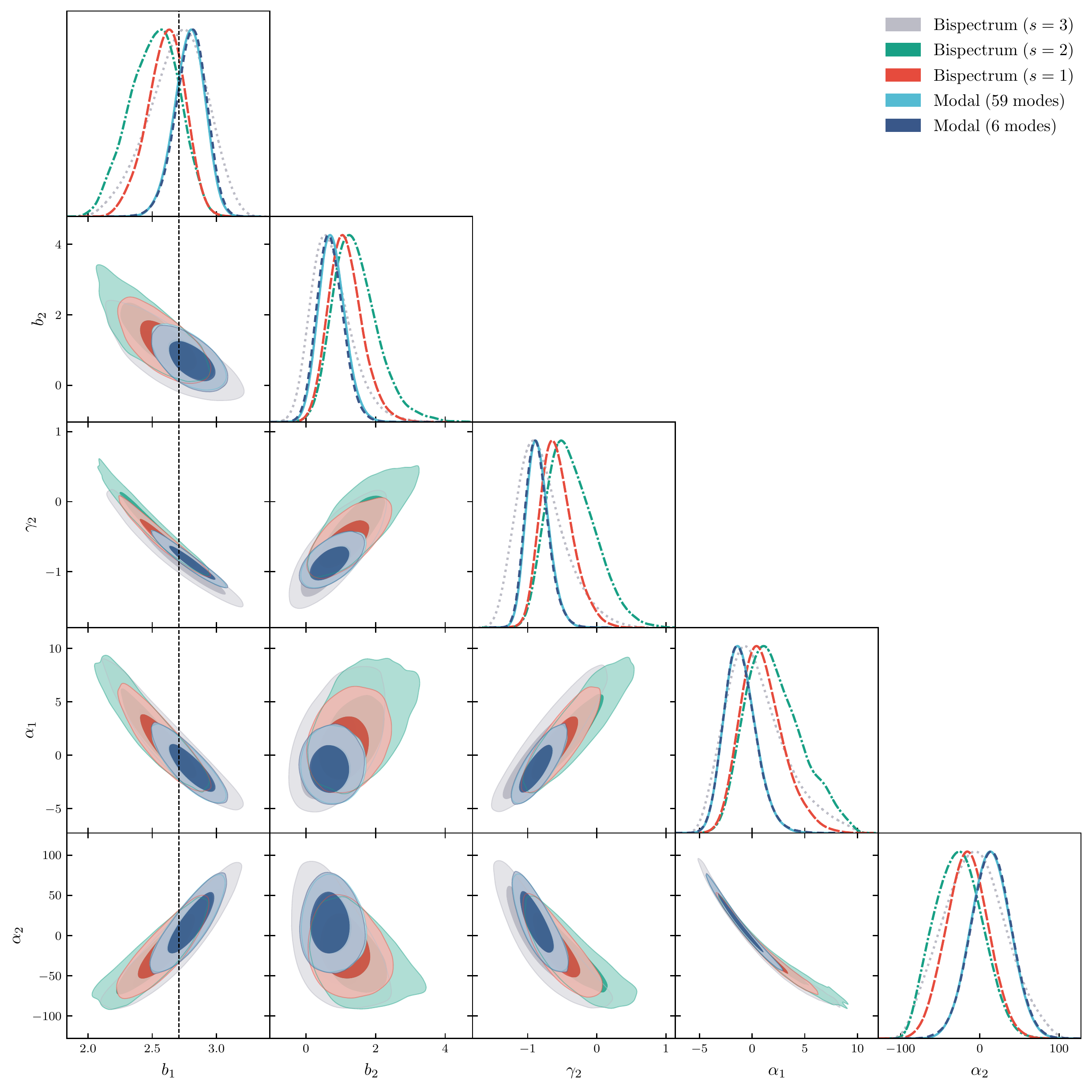}
\hfill
\caption[Modal bispectrum vs standard bispectrum benchmark comparison for $k_{\rm max} = 13.5 \, k_f$]{Benchmark comparison between modal bispectrum and standard bispectrum constraints for $k_{\rm max} = 13.5 \, k_f \approx 0.06 \, h \,\Mpc^{-1}$. The vertical dashed black line represents the constraint on $b_1$ from the large-scale ratio of cross-power spectra $P_{hm}/P_{mm}$. 
Standard bispectrum constraints with three different bin sizes are shown: $s=1$ (red), 2 (green), and 3 (gray).
The modal bispectrum constraints with six modes (dark blue) and 59 modes (light blue) are identical, indicating that the modal bispectrum constraints have already converged with six modes.}
\label{fig:benchmark}
\end{figure}

The benchmark constraints are shown in Fig.~\ref{fig:benchmark} for $k_{\rm max} = 13.5 \, k_f \approx 0.06 \, h\,\Mpc^{-1}$.
This $k_{\rm max}$ was chosen such that we conservatively consider scales over which the tree-level halo bispectrum model has been shown to accurately describe the data, and at this $k_{\rm max}$ we are able to compare with standard bispectrum results from all three bin widths in \cite{Oddo:2019run}, $s=1$, 2, and 3.

The modal constraints using six modes and 59 modes have the same posteriors, with the contours overlapping to the point of making the 59 modes case almost invisible, indicating that the parameter constraints have fully converged with the six custom modes.
This is fully consistent with the analysis in \cite{Oddo:2019run} which showed that the tree-level bispectrum model is a good fit to the data up to this $k_{\rm max}$.

We find that the modal bispectrum constraints are consistent with, though not identical to, the standard bispectrum constraints.
We note that the modal bispectrum estimator, while it is a measure of the bispectrum, is an independent estimation of it, so we do not require (in the sense of a test) that the constraints be identical.
Although all bin choices for the standard bispectrum estimator lead to consistent outcomes, the smallest bin width ($s=1$) takes better advantage of the shape-dependence of the bispectrum leading to slightly narrower constraints.
In comparison, the modal decomposition accounts for the shape-dependence of the bispectrum without loss of information due to the binning of wavenumbers.

\begin{figure}[t]
\centering 
\includegraphics[width=\textwidth]{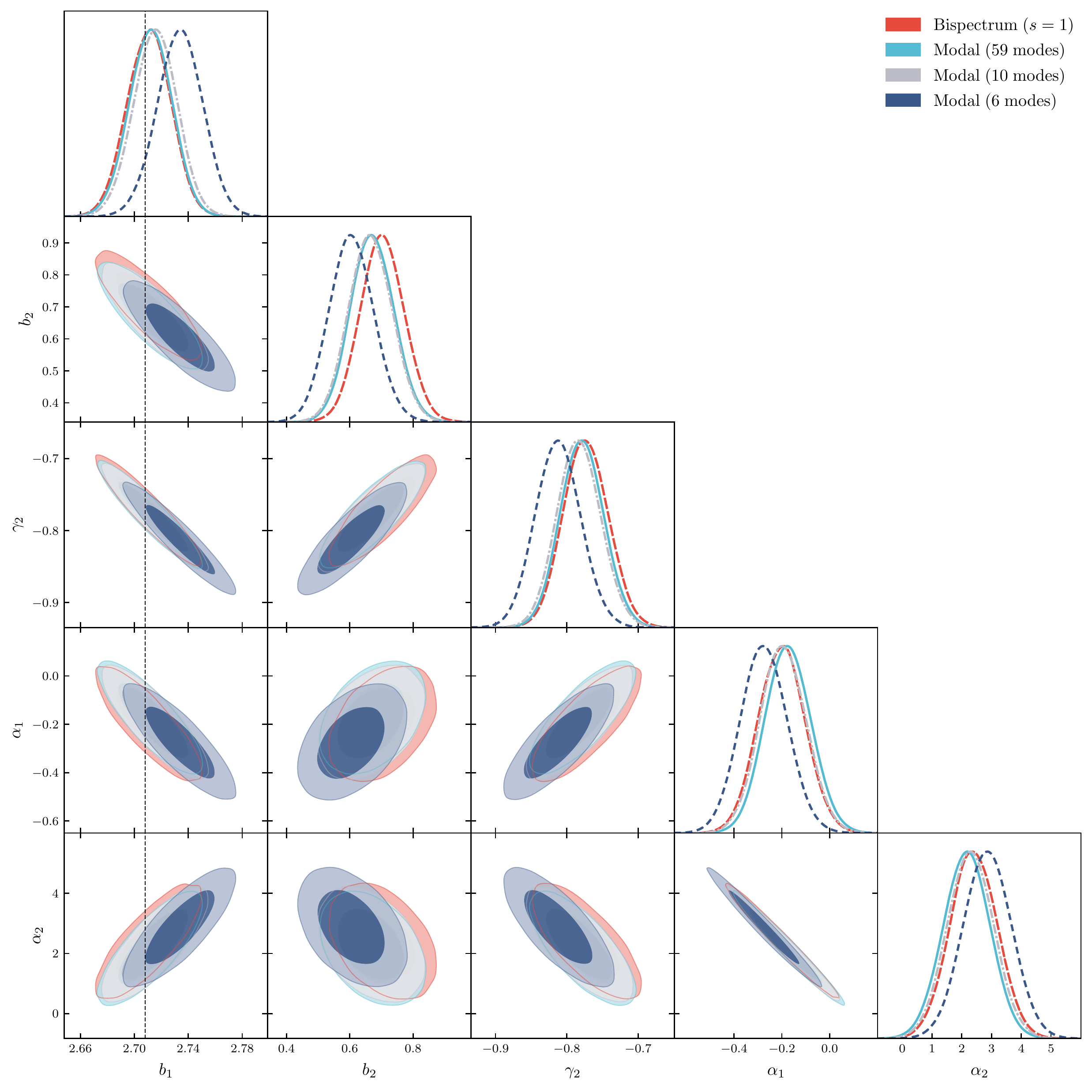}
\hfill
\caption[Modal bispectrum vs standard bispectrum benchmark comparison for $k_{\rm max} = 24.5 \, k_f$]{Comparison of modal bispectrum and standard bispectrum constraints at $k_{\rm max} = 24.5 \, k_f \approx 0.10 \, h\,{\rm Mpc}^{-1}$. At this $k_{\rm max}$, we can only compare with the $s=1$ binning of the standard bispectrum estimator (red).  
By comparing the constraints using six custom modes (dark blue), 10 modes (gray), and 59 modes (light blue), we find that 10 modes are sufficient for the modal constraints to converge, as there is no further change when 59 modes are used.
}
\label{fig:kmax} 
\end{figure}

In Fig.~\ref{fig:kmax} we perform the same comparison except with the $k$-range extended to $k_{\rm max} = 24.5 \, k_f \approx 0.10 \, h\,{\rm Mpc}^{-1}$.
In this case, by comparing the modal bispectrum constraints with different numbers of modes, we see that the six custom modes are no longer sufficient, but 10 modes has already converged, as there is no further benefit to using 59 modes.
The fact that there is further information in four additional modes, on top of the six custom modes, is a sign that the halo bispectrum deviates from the tree-level prediction on scales within this $k$-range, as studied in much more detail in \cite{Oddo:2019run}.
Although we know that the tree-level halo bispectrum model is no longer a good model for the data to these scales, we can still compare the constraints with those from the standard bispectrum.
At this $k_{\rm max}$, we can only compare the standard bispectrum constraints with the $s=1$ binning.
Fig.~\ref{fig:kmax} shows that the results from the modal bispectrum and standard bispectrum agree very well, with only a small bias relative to each other (most visible in the 1-dimensional posterior for $b_2$ which is biased by $\sim 0.5\,\sigma$).

Finally, we comment on the amount of compression that has been achieved in these benchmark comparisons. At $k_{\rm max} = 13.5 \, k_f$, the number of triangle bins that were used by the standard bispectrum estimator are 294 bins for $s=1$, 49 bins for $s=2$, and 19 bins for $s=3$, and we found that the modal bispectrum constraints had converged using only six custom modes.
At $k_{\rm max} = 24.5 \, k_f$, the standard bispectrum used 1,585 bins with $s=1$, yielding constraints that were very similar to the modal bispectrum using 10 modes.
This shows that the modal bispectrum is able to efficiently compress the information contained in the bispectrum into a data set that is 3 to 160 times smaller, while preserving most of the important cosmological information that we are interested in.


\subsection{Robustness checks in the modal implementation}
\label{subsec:checks}

In this section, we vary the settings in the modal analysis pipeline away from the benchmark settings to see how the modal constraints are sensitive to these choices.


\subsubsection*{Normal vs Legendre polynomials}

The six custom modes are defined to fit the tree-level bispectrum, and so their form is independent of whether we choose normal or (shifted) Legendre polynomials as our $q_n(k)$.
Therefore, we take $k_{\rm max} = 24.5 \, k_f$ and 10 modes, and compare the constraints between choosing normal polynomials vs Legendre polynomials.
The constraints from the Legendre polynomials are identical to the modal constraints with 10 modes and normal polynomials shown in Fig.~\ref{fig:kmax}, and plotted together only one set of posteriors would be visible, so to save space we do not show this comparison in a figure.


\subsubsection*{Custom modes}

\begin{figure}[t]
\centering 
\includegraphics[width=\textwidth]{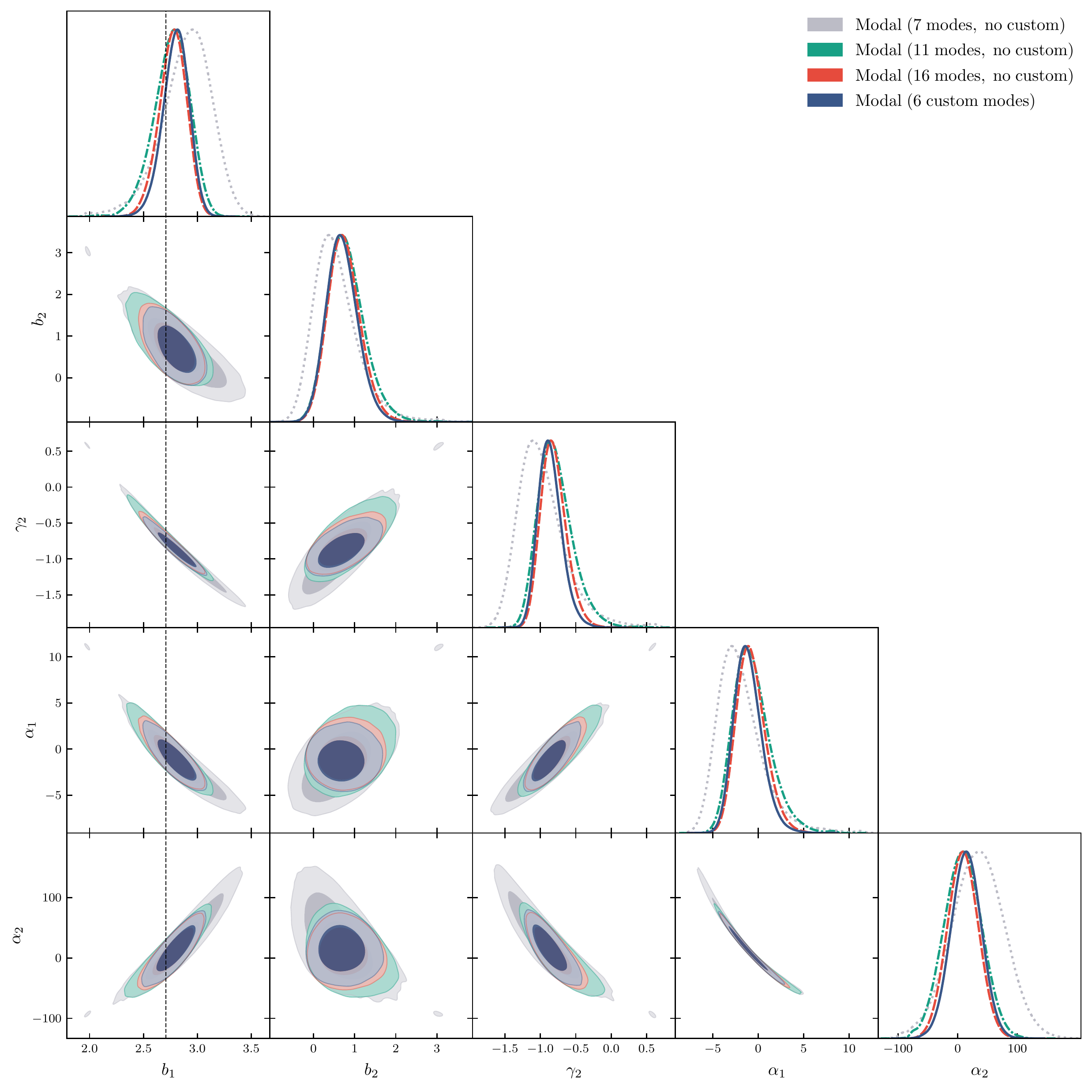}
\hfill
\caption[Impact of custom modes for $k_{\rm max} = 13.5 \, k_f$]{\label{fig:custom_kmax13pt5} Comparison of modal bispectrum constraints at $k_{\rm max} = 13.5 \, k_f \, \approx 0.06 \, h \,{\rm Mpc}^{-1}$ without and with custom modes.
We compare constraints without custom modes using 7 modes (gray), 11 modes (green), and 16 modes (red) with the converged benchmark constraints that used 6 custom modes (dark blue).
}
\end{figure}

\begin{figure}[t]
\centering 
\includegraphics[width=\textwidth]{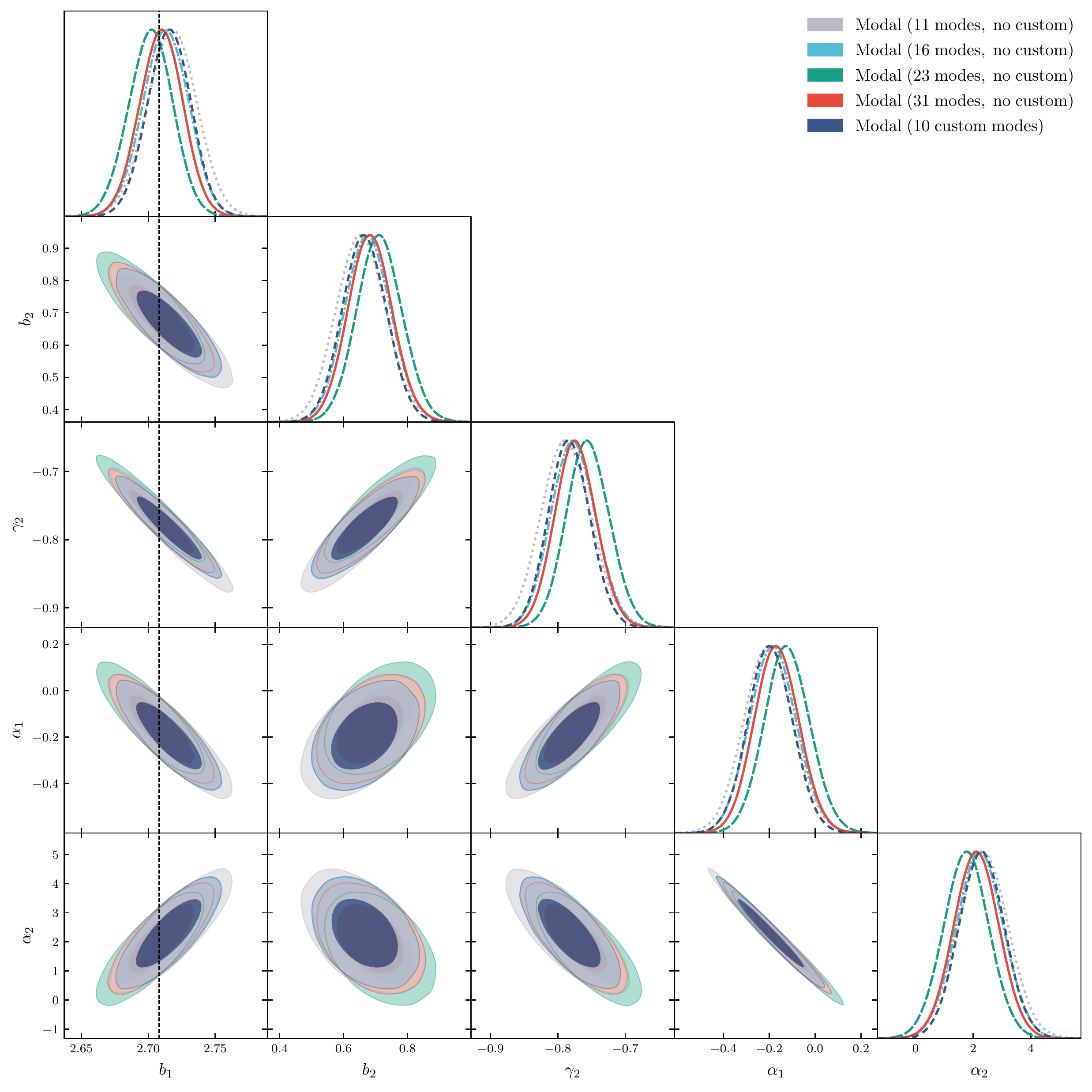}
\hfill
\caption[Impact of custom modes for $k_{\rm max} = 24.5 \, k_f$]{\label{fig:custom_kmax24pt5} Comparison of modal bispectrum constraints at $k_{\rm max} = 24.5 \, k_f \approx 0.10 \, h \,{\rm Mpc}^{-1}$ without and with custom modes.
We compare constraints without custom modes using 11 modes (gray), 16 modes (light blue), 23 modes (green), and 31 modes (red) with the converged benchmark constraints that used 10 custom modes (dark blue).
}
\end{figure}

What is the impact of including the custom modes? 
To see this, we perform the same analysis as in the benchmark case for $k_{\rm max} = 13.5\, k_f$ and $24.5\, k_f$, but do not include the six custom modes, and see how the modal expansion convergence is affected by not including custom modes.
The case for $k_{\rm max} = 13.5\, k_f$ is in Fig.~\ref{fig:custom_kmax13pt5}, and for $k_{\rm max} = 24.5\, k_f$ is in Fig.~\ref{fig:custom_kmax24pt5}.
For lower $k_{\rm max}$, we find that 16 modes are sufficient for the parameter errors to agree to within 8\% with the result from six custom modes, and the parameter means are shifted by $\lesssim 0.25 \, \sigma$ compared to the six custom modes case.
For higher $k_{\rm max}$, the expansion converges with 31 modes, compared to only 10 modes when six of these are the custom modes.
(It appears to converge also with 16 modes, but we find that this is not stable, because the 16 modes case changes when compared with 23 modes.)
The means are shifted by $\lesssim 0.3 \, \sigma$ and the errors are consistent to within 1\% compared to the benchmark case with custom modes.
This shows that the inclusion of custom modes which are constructed based on an informative theoretical model for the bispectrum can help to compress the data into the most efficient basis.

\subsubsection*{Inner product methods}

The benchmark results used a $\gamma$ matrix computed with the 3D FFT method using $N_g=256$ and $L=1500 \, h^{-1}\,{\rm Mpc}$. 
In this section, we compare constraints that have used three different numerical methods for computing $\gamma$: 3D FFT, voxels, and 1D FFT.
(As mentioned in Section \ref{subsec:gamma_methods}, we do not include in this comparison the $\gamma$ computed using the Vegas routine in Cuba, as this results in matrices that are not positive-definite and therefore cannot be used.)
All three methods are unique and have their own free parameters, which roughly correspond to the resolution of the inner product integration in $k$-space.
By comparing constraints obtained with $\gamma$ matrices computed from each method, we find that both the method and how its free parameters are set can affect the resulting constraints.

\begin{figure}[t]
\centering 
\includegraphics[width=\textwidth]{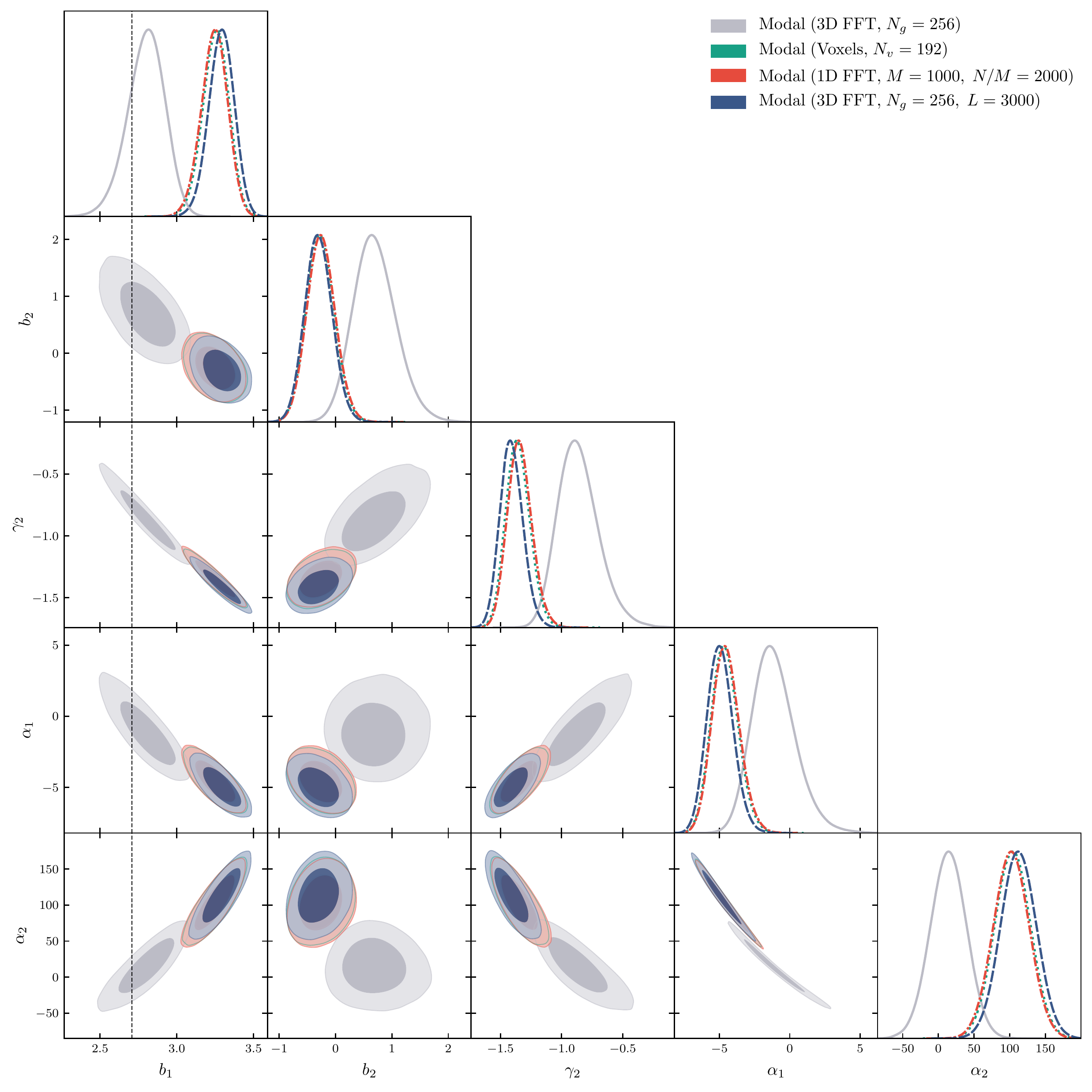}
\hfill
\caption[Comparison of inner product methods for $k_{\rm max} = 13.5 \, k_f$]{Comparison of modal bispectrum constraints at $k_{\rm max} = 13.5 \, k_f \approx 0.06 \, h\,{\rm Mpc}^{-1}$ using different inner product methods. The voxel (green) and 1D FFT  (red) methods have converged towards each other, and both disagree with the 3D FFT result (gray) that agreed with the standard bispectrum constraint in Fig.~\ref{fig:benchmark}. The 3D FFT case with $L=3000 \, h^{-1}\,{\rm Mpc}$ (dark blue) is also biased, illustrating that getting the discretization as in the measurements is important.}
\label{fig:gamma_kmax13pt5}
\end{figure}

\begin{figure}[t]
\centering 
\includegraphics[width=\textwidth]{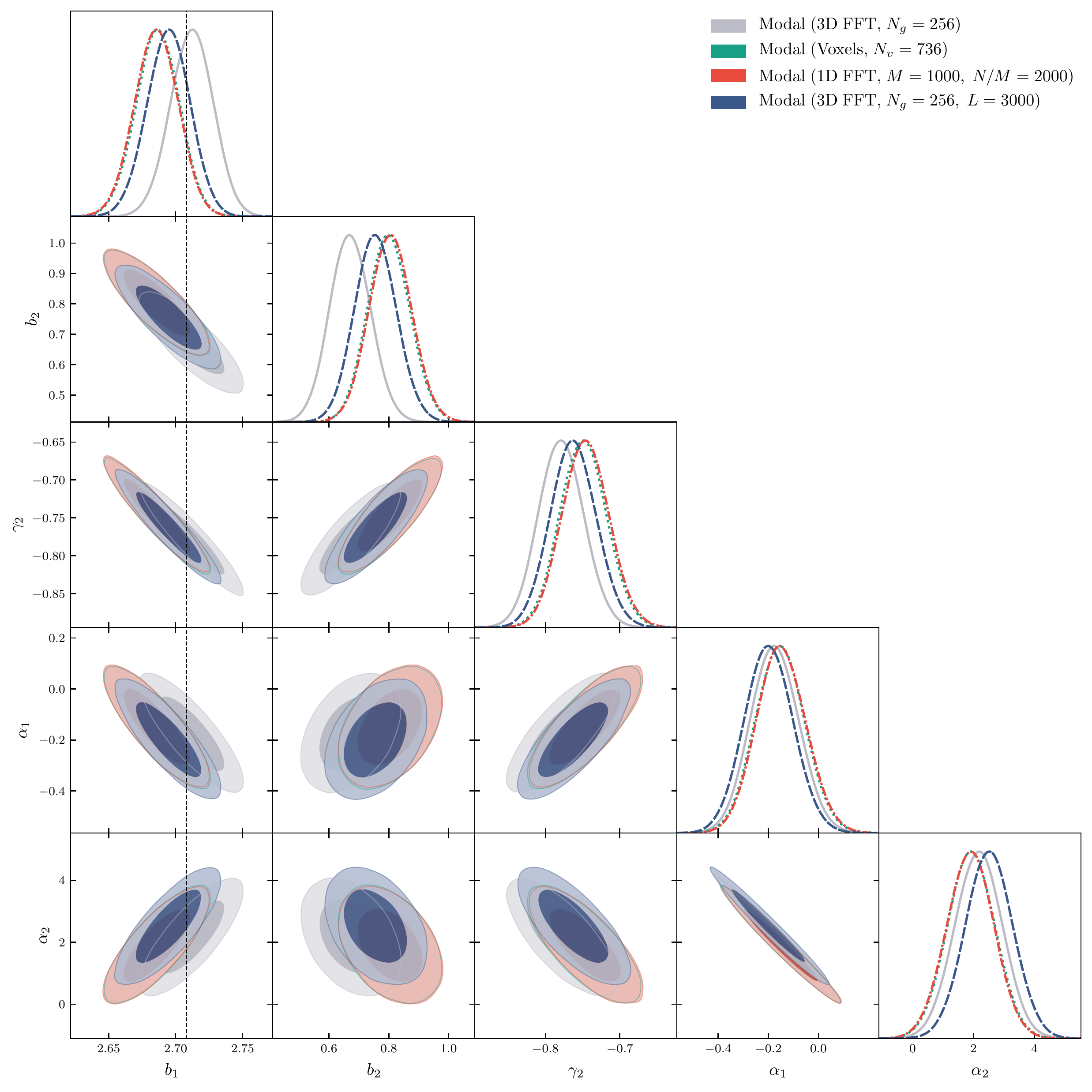}
\hfill
\caption[Comparison of inner product methods for $k_{\rm max} = 24.5 \, k_f$]{Comparison of modal bispectrum constraints at $k_{\rm max} = 24.5 \, k_f \approx 0.10 \, h\,{\rm Mpc}^{-1}$ using different inner product methods. As in Fig.~\ref{fig:gamma_kmax13pt5}, the voxel (green) and 1D FFT (red) methods have converged towards each other, and both are biased relative to the correct 3D FFT result (gray). The 3D FFT case with $L=3000 \, h^{-1}\,{\rm Mpc}$ (dark blue) is also biased, illustrating that getting the discretization as in the measurements is important.}
\label{fig:gamma_kmax24pt5} 
\end{figure}

Figs.~\ref{fig:gamma_kmax13pt5} and \ref{fig:gamma_kmax24pt5} compare the methods for $k_{\rm max} = 13.5 \, k_f$ and $24.5 \, k_f$, respectively.
The $N_v$ that is shown for the voxel method is the number of voxel cells in each dimension, and the $N$ and $M$ resolution parameters for the 1D FFT method (defined in Appendix \ref{app:1dfft}), are converged; doubling these values did not show any changes in the resulting constraints.
When $k_{\rm max}$ is small, we find that the voxel and 1D FFT methods converge on the same posterior for sufficiently high resolutions, but they strongly disagree with the 3D FFT result in both the position and size of the posterior.
For higher $k_{\rm max}$, the different methods produce posterior contours that agree in their size, but with non-negligible shifts (biases) between them.
The observation that the voxel and 1D FFT methods converge to each other for both $k_{\rm max}$, while disagreeing more strongly with the 3D FFT calculation at lower $k_{\rm max}$, reflects the fact that the inner product calculation captured by $\gamma$ must be treated using the same discretization scheme as the measurements  in order to obtain correct constraints that are consistent with the standard bispectrum analysis.
The voxel and 1D FFT methods are ways of calculating the inner product assuming it is a smooth continuous integral, and it is not straightforward to adapt these methods such that they account for the same discretization effects as the measurements, while the 3D FFT method is, by construction, computing the inner product in the same way that the measurements are taken. 

When $k_{\rm max}$ is low, the discretization of the Fourier grid during the estimation is more important to take into account, because fewer triangle configurations are being averaged (i.e. the tetrapyd is more sparsely sampled).
This interpretation also be confirmed another way: within the 3D FFT method, the resolution is increased if the size of the FFT box, $L$, is larger. If we increase the resolution by increasing $L$ to be twice as large as the simulation box, we find in both Figs.~\ref{fig:gamma_kmax13pt5} and \ref{fig:gamma_kmax24pt5} that the resulting posterior becomes more similar to the voxel and 1D FFT case, as we would expect. 

It may be the case that for a higher $k_{\rm max}$ than what we have used in this work, the different methods for computing $\gamma$ may yield results that are similar enough that the different methods can be interchangeable.
This is expected because more cosmological information is contained in non-linear scales and the continuous integration of the voxel and 1D FFT methods becomes a better approximation to the discretized inner product when $k_{\rm max}$ is higher.
However, we emphasize that for the range of scales that we have used in this work, $k_{\rm max} \lesssim 0.10 \, h\,{\rm Mpc}^{-1}$, the different methods for computing $\gamma$ are \textit{not} interchangeable, and the 3D FFT method is the only one which treats the inner product identically to how the measurements are performed, resulting in correct constraints.

The 3D FFT method is also preferable for the speed and ease of its calculation. 
The 3D FFTs are performed very quickly using the same FFT routines which are already necessary for the modal bispectrum (and standard bispectrum) measurements, while the other voxel and 1D FFT methods require different algorithms which must be coded independently and, in our implementation, are not as fast.
As $\gamma$ must only be computed once for a fixed $k_{\rm max}$, we do not anticipate that the computation of $\gamma$ which is unique to the modal bispectrum analysis increases the computational cost of using the modal method by a significant amount.


\subsubsection*{Dependence on FFT grid resolution}

Depending on $k_{\rm max}$, the FFT grid on which the 3D FFTs are calculated can have a configuration-space grid resolution much smaller than our default value of $N_g = 256$.
Since $L$ fixes the resolution of the Fourier grid to be $k_f \equiv 2\pi/L$, increasing $N_g$ increases the $k_{\rm max}$ that can be probed without too much aliasing contamination.
\cite{Jeong2010} and \cite{Sefusatti:2015aex} have suggested that the standard FFT bispectrum estimator, which takes a form very similar to the modal estimator, can probe up to $k_{\rm max} = k_f N_g/3 = 2 k_{\rm Ny}/3$, where $k_{\rm Ny}$ is the Nyquist wavenumber, unlike the power spectrum estimator which is valid up to $k_{\rm max} = k_f N_g/2 = k_{\rm Ny}$.
This is because the factor of $e^{i \bk_{123} \cdot \bx}$ in the estimator is invariant under shifts (in one dimension) of each $k_i$ to $k_i+k_f N_g/3$ \cite{Sefusatti:2015aex}.
On the other hand, the opposite has been argued by \cite{Watkinson:2017zbs} for the FFT bispectrum estimator and \cite{Hung:2019ygc} for the modal estimator---that these estimators are valid up to $k_{\rm max} = k_{\rm Ny}$.

\begin{figure}[t]
\centering 
\includegraphics[width=\textwidth]{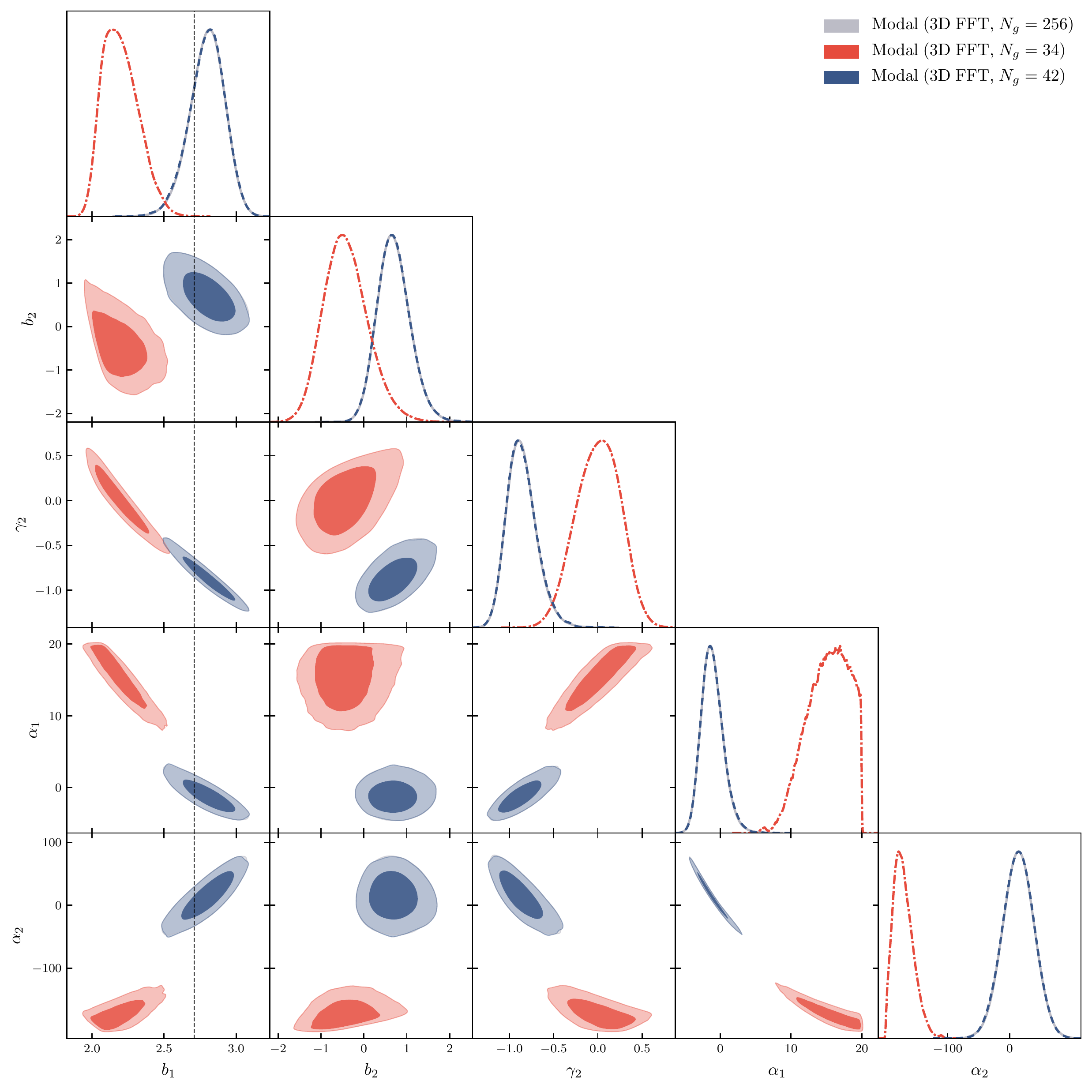}
\hfill
\caption[Dependence on FFT grid resolution $N_g$]{Comparison of modal bispectrum constraints at $k_{\rm max} = 13.5 \, k_f \approx 0.06 \, h\,{\rm Mpc}^{-1}$ using different FFT grid resolutions, $N_g$. The case with $N_g=42$ (dark blue) yields identical results to the benchmark case with $N_g=256$ (gray, hidden underneath the dark blue contours), because both satisfy $N_g >3 k_{\rm max}/k_f \approx 41$, while $N_g=34$ (red) is insufficient and leads to incorrect constraints.}
\label{fig:gamma_kmax13pt5_Ng} 
\end{figure}

Here we fix our $k$-range to have $k_{\rm max} = 13.5 \, k_f$ and test which criterion for $N_g$, either $N_g > 3 k_{\rm max}/k_f \approx 41$ or $N_g > 2 k_{\rm max}/k_f = 27$ is sufficient to return the same constraints from the modal estimator pipeline as the benchmark value of $N_g = 256$.
We note that changing $N_g$ requires the pipeline to be run from the beginning, starting with the construction of the configuration-space density grid, and including the calculation of $\gamma$ with the 3D FFT method.
Fig.~\ref{fig:gamma_kmax13pt5_Ng} compares the constraints for different values of $N_g = 34$, 42, and 256.
We find that using $N_g = 34$ leads to constraints that strongly disagree with the benchmark case of $N_g = 256$, while $N_g = 42$ gives identical constraints to the benchmark case, showing that the modal estimator is valid only up to $2k_{\rm Ny}/3$.
If this result is explained by the argument in \cite{Sefusatti:2015aex}, then we would expect this result to also hold for the FFT-based standard bispectrum estimator, which also has a factor of $e^{i \bk_{123} \cdot \bx}$.

\subsubsection*{Weighting}

In the benchmark analysis, the bispectrum was weighted by $w$ in eq.~\eqref{eq:w}, where the power spectrum $P(k_i)$ was the average total halo power spectrum measured from the Minerva simulations.
What is the effect of using a different weighting function? 

We note that changing the weighting only changes two parts of the modal pipeline.
First, the $q_n^{\rm tree}$ functions in eqs.~\eqref{eq:q0tree}--\eqref{eq:q5tree} will change, such that the factor of $\sqrt{k/P(k)}$ in each one will be different.
This will, however, not change the fact that the six custom modes, $Q_n^{\rm tree}$, are able to reconstruct the tree-level halo bispectrum model exactly.
The second change is that when $\llangle Q | w\hat{\mathcal{B}} \rrangle$ in eq.~\eqref{eq:QwB estimator} is estimated from simulations, the factor of $[ \sqrt{k_1 k_2 k_3}\sqrt{P(k_1)P(k_2)P(k_3)}]^{-1/2}$ in the integrand will change to $w/[k_1k_2k_3]$.

We have considered the case where $w$ takes the same form as in eq.~\eqref{eq:w}, but $P(k_i)$ is set to the linear \textit{matter} power spectrum $P_L$. 
This power spectrum is different to the benchmark weighting in that the power spectrum does not have any halo bias, non-linearities, or shot noise.
Therefore, we use this situation to reflect an analysis where the halo power spectrum in the weight is not perfectly modeled or measured.
We find that this difference does not have any effect on the resulting parameter constraints, implying that the modal bispectrum constraints are not strongly affected by the particular power spectrum that is used for the weighting.
In particular, it does not change how quickly the modal expansion converges.

Still, there is a reason to prefer the optimal weighting with the non-linear total halo power spectrum, which is that it is in this case that the covariance of the $\beta^R$ is best approximated by the Gaussian covariance expression in eq.~\eqref{eq:betaR gaussian covariance}.
If the linear power spectrum is used in the weighting, the $P(k_1)P(k_2)P(k_3)$ that appears in eq.~\eqref{eq:gaussian 6pt correlator} does not cancel out with the power spectra in the weight, such that the Gaussian covariance expression for $\beta^R$ is not eq.~\eqref{eq:betaR gaussian covariance}.

\begin{figure}[t]
\centering 
\includegraphics[width=\textwidth]{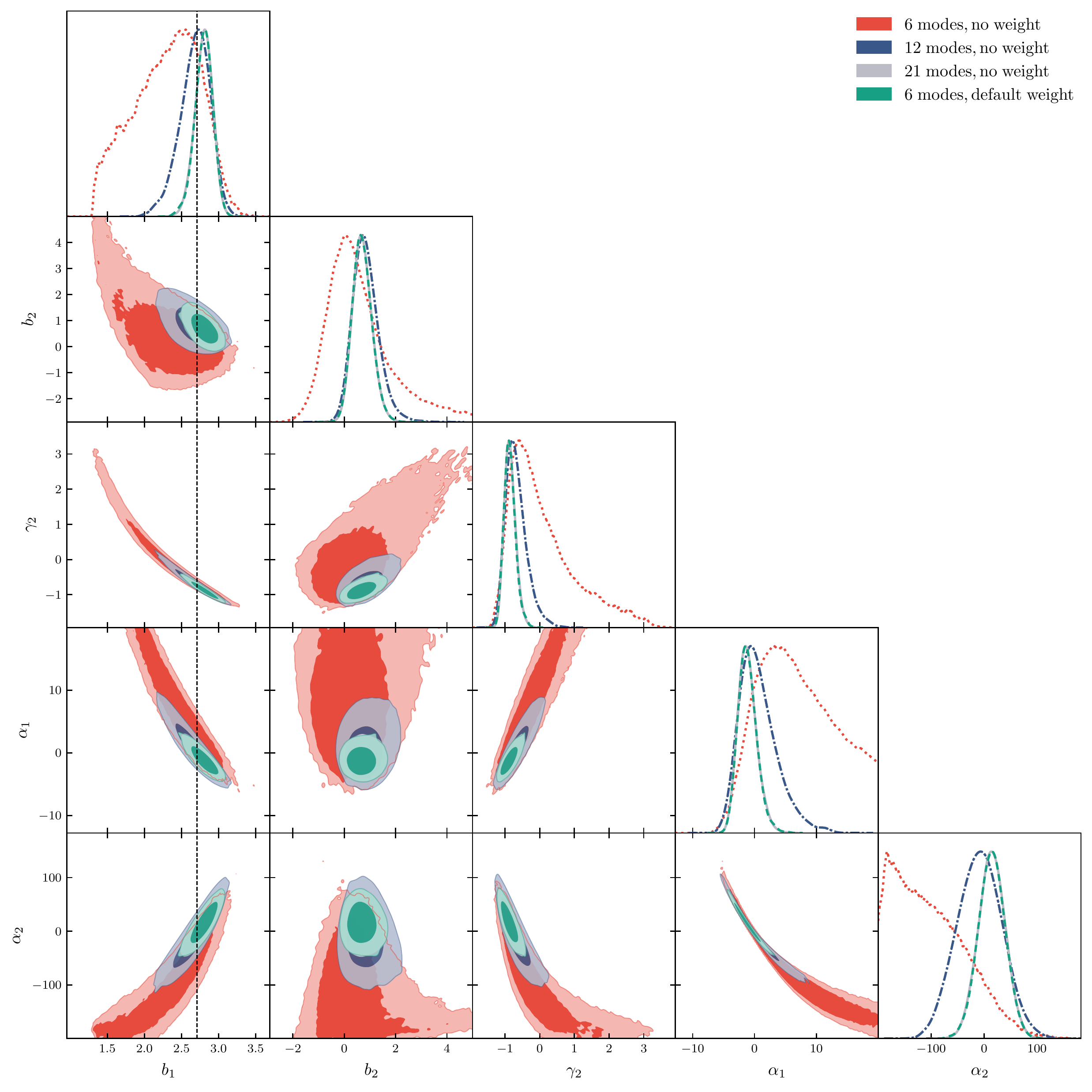}
\hfill
\caption[Dependence on weighting function]{Comparison of modal bispectrum constraints at $k_{\rm max} = 13.5 \, k_f \approx 0.06 \, h\,{\rm Mpc}^{-1}$ using the benchmark weighting function in eq.~\eqref{eq:w} (green) vs no weight, where $w=1$.
When the six custom modes are used in both cases, $w=1$ (red) leads to weaker parameter constraints, but this sub-optimal choice of weighting can be fully compensated for by using more modes.
For this $k_{\rm max}$, 12 modes (dark blue) is not sufficient, but 21 modes (gray, overlapping exactly with the green contours) are able to recover the same constraints as the more optimal weighting.
}
\label{fig:weighting_kmax13pt5} 
\end{figure}

This result does not necessarily mean that the parameter constraints are totally immune to especially sub-optimal choices for $w$.
We have checked that when $w=1$ is adopted, in other words, no weighting at all is used, the information in the bispectrum is less efficiently extracted.
This is shown in Fig.~\ref{fig:weighting_kmax13pt5}, which compares the constraints in the `no weight' case with the benchmark results for $k_{\rm max}=13.5 \, k_f$.
With no weighting, we find that the constraints using six custom modes is much weaker than, though still consistent with, the benchmark case of six custom modes with default weighting. 
This difference must originate from the choice of weighting, because in both cases the six custom modes, by construction, can exactly reproduce the tree-level bispectrum model that is a good description of the bispectrum up to this $k_{\rm max}$.
The fact that the no weighting constraints are weaker is consistent with the fact that less optimal weighting of the $(k_1,k_2,k_3)$ Fourier triangles should lead to less information being extracted.
However, even in this case, the modal expansion method can compensate for the less-than-optimal weighting if a larger number of modes are included.
For this $k_{\rm max}$, Fig.~\ref{fig:weighting_kmax13pt5} shows that the benchmark constraints can be recovered if 21 modes are included.\footnote{
If $w=1$, we note that both the last custom mode $Q_5^{\rm tree}$ and $Q_0$ will be constants that do not have any dependence on $(k_1,k_2,k_3)$, so a modal basis that includes both of them will correspond to having a non-positive definite $\gamma$ matrix. 
For this reason, when $w=1$ and custom modes are included, the basis sets with more than six modes leave out the $Q_0$ mode, but otherwise have the same ordering of $Q_n$ functions as in the rest of this work.
}

\subsection{Modal expansion correlators}
\label{subsec:correlators}

One way of measuring the accuracy of the modal expansion is to define and compute so-called \textit{correlators} that quantify different aspects of the accuracy of the reconstruction.
For example, the shape correlator $\mathcal{S}$, amplitude correlator $\mathcal{A}$, and total correlator $\mathcal{T}$ are \cite{Lazanu:2015rta,Hung:2019ygc}
\begin{eqnarray}
	\mathcal{S}(B_i,B_j) &\equiv& \dfrac{[B_i,B_j]}{\sqrt{[B_i,B_i][B_j,B_j]}} \\
	\mathcal{A}(B_i,B_j) &\equiv& \dfrac{\sqrt{[B_i,B_i]}}{\sqrt{[B_j,B_j]}} \\ 
	\mathcal{T}(B_i,B_j) &\equiv& 1 - \sqrt{1-2\mathcal{S}(B_i,B_j)\mathcal{A}(B_i,B_j)+\mathcal{A}(B_i,B_j)^2},
	\label{eq:correlators}
\end{eqnarray}
where the square brackets notation above from \cite{Hung:2019ygc} is 
\begin{equation}
	[B_i,B_j] \propto \sum_n \beta^{R(i)}_n \, \beta^{R(j)}_n,
\end{equation}
such that $[B_i,B_j]$ is proportional to our $\llangle wB_i | wB_j \rrangle$.
The shape correlator $\mathcal{S}$ takes values between -1 and 1 and is insensitive to constant multiplicative factors that change the bispectrum amplitude, while the amplitude correlator $\mathcal{A}$ can take any positive value.
The total correlator $\mathcal{T}$ is sensitive to both the shape and the amplitude of the bispectrum, such that if both the shape and amplitude are perfectly matched, then $\mathcal{S}=\mathcal{A}=\mathcal{T}=1$, and if either the shape or the amplitude are not perfectly reproduced then $\mathcal{T} < 1$.
These correlators have an intuitive quantitative meaning in cases where an amplitude parameter (like $f_{\rm NL}$, the amplitude of primordial non-Gaussianity, for example) is measured with Gaussian data covariances \cite{Hung:2019ygc}.

\begin{figure}[t]
\centering 
\includegraphics[width=\textwidth]{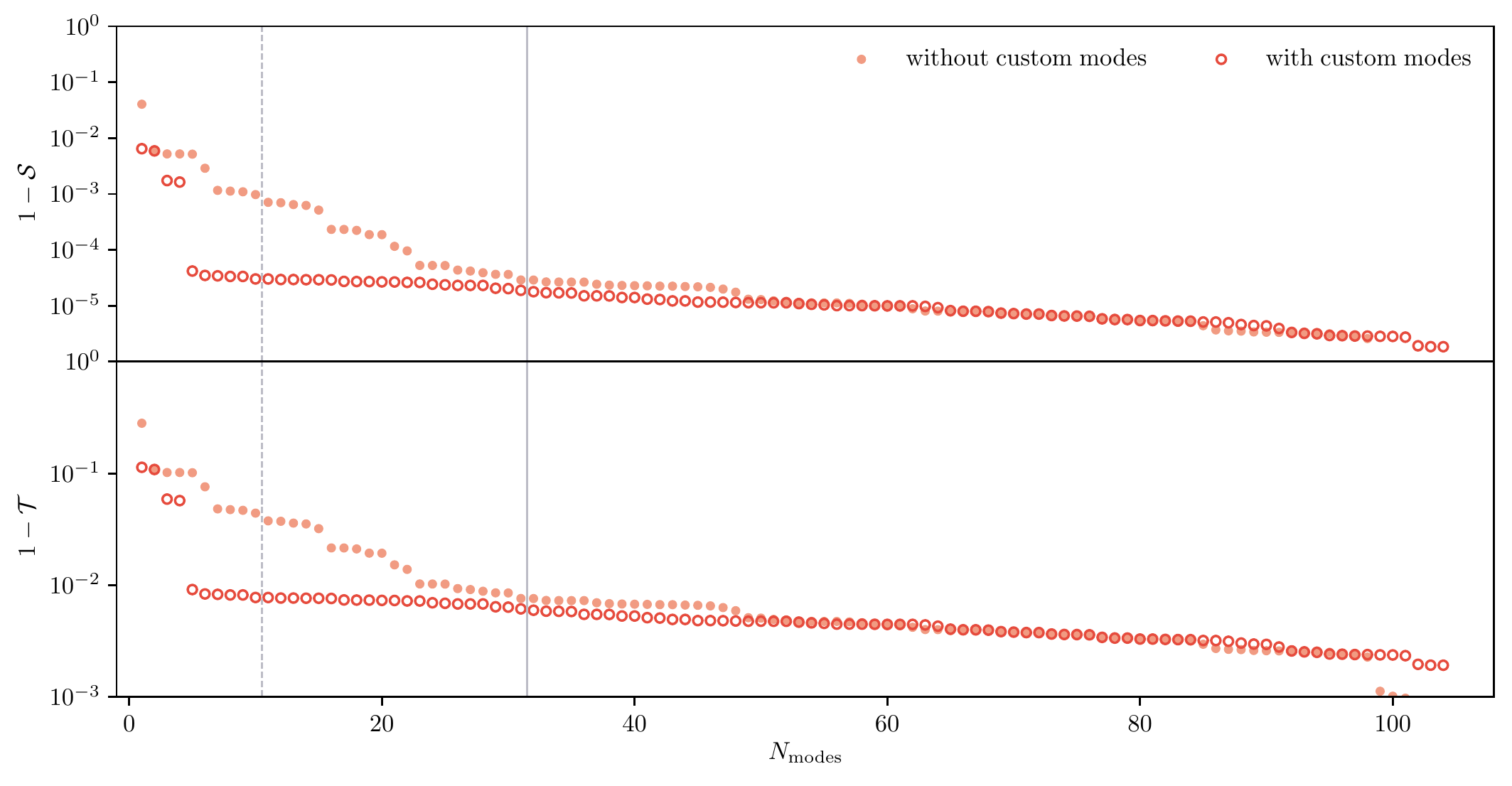}
\caption[Correlators as a function of $n_{\rm max}$]{Shape correlator $\mathcal{S}$ (top) and total correlator $\mathcal{T}$ (bottom) for $k_{\rm max}=24.5\,k_f$. 
Smaller values of $1-\mathcal{S}$ and $1-\mathcal{T}$ correspond to smaller differences in the reconstructed bispectrum compared to the case where the maximum number of modes are used.
The vertical gray lines mark $N_{\rm modes}=10$ and $31$, which we previously found was sufficient to obtain robust parameter constraints when custom modes are included or excluded, respectively.
}
\label{fig:correlators} 
\end{figure}

We show plots of $\mathcal{S}$ and $\mathcal{T}$ between the reconstructed bispectrum with $n$ modes, $B_i = B_{\rm rec}(N_{\rm modes}=n)$, and the reconstructed bispectrum with 108 modes, $B_j = B_{\rm rec}(N_{\rm modes}=108)$, in Fig.~\ref{fig:correlators}.
The figure is for $k_{\rm max}=24.5\,k_f \approx 0.10\, h\,{\rm Mpc}^{-1}$ and considers the mean modal bispectrum from 298 Minerva simulations.
The vertical gray lines at $N_{\rm modes}=10$ and $31$ mark the number of modes that we previously found in Sections \ref{subsec:benchmark} and \ref{subsec:checks} were sufficient for converged parameter constraints, with and without custom modes, respectively.
Both panels show that the correlators are already $\mathcal{S}, \mathcal{T} \gtrsim 0.99$ with six custom modes, and further improvements are gained slowly as more modes are accumulated.
The correlators approach unity more slowly in the absence of custom modes, but these too show small improvements after $\sim 25$ modes.
This behavior implies that the $\mathcal{S}$ and $\mathcal{T}$ correlators are not good indicators for predicting how many modes would be sufficient to use the modal bispectrum to constrain cosmological parameters of interest.
Given only the information in Fig.~\ref{fig:correlators}, it is not obvious how many modes will be needed for any specific purpose.
This is partly because the correlators do not take into account other information that will influence the number of sufficient modes, such as which parameters the modal pipeline will be used to measure.

In this work, we determined a sufficient number of modes by checking that parameter posteriors had converged.
However, this requires many steps, including measuring modal coefficients from a large number of simulations and running MCMC simulations multiple times for different $N_{\rm modes}$ to validate our results.
In the absence of these, one may consider calculating Fisher forecasts to estimate the number of modes that would be needed, to check that the modal bispectrum still provides a good compression of the information in the bispectrum.

Commonly in Fisher matrix analyses, the fiducial parameter values are kept fixed, and parameter errors are forecasted.
However, in this work we found that even though very few modes are needed to reproduce the size and degeneracy directions of the parameter contours, more modes are typically needed to reduce the bias in the positions of the contours in parameter space. 
(This implies that the modes are better at capturing the derivatives of the bispectrum with respect to our chosen parameters, than it is at capturing the mean bispectrum.)
Therefore, it is also prudent to estimate the bias using the Fisher formalism \cite{Tegmark1997,Knox1998}.

\begin{figure}[t]
\centering 
\includegraphics[width=\textwidth]{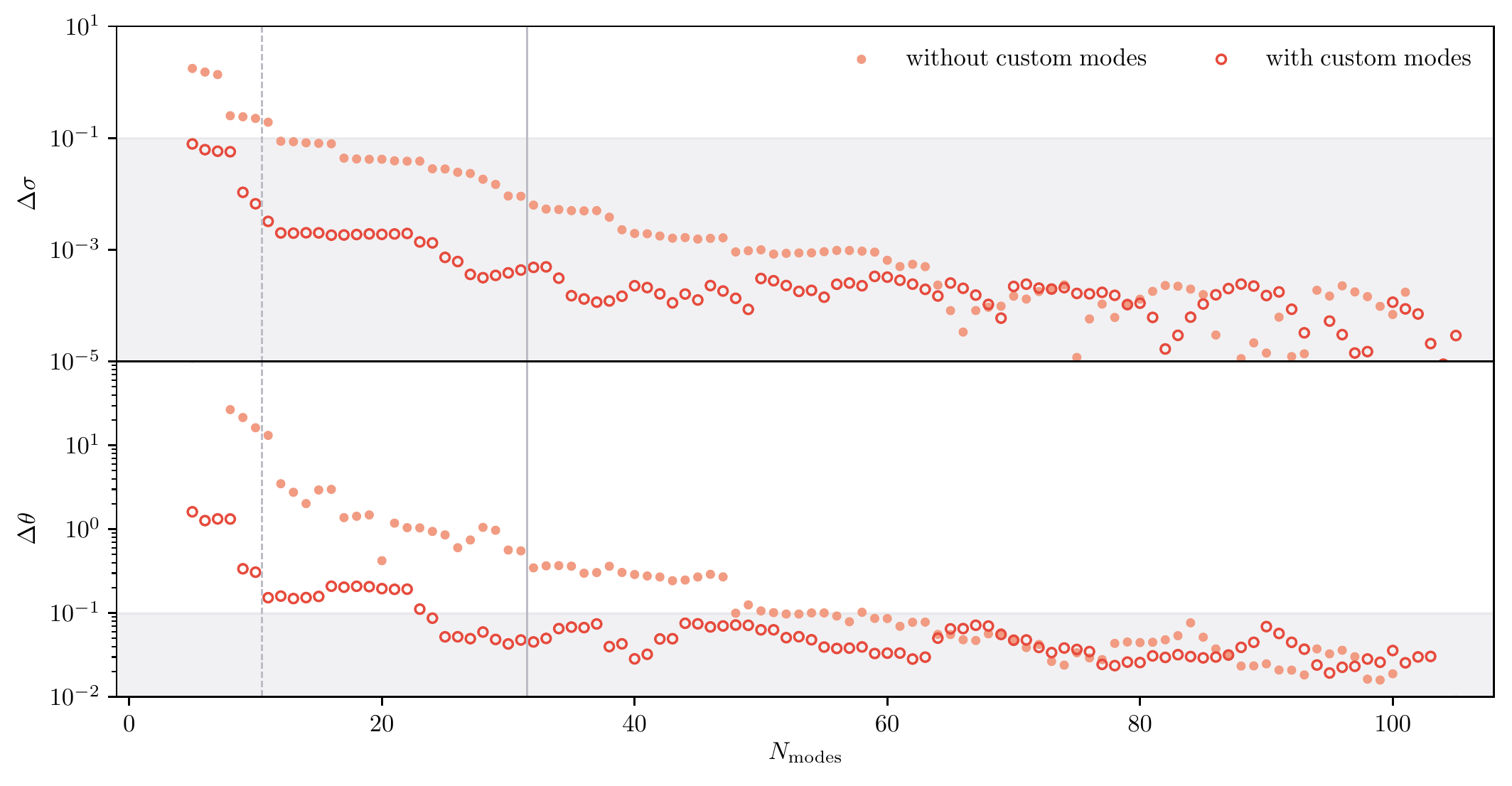}
\caption[Fisher forecasts for modal convergence]{Results from Fisher forecasts on the parameter error (top) and bias (bottom), defined in eqs.~\eqref{eq:Delta sigma Nmodes} and \eqref{eq:Delta theta Nmodes}, as a function of total modes used. 
The vertical gray lines mark $N_{\rm modes}=10$ and $31$, which we previously found was sufficient to obtain robust parameter constraints when custom modes are included or excluded, respectively.
The shaded gray regions are where the error is within 10\%, and the bias is within $0.1\,\sigma$, of the result with max $N_{\rm modes}$.
}
\label{fig:fisher} 
\end{figure}

The Fisher matrix corresponding to the modal pipeline is
\begin{equation}
	F_{ij} = \sum_{n,m}^{N_{\rm modes}-1} \frac{\partial \beta^R_n}{\partial \theta_i} \; \hat{C}(N_{\rm modes})^{-1}_{nm} \; \frac{\partial \beta^R_m}{\partial \theta_j},
\end{equation}
where the partial derivatives are evaluated for our tree-level model at a chosen fiducial.
The parameter covariance matrix is then $F^{-1}$.
The bias in the parameters due to an unaccounted for systematic error can be estimated as (e.g.~\cite{Amara2008})
\begin{eqnarray}
	b(\theta_i) &=& (F^{-1})_{ij} \, b_j \\
	b_j &=& \sum_{n,m}^{{\rm max}\;N_{\rm modes}-1} \beta^{R,{\rm sys}}_n \; \hat{C}({\rm max}\;N_{\rm modes})^{-1}_{nm} \; \frac{\partial \beta^R_m}{\partial \theta_j}
	\label{eq:bias bj}
\end{eqnarray}
where $\beta^{R,{\rm sys}}_n$ is a source of residual systematic uncertainty. In our case, to calculate the bias due to truncating at a certain number of modes,
\begin{eqnarray}
	\beta^{R, {\rm true}}_n &=& \beta^{R, {\rm obs}}_n \quad \ \ {\rm for}\ n < {\rm max}\;N_{\rm modes} \\
	\beta^{R, {\rm truncated}}_n &=& 
	\begin{cases}
    \beta^{R, {\rm obs}}_n & {\rm for}\ n < N_{\rm modes} \\
    0 & \text{otherwise}
    \end{cases} \\
	\beta^{R, {\rm sys}}_n &=& \beta^{R, {\rm true}}_n - \beta^{R, {\rm truncated}}_n \nonumber \\
	&=& \begin{cases}
 	0 & {\rm for}\ n < N_{\rm modes} \\
 	\beta^{R, {\rm obs}}_n & {\rm for}\ n \geq N_{\rm modes},
 	\end{cases}
\end{eqnarray}
where $\beta^{R, {\rm obs}}_n$ is the average measured from the Minerva simulations.
We set ${\rm max}\;N_{\rm modes}=108$ if custom modes are included, and ${\rm max}\;N_{\rm modes}=102$ if they are not.
Taking the case where $k_{\rm max}=24.5 \, k_f$, we show the Fisher forecasted errors and bias in Fig.~\ref{fig:fisher}.
Explicitly, we show $|\Delta\sigma|$ and $|\Delta\theta|$, where
\begin{eqnarray}
	\Delta\theta &\equiv& \frac{b(\theta)}{\sigma_\theta({\rm max}\;N_{\rm modes})}
	\label{eq:Delta theta Nmodes} \\
	\Delta\sigma &\equiv& \frac{\sigma_\theta(N_{\rm modes})}{\sigma_\theta({\rm max}\;N_{\rm modes})}-1,
	\label{eq:Delta sigma Nmodes}
\end{eqnarray}
and $\sigma_\theta$ is the Fisher forecasted error for parameter $\theta$.
To keep the plot simple, at each $N_{\rm modes}$ we have plotted the $|\Delta\sigma|$ and $|\Delta\theta|$ that is the largest among the five parameters.
This illustrates a simple case where we already have MCMC simulations, and we are simply verifying, after the fact, that Fisher forecasts can provide similar indications of modal expansion convergence.

However, we note that the bias due to a truncation of the modal expansion is a purely non-Gaussian effect that cannot be estimated without the non-Gaussian covariance matrix.
This is because the Gaussian covariance is diagonal, such that the $b_j$ in eq.~\eqref{eq:bias bj} would only be sensitive to the systematics in those $\beta^{R,{\rm sys}}_n$ modes that also explicitly vary with the $\theta$ parameters of the modeling.
Still, once a non-Gaussian covariance matrix is obtained, the Fisher formalism may allow for a forecast of how many modes are necessary for the errors and bias to converge to a desired level. 
This may be useful if a non-Gaussian covariance matrix is available, but one wants to estimate roughly how many modes may be needed without running potentially expensive MCMC simulations for multiple scenarios.


\subsection{Covariances}
\label{subsec:cov}

All of the results we have discussed so far used covariance matrices estimated from the full set of 10,000 Pinocchio mocks.
In this section, we explore how the modal bispectrum constraints are sensitive to the covariance matrix that is used. 
Unless otherwise mentioned explicitly, the results in this subsection use $k_{\rm max} = 24.5 \, k_f \approx 0.10\, h \,{\rm Mpc}^{-1}$ and 10 modes.

\begin{figure}[t]
\centering 
\includegraphics[width=\textwidth]{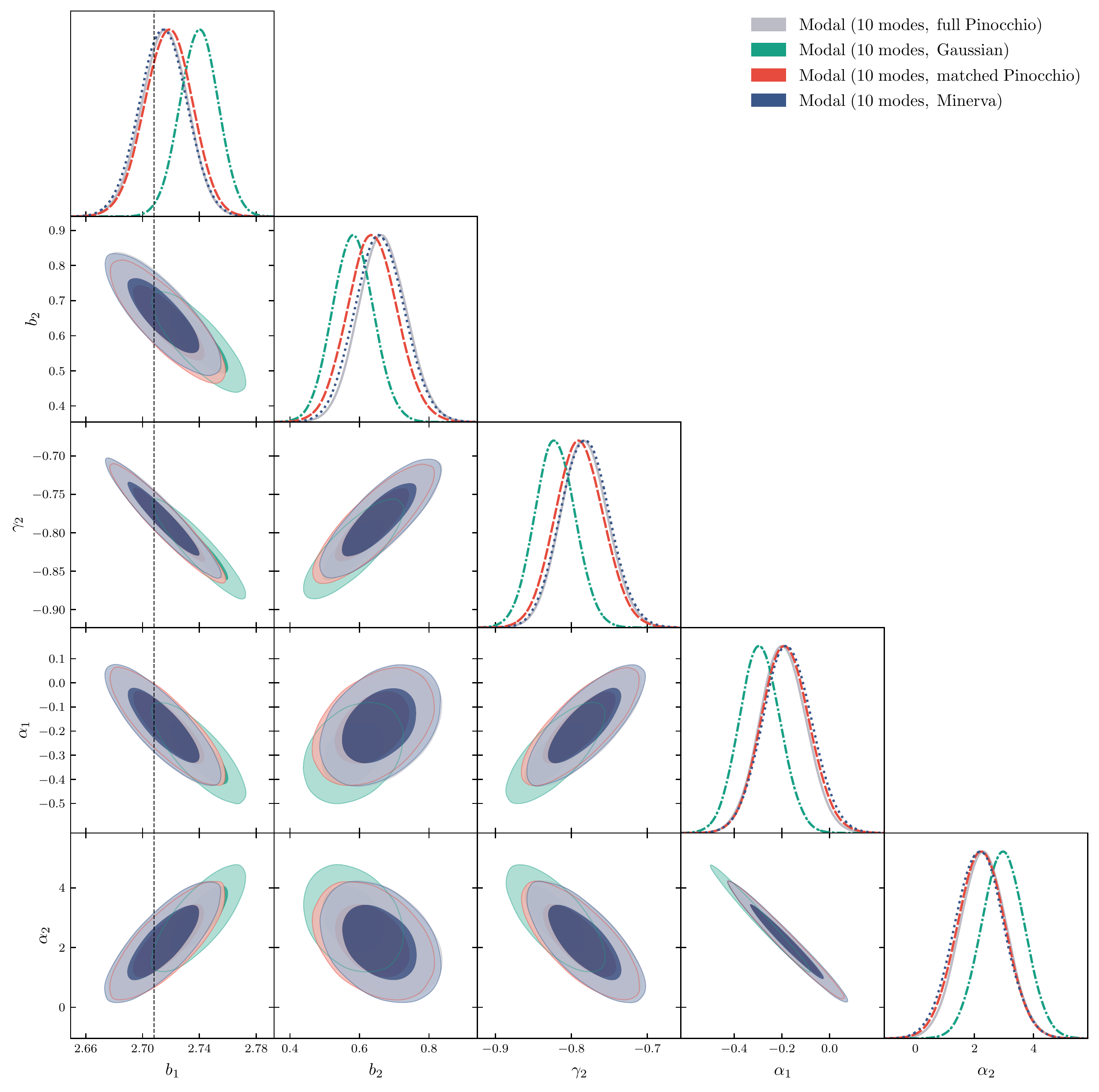}
\caption[Impact of difference covariance matrix estimates]{Comparison of parameter constraints when different covariance matrices are used to obtain constraints at $k_{\rm max}=24.5 \, k_f \approx 0.10\, h\,{\rm Mpc}^{-1}$ using 10 modes.
The Gaussian covariance matrix (green) leads to biased and underestimated parameter constraints, but the covariance matrices estimated from 298 Minerva simulations (dark blue) and 298 Pinocchio simulations with matched initial conditions (red) are in good agreement with the benchmark modal constraints that used the full set of 10,000 Pinocchio simulations (gray, nearly identical to the dark blue and red contours).}
\label{fig:covariance comparison} 
\end{figure}

In the limit of Gaussian covariance, the covariance matrix for the $\beta^R_n$ modal coefficients is diagonal, with the same variance for each $n$.
When $k_{\rm max}=13.5 \, k_f$ with six modes, we find that the Gaussian covariance approximation is very accurate, giving the same constraints as the fully non-Gaussian covariance matrix estimated from 10,000 Pinocchio mocks.
However, when $k_{\rm max}=24.5 \, k_f$ with 10 modes, as shown in Fig.~\ref{fig:covariance comparison}, the Gaussian covariance underestimates the parameter errors by up to 20\%, and the constraints are biased by up to $2\,\sigma$, depending on the parameter.
Fig.~\ref{fig:covariance comparison} also shows that constraints using covariance matrices estimated from 298 Minerva simulations, 298 Pinocchio mocks with matched initial conditions, and the full set of 10,000 Pinocchio mocks are in good agreement: parameter errors agree to within 10\% and  biases are small, less than $\sim 0.4\,\sigma$.

\begin{figure}[t]
\centering 
\includegraphics[width=\textwidth,keepaspectratio]{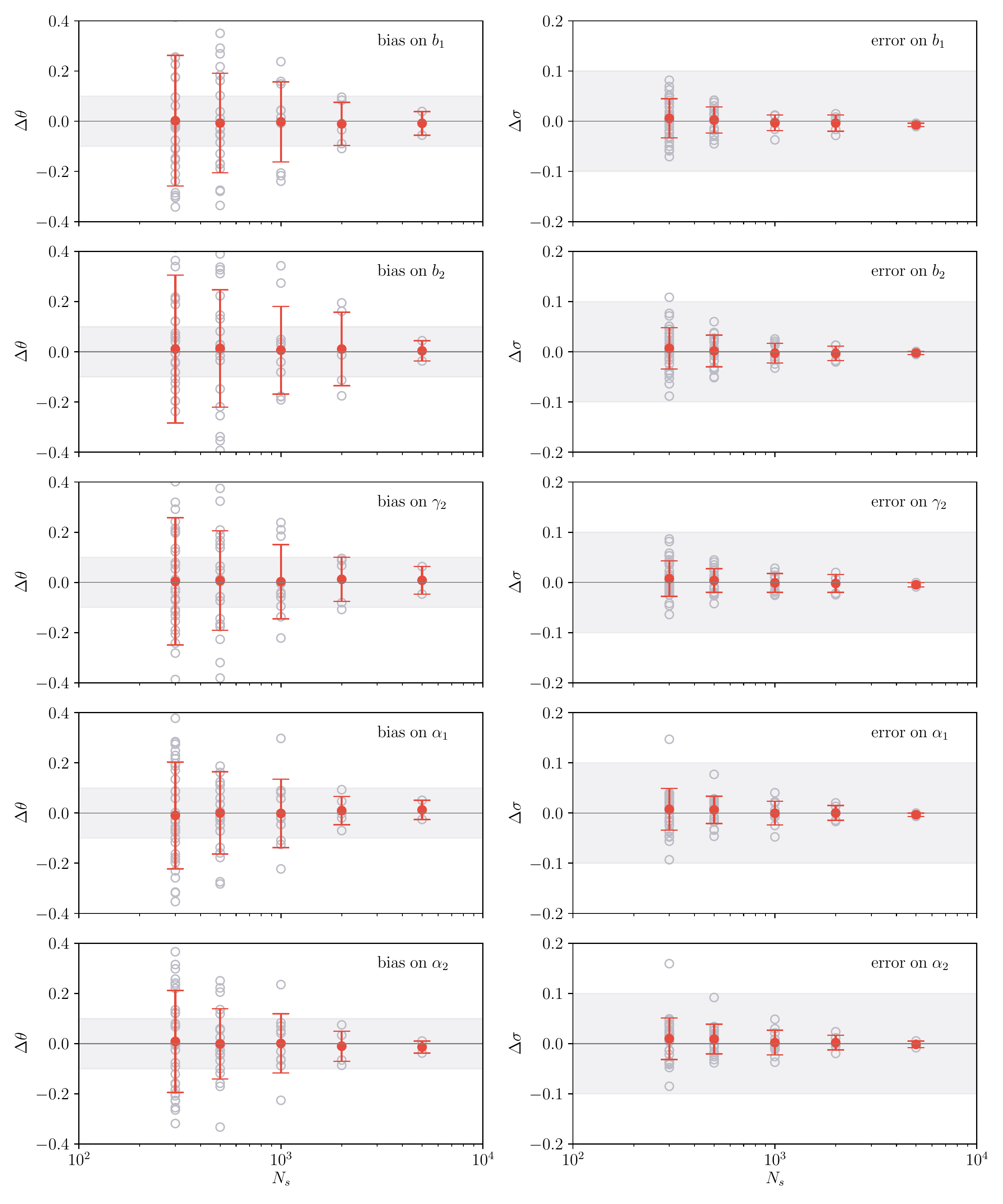}
\caption[Convergence of the covariance matrix]{Comparison of the bias and change in parameter errors, for $k_{\rm max}=24.5 \, k_f$ with 10 modes, as a function of how many Pinocchio mocks, $N_s$, are used to estimate the covariance. 
$\Delta\theta$ and $\Delta\sigma$ are defined in eqs.~\eqref{eq:Delta theta Ns} and \eqref{eq:Delta sigma Ns}.
The gray circles mark the result from each set of $N_s$ mocks, while the red error bars are the $1\,\sigma$ scatter of the gray circles. The gray shaded bands mark the regions where deviations are within $0.1\,\sigma$ of the benchmark result using $N_s=10^4$.}
\label{fig:covariance convergence} 
\end{figure}

In Fig.~\ref{fig:covariance convergence}, we show how the parameter means and errors can depend on the number of mocks used to estimate the covariance matrix.
From the 10,000 Pinocchio mocks, we take subsets of the mocks divided into groups of $N_s$ and compare the resulting constraints.
This is along the lines of \cite{Blot:2015cvj}, which considered the impact of covariance matrix errors on cosmological parameter constraints from the power spectrum.
Similarly to eqs.~\eqref{eq:Delta theta Nmodes} and \eqref{eq:Delta sigma Nmodes}, we define and show
\begin{eqnarray}
\Delta\theta &\equiv& \frac{\overline{\theta}(N_s) - \overline{\theta}(N_s=10^4)}{\sigma_\theta(N_s=10^4)}
	\label{eq:Delta theta Ns} \\
	\Delta\sigma &\equiv& \frac{\sigma_\theta(N_s)}{\sigma_\theta(N_s=10^4)}-1
	\label{eq:Delta sigma Ns}.
\end{eqnarray}
Each subset of mocks corresponds to a single gray circle in each panel of Fig.~\ref{fig:covariance convergence}, while the red points and error bars show the mean and standard deviation of the gray circles at one value of $N_s$.
This comparison shows that the parameter errors are recovered to within 10\% with only 300 mocks, but many more mocks are typically necessary to reduce the bias to the same level. For example, $N_s > 2000$ would be needed to reduce the bias to $\lesssim 0.1\,\sigma$.
One caveat to this result, however, is that the red points in the plot are not independent, since they are dividing up the same realizations, just in different groups. 
This fact will tend to make the different $N_s$ appear more consistent with the case we are comparing with, using all 10,000 mocks.

\subsection{Gaussian vs Sellentin-Heavens likelihood}
\label{subsec:shlike}

\begin{figure}[t]
\centering 
\includegraphics[height=0.8\textheight,keepaspectratio]{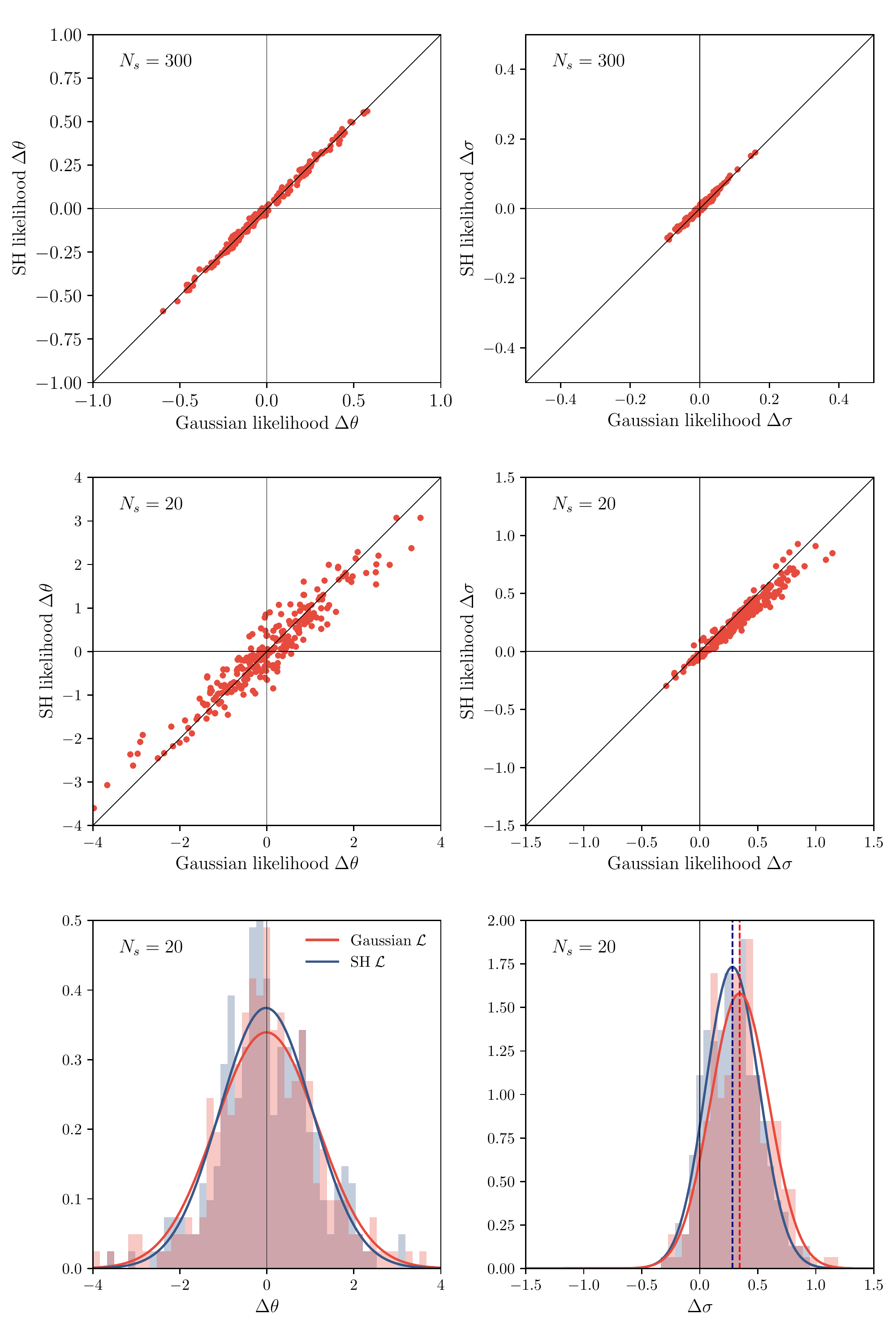}
\caption[Gaussian vs Sellentin-Heavens likelihood]{Comparison of the bias (left column) and errors (right column) obtained with the Gaussian and SH likelihoods when covariance matrices are estimated using $N_s=300$ (top row) or $N_s=20$ (middle row) mocks. 
$\Delta\theta$ and $\Delta\sigma$ are defined in eqs.~\eqref{eq:Delta theta Ns} and \eqref{eq:Delta sigma Ns}.
The bottom row shows histograms of posterior biases and sizes for the likelihoods when $N_s=20$. 
The solid lines are Gaussian fits to the histograms and show that the Sellentin-Heavens likelihood on average is slightly closer to the answer given by the full set of $10^4$ mocks.
}
\label{fig:shlike}
\end{figure}

We compared the two likelihoods when all $N_s = 10^4$ mocks are used to estimate the covariance, and we find that they result in indistinguishable parameter constraints, which is the expected behavior when $N_s$ is very large.
For some smaller value of $N_s$, we expect that the two likelihoods will show different results. 
For $k_{\rm max}=24.5 \, k_f$ with 10 modes, we show this comparison for two values of $N_s$, $N_s=20$ and 300, in Fig.~\ref{fig:shlike}.
For $N_s=20$, we simulate 50 analyses, and for $N_s=300$ we simulate 33. 
$\Delta\theta$ and $\Delta\sigma$ are defined in eqs.~\eqref{eq:Delta theta Ns} and \eqref{eq:Delta sigma Ns}, and for each MCMC simulation, we plot five points for $\Delta\theta$ and $\Delta\sigma$, one point for each parameter.

The top row of Fig.~\ref{fig:shlike} shows that $N_s=300$ is sufficiently high to make differences between the two likelihoods negligible when constraints are compared between individual sets of 300 mocks.
When $N_s=20$ (middle row), which is much closer to the number of data bins, $N_{\rm modes}=10$, the two likelihoods can produce posteriors that are noticeably different.
For any one analysis using 20 mocks, the two posteriors will be biased relative to each other and can produce parameter errors that are too big or too small relative to the true answer, which we assume is the result from $10^4$ mocks.
On average though, both likelihoods have posteriors that are unbiased relative to the true answer, but the parameter error from both likelihoods will be larger than the case with $N_s = 10^4$. 
This is expected, because the parameter constraints should be worse when the covariance is less well estimated from fewer mocks.
However, with the SH likelihood, the result on average is closer to the truth: the scatter in the bias $|\Delta \theta|$ is smaller (bottom left panel), the average parameter errors are closer to the truth (bottom right panel), and the distribution of parameter errors are more tightly scattered around the true value (also bottom right panel).

Despite this, we find that in practice it does not matter which likelihood is implemented, because in either case enough mocks would have to be used to ensure that the parameter constraints are not dominated by the covariance matrix error.
In this work, we find that once $N_s$ is large enough for either likelihood to be stable to within a few tens of per cent (as shown in Fig.~\ref{fig:covariance convergence}),
the two likelihoods will produce identical results.

\section{Conclusions}
\label{sec:conclusions}

In this work, we have implemented an MCMC analysis using the compressed modal bispectrum for the first time. 
By using the same data, modeling, and analysis choices as \cite{Oddo:2019run}, we are able to rigorously compare the constraints from the standard bispectrum estimator and the modal bispectrum estimator within a controlled setting.
Specifically, we use the real-space tree-level halo bispectrum model to constrain the halo bias and shot noise parameters measured in the Minerva $N$-body simulations, which represents an idealized survey with volume $\approx 1{,}000 \, h^{-3}\,{\rm Gpc}^3$.

Our key result is that the modal bispectrum provides a very efficient compression of the information in the bispectrum, while requiring minimal new calculations compared to the standard bispectrum analysis;
the critical components of the pipeline are the the modal estimator in eqs.~\eqref{eq:QwB estimator with Mrx} and \eqref{eq:Mrx} and the inner product matrix, $\gamma$, and both can be computed with minor modifications to the standard bispectrum estimator.
We find that for $k_{\rm max} = 13.5 \, k_f \approx 0.06\, h\,{\rm Mpc}^{-1}$ $(24.5\, k_f \approx 0.10\, h\,{\rm Mpc}^{-1})$, the constraints on halo bias and shot noise parameters converge with only 6 (10) modal coefficients, yielding very similar constraints compared to the standard bispectrum analysis in \cite{Oddo:2019run} that used $\sim 20$ to 1,600 triangle bins.
We showed that this convergence of the constraints with $N_{\rm modes}$ is only qualitatively reflected by the shape and total correlators (in Section \ref{subsec:correlators}), but Fisher forecasts can estimate the $N_{\rm modes}$ needed for parameter constraints to converge to a desired level.

We tested the robustness of the modal bispectrum constraints to different user choices within the modal pipeline implementation. 
We find that the choice between the normal polynomials or shifted Legendre polynomials for constructing the $Q_n$ basis functions has no impact on the results, but using a near-optimal weighting function $w$ to weight Fourier triangles and including some customized basis functions, like $Q_n^{\rm tree}$, that are more tuned to the parameters being constrained can help minimize the number of modes needed for more efficient compression.
We also compared different methods of computing the inner product matrix, $\gamma_{nm} \equiv \llangle Q_n|Q_m \rrangle$, a critical piece of the pipeline, and find that only the 3D FFT method is always correct, though other methods appear to be approximately correct for higher $k_{\rm max}$. 
This is because the 3D FFT method calculates the inner product between basis functions, $\llangle Q_n|Q_m \rrangle$, on the Fourier-space grid in the same way that the modal estimator calculates the inner product between basis functions and the data, $\llangle Q|w\mathcal{B} \rrangle$, treating data and theory in the most consistent way possible.
The voxel and 1D FFT methods, on the other hand, take the continuous limit of the inner product, which becomes a good approximation when the Fourier grid is very fine.

We also noted that, while the modal bispectrum and standard bispectrum estimators are both summary statistics of the true bispectrum, they are performing different operations on the density grid in Fourier space, $\delta(\bk)$. 
Thus they are not always interchangeable and care should be taken when comparing the two. 
To illustrate this, we have shown that they agree on the mean bispectrum averaged over many simulations (i.e.~the mean standard bispectrum estimator vs the mean reconstructed bispectrum), but for one realization they give different answers for the triangle-dependence of the measured bispectrum. 
Additionally, the two estimators have different error properties, and the covariance of $B_{\rm rec}$ cannot be used as a substitute for the covariance of measurements made with the standard bispectrum estimator.

The highly efficient compression achieved by the modal bispectrum and the large number of simulations available have allowed us to explore how the modal estimator constraints depend on the the number of simulations used to estimate the covariance, $N_s$, and whether $N$-body simulations or Pinocchio approximate mocks are used.
Such calculations can usually only be done in a limited way for bispectrum data sets using the standard estimator because of its much larger size.
We find that the covariance matrices from 298 $N$-body simulations, 298 Pinocchio mocks with matched initial conditions, and the full set of 10,000 mocks lead to constraints that are biased by up to $\sim 0.4\,\sigma$ relative to each other, and to reduce this bias to $\lesssim 0.1\,\sigma$ would require $N_s > 2{,}000$ mocks.
We also show that the Gaussian and Sellentin-Heavens likelihood functions only show different results when $N_s$ is extremely low.
However, since $N_s$ should be large enough such that the error in the covariance matrix estimate is subdominant with either likelihood function (at which point the two likelihoods give identical parameter constraints), in practice either likelihood could be used, though in principle the Sellentin-Heavens likelihood is more theoretically motivated from a Bayesian perspective.

This work has shown that the modal method remains a promising avenue for accessing cosmological information in the bispectrum through a compressed data set, and we have developed a better understanding of how to implement and interpret the estimator and its results.
However, the modal bispectrum pipeline presented here would require further work before it could be applied to a realistic galaxy catalog, including redshift-space distortions, and potentially probing a larger range of scales (higher $k_{\rm max}$), where a theoretical model beyond the tree-level SPT bispectrum would be necessary.
A larger $k_{\rm max}$ will most likely require more modes, ideally including more custom modes that would be theoretically motivated.
If a theoretical model beyond the separable tree-level SPT one is used, computational methods in the pipeline will need to be adapted so that the inner product $\llangle Q| wB^{\rm theory} \rrangle$ could still be computed quickly.
(The analogous problem in the standard bispectrum analysis is handled by evaluating the theoretical bispectrum at an effective triangle in each triangle bin \cite{Oddo:2019run}.)
It is likely that if $k_{\rm max}$ is sufficiently high, the calculation of the inner product could be well-approximated by another method that does not require evaluating the theory on the Fourier-space grid.
Additionally, the modal bispectrum method presented here would need to be extended to capture anisotropies coming from redshift-space distortions \cite{Regan:2017vgi}, which are always present in real observations and also act as a source of more cosmological information.
We plan to investigate these outstanding issues in future work.

\acknowledgments

We are grateful to Pierluigi Monaco for providing the Pinocchio mock halo catalogs and to Claudio Dalla Vecchia and Ariel S\'{a}nchez for providing the Minerva $N$-body simulations.
We thank Dionysios Karagiannis for suggesting the use of the 1D FFT inner product method.
We also wish to thank the Institute for Fundamental Physics of the Universe (IFPU) in Trieste, Italy for hosting the workshop of the Euclid Galaxy Clustering Higher-order Statistics Work Package where part of this work was done.

JB is supported by the Sinergia Grant No.~173716 from the Swiss National Science Foundation. ES acknowledges support from PRIN MIUR 2015 Cosmology and Fundamental Physics: illuminating the Dark Universe with Euclid.

The modal bispectrum analysis was performed on the Baobab cluster at the University of Geneva.

\bibliographystyle{JHEP}
\bibliography{references}

\providecommand{\href}[2]{#2}\begingroup\raggedright\begin{thebibliography}{10}

\bibitem{Gil-Marin:2014sta}
H.~Gil-Marín, J.~Noreña, L.~Verde, W.~J. Percival, C.~Wagner, M.~Manera
  et~al., \emph{{The power spectrum and bispectrum of SDSS DR11 BOSS galaxies
  -- I. Bias and gravity}},
  \href{https://doi.org/10.1093/mnras/stv961}{\emph{Mon. Not. Roy. Astron.
  Soc.} {\bfseries 451} (2015) 539}
  [\href{https://arxiv.org/abs/1407.5668}{{\ttfamily 1407.5668}}].

\bibitem{Gil-Marin:2014baa}
H.~Gil-Marín, L.~Verde, J.~Noreña, A.~J. Cuesta, L.~Samushia, W.~J. Percival
  et~al., \emph{{The power spectrum and bispectrum of SDSS DR11 BOSS galaxies
  -- II. Cosmological interpretation}},
  \href{https://doi.org/10.1093/mnras/stv1359}{\emph{Mon. Not. Roy. Astron.
  Soc.} {\bfseries 452} (2015) 1914}
  [\href{https://arxiv.org/abs/1408.0027}{{\ttfamily 1408.0027}}].

\bibitem{Gil-Marin:2016wya}
H.~Gil-Marín, W.~J. Percival, L.~Verde, J.~R. Brownstein, C.-H. Chuang, F.-S.
  Kitaura et~al., \emph{{The clustering of galaxies in the SDSS-III Baryon
  Oscillation Spectroscopic Survey: RSD measurement from the power spectrum and
  bispectrum of the DR12 BOSS galaxies}},
  \href{https://doi.org/10.1093/mnras/stw2679}{\emph{Mon. Not. Roy. Astron.
  Soc.} {\bfseries 465} (2017) 1757}
  [\href{https://arxiv.org/abs/1606.00439}{{\ttfamily 1606.00439}}].

\bibitem{Slepian:2016kfz}
Z.~Slepian et~al., \emph{{Detection of baryon acoustic oscillation features in
  the large-scale three-point correlation function of SDSS BOSS DR12 CMASS
  galaxies}}, \href{https://doi.org/10.1093/mnras/stx488}{\emph{Mon. Not. Roy.
  Astron. Soc.} {\bfseries 469} (2017) 1738}
  [\href{https://arxiv.org/abs/1607.06097}{{\ttfamily 1607.06097}}].

\bibitem{Pearson:2017wtw}
D.~W. Pearson and L.~Samushia, \emph{{A Detection of the Baryon Acoustic
  Oscillation features in the SDSS BOSS DR12 Galaxy Bispectrum}},
  \href{https://doi.org/10.1093/mnras/sty1266}{\emph{Mon. Not. Roy. Astron.
  Soc.} {\bfseries 478} (2018) 4500}
  [\href{https://arxiv.org/abs/1712.04970}{{\ttfamily 1712.04970}}].

\bibitem{PearsonSamushia2018errata}
D.~W. Pearson and L.~Samushia, \emph{{Erratum: A Detection of the Baryon
  Acoustic Oscillation Features in the SDSS BOSS DR12 Galaxy Bispectrum}},
  \href{https://doi.org/10.1093/mnras/sty3173}{\emph{Mon. Not. Roy. Astron.
  Soc.} {\bfseries 483} (2018) 915}.

\bibitem{Sugiyama:2018yzo}
N.~S. Sugiyama, S.~Saito, F.~Beutler and H.-J. Seo, \emph{{A complete FFT-based
  decomposition formalism for the redshift-space bispectrum}},
  \href{https://doi.org/10.1093/mnras/sty3249}{\emph{Mon. Not. Roy. Astron.
  Soc.} {\bfseries 484} (2019) 364}
  [\href{https://arxiv.org/abs/1803.02132}{{\ttfamily 1803.02132}}].

\bibitem{Levi:2013gra}
{\scshape DESI} collaboration, \emph{{The DESI Experiment, a whitepaper for
  Snowmass 2013}},  \href{https://arxiv.org/abs/1308.0847}{{\ttfamily
  1308.0847}}.

\bibitem{Laureijs:2011gra}
{\scshape EUCLID} collaboration, \emph{{Euclid Definition Study Report}},
  \href{https://arxiv.org/abs/1110.3193}{{\ttfamily 1110.3193}}.

\bibitem{Dore:2014cca}
O.~Dor\'e et~al., \emph{{Cosmology with the SPHEREX All-Sky Spectral Survey}},
  \href{https://arxiv.org/abs/1412.4872}{{\ttfamily 1412.4872}}.

\bibitem{Spergel:2015sza}
D.~Spergel et~al., \emph{{Wide-Field InfrarRed Survey Telescope-Astrophysics
  Focused Telescope Assets WFIRST-AFTA 2015 Report}},
  \href{https://arxiv.org/abs/1503.03757}{{\ttfamily 1503.03757}}.

\bibitem{Chan:2016ehg}
K.~C. Chan and L.~Blot, \emph{{Assessment of the Information Content of the
  Power Spectrum and Bispectrum}},
  \href{https://doi.org/10.1103/PhysRevD.96.023528}{\emph{Phys. Rev. D}
  {\bfseries 96} (2017) 023528}
  [\href{https://arxiv.org/abs/1610.06585}{{\ttfamily 1610.06585}}].

\bibitem{Byun:2017fkz}
J.~Byun, A.~Eggemeier, D.~Regan, D.~Seery and R.~E. Smith, \emph{{Towards
  optimal cosmological parameter recovery from compressed bispectrum
  statistics}}, \href{https://doi.org/10.1093/mnras/stx1681}{\emph{Mon. Not.
  Roy. Astron. Soc.} {\bfseries 471} (2017) 1581}
  [\href{https://arxiv.org/abs/1705.04392}{{\ttfamily 1705.04392}}].

\bibitem{Song:2015gca}
Y.-S. Song, A.~Taruya and A.~Oka, \emph{{Cosmology with anisotropic galaxy
  clustering from the combination of power spectrum and bispectrum}},
  \href{https://doi.org/10.1088/1475-7516/2015/08/007}{\emph{JCAP} {\bfseries
  08} (2015) 007} [\href{https://arxiv.org/abs/1502.03099}{{\ttfamily
  1502.03099}}].

\bibitem{Gagrani:2016rfy}
P.~Gagrani and L.~Samushia, \emph{{Information Content of the Angular
  Multipoles of Redshift-Space Galaxy Bispectrum}},
  \href{https://doi.org/10.1093/mnras/stx135}{\emph{Mon. Not. Roy. Astron.
  Soc.} {\bfseries 467} (2017) 928}
  [\href{https://arxiv.org/abs/1610.03488}{{\ttfamily 1610.03488}}].

\bibitem{Yankelevich:2018uaz}
V.~Yankelevich and C.~Porciani, \emph{{Cosmological information in the
  redshift-space bispectrum}},
  \href{https://doi.org/10.1093/mnras/sty3143}{\emph{Mon. Not. Roy. Astron.
  Soc.} {\bfseries 483} (2019) 2078}
  [\href{https://arxiv.org/abs/1807.07076}{{\ttfamily 1807.07076}}].

\bibitem{Gualdi:2020ymf}
D.~Gualdi and L.~Verde, \emph{{Galaxy redshift-space bispectrum: the Importance
  of Being Anisotropic}},
  \href{https://doi.org/10.1088/1475-7516/2020/06/041}{\emph{JCAP} {\bfseries
  06} (2020) 041} [\href{https://arxiv.org/abs/2003.12075}{{\ttfamily
  2003.12075}}].

\bibitem{Agarwal:2020lov}
N.~Agarwal, V.~Desjacques, D.~Jeong and F.~Schmidt, \emph{{Information content
  in the redshift-space galaxy power spectrum and bispectrum}},
  \href{https://arxiv.org/abs/2007.04340}{{\ttfamily 2007.04340}}.

\bibitem{Yamauchi:2017ibz}
D.~Yamauchi, S.~Yokoyama and H.~Tashiro, \emph{{Constraining modified theories
  of gravity with the galaxy bispectrum}},
  \href{https://doi.org/10.1103/PhysRevD.96.123516}{\emph{Phys. Rev. D}
  {\bfseries 96} (2017) 123516}
  [\href{https://arxiv.org/abs/1709.03243}{{\ttfamily 1709.03243}}].

\bibitem{Bose:2018zpk}
B.~Bose and A.~Taruya, \emph{{The one-loop matter bispectrum as a probe of
  gravity and dark energy}},
  \href{https://doi.org/10.1088/1475-7516/2018/10/019}{\emph{JCAP} {\bfseries
  10} (2018) 019} [\href{https://arxiv.org/abs/1808.01120}{{\ttfamily
  1808.01120}}].

\bibitem{Bose:2019wuz}
B.~Bose, J.~Byun, F.~Lacasa, A.~Moradinezhad~Dizgah and L.~Lombriser,
  \emph{{Modelling the matter bispectrum at small scales in modified gravity}},
  \href{https://doi.org/10.1088/1475-7516/2020/02/025}{\emph{JCAP} {\bfseries
  02} (2020) 025} [\href{https://arxiv.org/abs/1909.02504}{{\ttfamily
  1909.02504}}].

\bibitem{Tellarini:2016sgp}
M.~Tellarini, A.~J. Ross, G.~Tasinato and D.~Wands, \emph{{Galaxy bispectrum,
  primordial non-Gaussianity and redshift space distortions}},
  \href{https://doi.org/10.1088/1475-7516/2016/06/014}{\emph{JCAP} {\bfseries
  06} (2016) 014} [\href{https://arxiv.org/abs/1603.06814}{{\ttfamily
  1603.06814}}].

\bibitem{Karagiannis:2018jdt}
D.~Karagiannis, A.~Lazanu, M.~Liguori, A.~Raccanelli, N.~Bartolo and L.~Verde,
  \emph{{Constraining primordial non-Gaussianity with bispectrum and power
  spectrum from upcoming optical and radio surveys}},
  \href{https://doi.org/10.1093/mnras/sty1029}{\emph{Mon. Not. Roy. Astron.
  Soc.} {\bfseries 478} (2018) 1341}
  [\href{https://arxiv.org/abs/1801.09280}{{\ttfamily 1801.09280}}].

\bibitem{Ruggeri:2017dda}
R.~Ruggeri, E.~Castorina, C.~Carbone and E.~Sefusatti, \emph{{DEMNUni: Massive
  neutrinos and the bispectrum of large scale structures}},
  \href{https://doi.org/10.1088/1475-7516/2018/03/003}{\emph{JCAP} {\bfseries
  03} (2018) 003} [\href{https://arxiv.org/abs/1712.02334}{{\ttfamily
  1712.02334}}].

\bibitem{Hahn:2019zob}
C.~Hahn, F.~Villaescusa-Navarro, E.~Castorina and R.~Scoccimarro,
  \emph{{Constraining $M_\nu$ with the bispectrum. Part I. Breaking parameter
  degeneracies}},
  \href{https://doi.org/10.1088/1475-7516/2020/03/040}{\emph{JCAP} {\bfseries
  03} (2020) 040} [\href{https://arxiv.org/abs/1909.11107}{{\ttfamily
  1909.11107}}].

\bibitem{Monaco:2016pys}
P.~Monaco, \emph{{Approximate methods for the generation of dark matter halo
  catalogs in the age of precision cosmology}},
  \href{https://doi.org/10.3390/galaxies4040053}{\emph{Galaxies} {\bfseries 4}
  (2016) 53} [\href{https://arxiv.org/abs/1605.07752}{{\ttfamily 1605.07752}}].

\bibitem{Colavincenzo:2018cgf}
M.~Colavincenzo et~al., \emph{{Comparing approximate methods for mock
  catalogues and covariance matrices -- III: bispectrum}},
  \href{https://doi.org/10.1093/mnras/sty2964}{\emph{Mon. Not. Roy. Astron.
  Soc.} {\bfseries 482} (2019) 4883}
  [\href{https://arxiv.org/abs/1806.09499}{{\ttfamily 1806.09499}}].

\bibitem{Joachimi:2016xhk}
B.~Joachimi, \emph{{Non-linear shrinkage estimation of large-scale structure
  covariance}}, \href{https://doi.org/10.1093/mnrasl/slw240}{\emph{Mon. Not.
  Roy. Astron. Soc.} {\bfseries 466} (2017) L83}
  [\href{https://arxiv.org/abs/1612.00752}{{\ttfamily 1612.00752}}].

\bibitem{FriedrichEifler2018}
O.~{Friedrich} and T.~{Eifler}, \emph{{Precision matrix expansion - efficient
  use of numerical simulations in estimating errors on cosmological
  parameters}}, \href{https://doi.org/10.1093/mnras/stx2566}{\emph{Mon. Not.
  Roy. Astron. Soc.} {\bfseries 473} (2018) 4150}
  [\href{https://arxiv.org/abs/1703.07786}{{\ttfamily 1703.07786}}].

\bibitem{Hall:2018umb}
A.~Hall and A.~Taylor, \emph{{A Bayesian method for combining theoretical and
  simulated covariance matrices for large-scale structure surveys}},
  \href{https://doi.org/10.1093/mnras/sty3102}{\emph{Mon. Not. Roy. Astron.
  Soc.} {\bfseries 483} (2019) 189}
  [\href{https://arxiv.org/abs/1807.06875}{{\ttfamily 1807.06875}}].

\bibitem{Pearson:2015gca}
D.~W. Pearson and L.~Samushia, \emph{{Estimating the power spectrum covariance
  matrix with fewer mock samples}},
  \href{https://doi.org/10.1093/mnras/stw062}{\emph{Mon. Not. Roy. Astron.
  Soc.} {\bfseries 457} (2016) 993}
  [\href{https://arxiv.org/abs/1509.00064}{{\ttfamily 1509.00064}}].

\bibitem{Howlett:2017vwp}
C.~Howlett and W.~J. Percival, \emph{{Galaxy two-point covariance matrix
  estimation for next generation surveys}},
  \href{https://doi.org/10.1093/mnras/stx2342}{\emph{Mon. Not. Roy. Astron.
  Soc.} {\bfseries 472} (2017) 4935}
  [\href{https://arxiv.org/abs/1709.03057}{{\ttfamily 1709.03057}}].

\bibitem{Mohammed:2016sre}
I.~Mohammed, U.~Seljak and Z.~Vlah, \emph{{Perturbative approach to covariance
  matrix of the matter power spectrum}},
  \href{https://doi.org/10.1093/mnras/stw3196}{\emph{Mon. Not. Roy. Astron.
  Soc.} {\bfseries 466} (2017) 780}
  [\href{https://arxiv.org/abs/1607.00043}{{\ttfamily 1607.00043}}].

\bibitem{Sugiyama:2019ike}
N.~S. Sugiyama, S.~Saito, F.~Beutler and H.-J. Seo, \emph{{Perturbation theory
  approach to predict the covariance matrices of the galaxy power spectrum and
  bispectrum in redshift space}},
  \href{https://doi.org/10.1093/mnras/staa1940}{\emph{Mon. Not. Roy. Astron.
  Soc.} {\bfseries 497} (2020) 1684}
  [\href{https://arxiv.org/abs/1908.06234}{{\ttfamily 1908.06234}}].

\bibitem{Wadekar:2019rdu}
D.~Wadekar and R.~Scoccimarro, \emph{{The Galaxy Power Spectrum Multipoles
  Covariance in Perturbation Theory}},
  \href{https://arxiv.org/abs/1910.02914}{{\ttfamily 1910.02914}}.

\bibitem{Taruya:2020qoy}
A.~Taruya, T.~Nishimichi and D.~Jeong, \emph{{The covariance of the matter
  power spectrum including the survey window function effect: N-body
  simulations vs. fifth-order perturbation theory on grid}},
  \href{https://arxiv.org/abs/2007.05504}{{\ttfamily 2007.05504}}.

\bibitem{Gualdi:2017iey}
D.~Gualdi, M.~Manera, B.~Joachimi and O.~Lahav, \emph{{Maximal compression of
  the redshift space galaxy power spectrum and bispectrum}},
  \href{https://doi.org/10.1093/mnras/sty261}{\emph{Mon. Not. Roy. Astron.
  Soc.} {\bfseries 476} (2018) 4045}
  [\href{https://arxiv.org/abs/1709.03600}{{\ttfamily 1709.03600}}].

\bibitem{Gualdi:2018pyw}
D.~Gualdi, H.~Gil-Marín, R.~L. Schuhmann, M.~Manera, B.~Joachimi and O.~Lahav,
  \emph{{Enhancing BOSS bispectrum cosmological constraints with maximal
  compression}}, \href{https://doi.org/10.1093/mnras/stz051}{\emph{Mon. Not.
  Roy. Astron. Soc.} {\bfseries 484} (2019) 3713}
  [\href{https://arxiv.org/abs/1806.02853}{{\ttfamily 1806.02853}}].

\bibitem{Heavens:1999am}
A.~Heavens, R.~Jimenez and O.~Lahav, \emph{{Massive lossless data compression
  and multiple parameter estimation from galaxy spectra}},
  \href{https://doi.org/10.1046/j.1365-8711.2000.03692.x}{\emph{Mon. Not. Roy.
  Astron. Soc.} {\bfseries 317} (2000) 965}
  [\href{https://arxiv.org/abs/astro-ph/9911102}{{\ttfamily
  astro-ph/9911102}}].

\bibitem{Heavens:2020spq}
A.~F. Heavens, E.~Sellentin and A.~H. Jaffe, \emph{{Extreme data compression
  while searching for new physics}},
  \href{https://doi.org/10.1093/mnras/staa2589}{\emph{Mon. Not. Roy. Astron.
  Soc.} {\bfseries 498} (2020) 3440}
  [\href{https://arxiv.org/abs/2006.06706}{{\ttfamily 2006.06706}}].

\bibitem{Philcox:2020zyp}
O.~H. Philcox, M.~M. Ivanov, M.~Zaldarriaga, M.~Simonovic and M.~Schmittfull,
  \emph{{Fewer Mocks and Less Noise: Reducing the Dimensionality of
  Cosmological Observables with Subspace Projections}},
  \href{https://arxiv.org/abs/2009.03311}{{\ttfamily 2009.03311}}.

\bibitem{Gualdi:2019ybt}
D.~Gualdi, H.~Gil-Marín, M.~Manera, B.~Joachimi and O.~Lahav,
  \emph{{Geometrical compression: a new method to enhance the BOSS galaxy
  bispectrum monopole constraints}},
  \href{https://doi.org/10.1093/mnrasl/sly242}{\emph{Mon. Not. Roy. Astron.
  Soc.} {\bfseries 484} (2019) L29}
  [\href{https://arxiv.org/abs/1901.00987}{{\ttfamily 1901.00987}}].

\bibitem{Gualdi:2019sfc}
D.~Gualdi, H.~Gil-Marín, M.~Manera, B.~Joachimi and O.~Lahav, \emph{{GEOMAX:
  beyond linear compression for three-point galaxy clustering statistics}},
  \href{https://doi.org/10.1093/mnras/staa1941}{\emph{Mon. Not. Roy. Astron.
  Soc.} {\bfseries 497} (2020) 776}
  [\href{https://arxiv.org/abs/1912.01011}{{\ttfamily 1912.01011}}].

\bibitem{Pratten:2011kh}
G.~Pratten and D.~Munshi, \emph{{Non-Gaussianity in Large Scale Structure and
  Minkowski Functionals}},
  \href{https://doi.org/10.1111/j.1365-2966.2012.21103.x}{\emph{Mon. Not. Roy.
  Astron. Soc.} {\bfseries 423} (2012) 3209}
  [\href{https://arxiv.org/abs/1108.1985}{{\ttfamily 1108.1985}}].

\bibitem{Schmittfull:2014tca}
M.~Schmittfull, T.~Baldauf and U.~s. Seljak, \emph{{Near optimal bispectrum
  estimators for large-scale structure}},
  \href{https://doi.org/10.1103/PhysRevD.91.043530}{\emph{Phys. Rev. D}
  {\bfseries 91} (2015) 043530}
  [\href{https://arxiv.org/abs/1411.6595}{{\ttfamily 1411.6595}}].

\bibitem{MoradinezhadDizgah:2019xun}
A.~Moradinezhad~Dizgah, H.~Lee, M.~Schmittfull and C.~Dvorkin, \emph{{Capturing
  non-Gaussianity of the large-scale structure with weighted skew-spectra}},
  \href{https://doi.org/10.1088/1475-7516/2020/04/011}{\emph{JCAP} {\bfseries
  04} (2020) 011} [\href{https://arxiv.org/abs/1911.05763}{{\ttfamily
  1911.05763}}].

\bibitem{Chiang:2014oga}
C.-T. Chiang, C.~Wagner, F.~Schmidt and E.~Komatsu, \emph{{Position-dependent
  power spectrum of the large-scale structure: a novel method to measure the
  squeezed-limit bispectrum}},
  \href{https://doi.org/10.1088/1475-7516/2014/05/048}{\emph{JCAP} {\bfseries
  05} (2014) 048} [\href{https://arxiv.org/abs/1403.3411}{{\ttfamily
  1403.3411}}].

\bibitem{Chiang:2015eza}
C.-T. Chiang, C.~Wagner, A.~G. Sánchez, F.~Schmidt and E.~Komatsu,
  \emph{{Position-dependent correlation function from the SDSS-III Baryon
  Oscillation Spectroscopic Survey Data Release 10 CMASS Sample}},
  \href{https://doi.org/10.1088/1475-7516/2015/9/028}{\emph{JCAP} {\bfseries
  09} (2015) 028} [\href{https://arxiv.org/abs/1504.03322}{{\ttfamily
  1504.03322}}].

\bibitem{Chiang:2015pwa}
C.-T. Chiang, \emph{{Position-dependent power spectrum: a new observable in the
  large-scale structure}}, Ph.D. thesis, Munich U., 2015.
\newblock \href{https://arxiv.org/abs/1508.03256}{{\ttfamily 1508.03256}}.

\bibitem{Obreschkow:2012yb}
D.~Obreschkow, C.~Power, M.~Bruderer and C.~Bonvin, \emph{{A Robust Measure of
  Cosmic Structure beyond the Power-Spectrum: Cosmic Filaments and the
  Temperature of Dark Matter}},
  \href{https://doi.org/10.1088/0004-637X/762/2/115}{\emph{Astrophys. J.}
  {\bfseries 762} (2013) 115}
  [\href{https://arxiv.org/abs/1211.5213}{{\ttfamily 1211.5213}}].

\bibitem{Wolstenhulme:2014cla}
R.~Wolstenhulme, C.~Bonvin and D.~Obreschkow, \emph{{Three-point Phase
  Correlations: a new Measure of Nonlinear Large-scale Structure}},
  \href{https://doi.org/10.1088/0004-637X/804/2/132}{\emph{Astrophys. J.}
  {\bfseries 804} (2015) 132}
  [\href{https://arxiv.org/abs/1409.3007}{{\ttfamily 1409.3007}}].

\bibitem{Eggemeier:2015ifa}
A.~Eggemeier, T.~Battefeld, R.~E. Smith and J.~Niemeyer, \emph{{The Anisotropic
  Line Correlation Function as a Probe of Anisotropies in Galaxy Surveys}},
  \href{https://doi.org/10.1093/mnras/stv1602}{\emph{Mon. Not. Roy. Astron.
  Soc.} {\bfseries 453} (2015) 797}
  [\href{https://arxiv.org/abs/1504.04036}{{\ttfamily 1504.04036}}].

\bibitem{Eggemeier:2016asq}
A.~Eggemeier and R.~E. Smith, \emph{{Cosmology with Phase Statistics: Parameter
  Forecasts and Detectability of BAO}},
  \href{https://doi.org/10.1093/mnras/stw3249}{\emph{Mon. Not. Roy. Astron.
  Soc.} {\bfseries 466} (2017) 2496}
  [\href{https://arxiv.org/abs/1611.01160}{{\ttfamily 1611.01160}}].

\bibitem{Franco:2018yag}
F.~O. Franco, C.~Bonvin, D.~Obreschkow, K.~Ali and J.~Byun, \emph{{Probing
  redshift-space distortions with phase correlations}},
  \href{https://doi.org/10.1103/PhysRevD.99.103530}{\emph{Phys. Rev.}
  {\bfseries D99} (2019) 103530}
  [\href{https://arxiv.org/abs/1805.10178}{{\ttfamily 1805.10178}}].

\bibitem{Ali:2018sdk}
K.~Ali, D.~Obreschkow, C.~Howlett, C.~Bonvin, C.~Llinares, F.~O. Franco et~al.,
  \emph{{Cosmological Constraints from Fourier Phase Statistics}},
  \href{https://doi.org/10.1093/mnras/sty1696}{\emph{Mon. Not. Roy. Astron.
  Soc.} {\bfseries 479} (2018) 2743}
  [\href{https://arxiv.org/abs/1806.10276}{{\ttfamily 1806.10276}}].

\bibitem{Byun:2020hun}
J.~Byun, F.~O. Franco, C.~Howlett, C.~Bonvin and D.~Obreschkow,
  \emph{{Constraining the growth rate of structure with phase correlations}},
  \href{https://doi.org/10.1093/mnras/staa2020}{\emph{Mon. Not. Roy. Astron.
  Soc.} {\bfseries 497} (2020) 1765}
  [\href{https://arxiv.org/abs/2005.06325}{{\ttfamily 2005.06325}}].

\bibitem{Fergusson:2009nv}
J.~Fergusson, M.~Liguori and E.~Shellard, \emph{{General CMB and Primordial
  Bispectrum Estimation I: Mode Expansion, Map-Making and Measures of f\_NL}},
  \href{https://doi.org/10.1103/PhysRevD.82.023502}{\emph{Phys. Rev. D}
  {\bfseries 82} (2010) 023502}
  [\href{https://arxiv.org/abs/0912.5516}{{\ttfamily 0912.5516}}].

\bibitem{Fergusson:2010dm}
J.~Fergusson, M.~Liguori and E.~Shellard, \emph{{The CMB Bispectrum}},
  \href{https://doi.org/10.1088/1475-7516/2012/12/032}{\emph{JCAP} {\bfseries
  12} (2012) 032} [\href{https://arxiv.org/abs/1006.1642}{{\ttfamily
  1006.1642}}].

\bibitem{Ade:2013ydc}
{\scshape Planck} collaboration, \emph{{Planck 2013 Results. XXIV. Constraints
  on primordial non-Gaussianity}},
  \href{https://doi.org/10.1051/0004-6361/201321554}{\emph{Astron. Astrophys.}
  {\bfseries 571} (2014) A24}
  [\href{https://arxiv.org/abs/1303.5084}{{\ttfamily 1303.5084}}].

\bibitem{Fergusson:2010ia}
J.~R. Fergusson, D.~M. Regan and E.~P.~S. Shellard, \emph{{Rapid Separable
  Analysis of Higher Order Correlators in Large Scale Structure}},
  \href{https://doi.org/10.1103/PhysRevD.86.063511}{\emph{Phys. Rev.}
  {\bfseries D86} (2012) 063511}
  [\href{https://arxiv.org/abs/1008.1730}{{\ttfamily 1008.1730}}].

\bibitem{Regan:2011zq}
D.~M. Regan, M.~M. Schmittfull, E.~P.~S. Shellard and J.~R. Fergusson,
  \emph{{Universal Non-Gaussian Initial Conditions for N-body Simulations}},
  \href{https://doi.org/10.1103/PhysRevD.86.123524}{\emph{Phys. Rev.}
  {\bfseries D86} (2012) 123524}
  [\href{https://arxiv.org/abs/1108.3813}{{\ttfamily 1108.3813}}].

\bibitem{Schmittfull:2012hq}
M.~M. Schmittfull, D.~M. Regan and E.~P.~S. Shellard, \emph{{Fast Estimation of
  Gravitational and Primordial Bispectra in Large Scale Structures}},
  \href{https://doi.org/10.1103/PhysRevD.88.063512}{\emph{Phys. Rev.}
  {\bfseries D88} (2013) 063512}
  [\href{https://arxiv.org/abs/1207.5678}{{\ttfamily 1207.5678}}].

\bibitem{Lazanu:2015rta}
A.~Lazanu, T.~Giannantonio, M.~Schmittfull and E.~P.~S. Shellard, \emph{{Matter
  bispectrum of large-scale structure: Three-dimensional comparison between
  theoretical models and numerical simulations}},
  \href{https://doi.org/10.1103/PhysRevD.93.083517}{\emph{Phys. Rev.}
  {\bfseries D93} (2016) 083517}
  [\href{https://arxiv.org/abs/1510.04075}{{\ttfamily 1510.04075}}].

\bibitem{Lazanu:2015bqo}
A.~Lazanu, T.~Giannantonio, M.~Schmittfull and E.~Shellard, \emph{{Matter
  bispectrum of large-scale structure with Gaussian and non-Gaussian initial
  conditions: Halo models, perturbation theory, and a three-shape model}},
  \href{https://doi.org/10.1103/PhysRevD.95.083511}{\emph{Phys. Rev. D}
  {\bfseries 95} (2017) 083511}
  [\href{https://arxiv.org/abs/1511.02022}{{\ttfamily 1511.02022}}].

\bibitem{Hung:2019ygc}
J.~Hung, J.~R. Fergusson and E.~P.~S. Shellard, \emph{{Advancing the matter
  bispectrum estimation of large-scale structure: a comparison of dark matter
  codes}},  \href{https://arxiv.org/abs/1902.01830}{{\ttfamily 1902.01830}}.

\bibitem{Hung:2019nma}
J.~Hung, M.~Manera and E.~Shellard, \emph{{Advancing the matter bispectrum
  estimation of large-scale structure: fast prescriptions for galaxy mock
  catalogues}},  \href{https://arxiv.org/abs/1909.03248}{{\ttfamily
  1909.03248}}.

\bibitem{Regan:2017vgi}
D.~Regan, \emph{{An Inventory of Bispectrum Estimators for Redshift Space
  Distortions}},
  \href{https://doi.org/10.1088/1475-7516/2017/12/020}{\emph{JCAP} {\bfseries
  12} (2017) 020} [\href{https://arxiv.org/abs/1708.05303}{{\ttfamily
  1708.05303}}].

\bibitem{Oddo:2019run}
A.~Oddo, E.~Sefusatti, C.~Porciani, P.~Monaco and A.~G. Sánchez, \emph{{Toward
  a robust inference method for the galaxy bispectrum: likelihood function and
  model selection}},
  \href{https://doi.org/10.1088/1475-7516/2020/03/056}{\emph{JCAP} {\bfseries
  03} (2020) 056} [\href{https://arxiv.org/abs/1908.01774}{{\ttfamily
  1908.01774}}].

\bibitem{Babich:2005en}
D.~Babich, \emph{{Optimal estimation of non-Gaussianity}},
  \href{https://doi.org/10.1103/PhysRevD.72.043003}{\emph{Phys. Rev. D}
  {\bfseries 72} (2005) 043003}
  [\href{https://arxiv.org/abs/astro-ph/0503375}{{\ttfamily
  astro-ph/0503375}}].

\bibitem{Hahn:2004fe}
T.~Hahn, \emph{{CUBA: A Library for multidimensional numerical integration}},
  \href{https://doi.org/10.1016/j.cpc.2005.01.010}{\emph{Comput. Phys. Commun.}
  {\bfseries 168} (2005) 78}
  [\href{https://arxiv.org/abs/hep-ph/0404043}{{\ttfamily hep-ph/0404043}}].

\bibitem{Hahn:2014fua}
T.~Hahn, \emph{{Concurrent Cuba}},
  \href{https://doi.org/10.1088/1742-6596/608/1/012066}{\emph{J. Phys. Conf.
  Ser.} {\bfseries 608} (2015) 012066}
  [\href{https://arxiv.org/abs/1408.6373}{{\ttfamily 1408.6373}}].

\bibitem{Press1996}
W.~H. Press, S.~A. Teukolsky, W.~T. Vetterling and B.~P. Flannery,
  \emph{{Numerical Recipes in Fortran 77: the Art of Scientific Computing.
  Second Edition}}, vol.~1. 1996.

\bibitem{Grieb:2015bia}
J.~N. Grieb, A.~G. S\'anchez, S.~Salazar-Albornoz and C.~Dalla~Vecchia,
  \emph{{Gaussian covariance matrices for anisotropic galaxy clustering
  measurements}}, \href{https://doi.org/10.1093/mnras/stw065}{\emph{Mon. Not.
  Roy. Astron. Soc.} {\bfseries 457} (2016) 1577}
  [\href{https://arxiv.org/abs/1509.04293}{{\ttfamily 1509.04293}}].

\bibitem{Monaco:2001jg}
P.~Monaco, T.~Theuns and G.~Taffoni, \emph{{Pinocchio: pinpointing
  orbit-crossing collapsed hierarchical objects in a linear density field}},
  \href{https://doi.org/10.1046/j.1365-8711.2002.05162.x}{\emph{Mon. Not. Roy.
  Astron. Soc.} {\bfseries 331} (2002) 587}
  [\href{https://arxiv.org/abs/astro-ph/0109323}{{\ttfamily
  astro-ph/0109323}}].

\bibitem{Monaco:2013qta}
P.~Monaco, E.~Sefusatti, S.~Borgani, M.~Crocce, P.~Fosalba, R.~Sheth et~al.,
  \emph{{An accurate tool for the fast generation of dark matter halo
  catalogs}}, \href{https://doi.org/10.1093/mnras/stt907}{\emph{Mon. Not. Roy.
  Astron. Soc.} {\bfseries 433} (2013) 2389}
  [\href{https://arxiv.org/abs/1305.1505}{{\ttfamily 1305.1505}}].

\bibitem{Munari:2016aut}
E.~Munari, P.~Monaco, E.~Sefusatti, E.~Castorina, F.~G. Mohammad, S.~Anselmi
  et~al., \emph{{Improving fast generation of halo catalogues with higher order
  Lagrangian perturbation theory}},
  \href{https://doi.org/10.1093/mnras/stw3085}{\emph{Mon. Not. Roy. Astron.
  Soc.} {\bfseries 465} (2017) 4658}
  [\href{https://arxiv.org/abs/1605.04788}{{\ttfamily 1605.04788}}].

\bibitem{Sefusatti:2015aex}
E.~Sefusatti, M.~Crocce, R.~Scoccimarro and H.~Couchman, \emph{{Accurate
  Estimators of Correlation Functions in Fourier Space}},
  \href{https://doi.org/10.1093/mnras/stw1229}{\emph{Mon. Not. Roy. Astron.
  Soc.} {\bfseries 460} (2016) 3624}
  [\href{https://arxiv.org/abs/1512.07295}{{\ttfamily 1512.07295}}].

\bibitem{Sellentin:2015waz}
E.~Sellentin and A.~F. Heavens, \emph{{Parameter inference with estimated
  covariance matrices}},
  \href{https://doi.org/10.1093/mnrasl/slv190}{\emph{Mon. Not. Roy. Astron.
  Soc.} {\bfseries 456} (2016) L132}
  [\href{https://arxiv.org/abs/1511.05969}{{\ttfamily 1511.05969}}].

\bibitem{Kaufmann1967}
G.~M. Kaufmann, \emph{Some Bayesian Moment Formulae, Report No. 6710. Centre
  for Operations Research and Econometrics}. Catholic University of Louvain,
  Heverlee, 1967.

\bibitem{Anderson2003}
T.~W. Anderson, \emph{An Introduction to Multivariate Statistical Analysis}.
  Wiley, 2003.

\bibitem{Hartlap2007}
J.~Hartlap, P.~Simon and P.~Schneider, \emph{{Why your model parameter
  confidences might be too optimistic: Unbiased estimation of the inverse
  covariance matrix}},
  \href{https://doi.org/10.1051/0004-6361:20066170}{\emph{Astron. Astrophys.}
  (2006) } [\href{https://arxiv.org/abs/astro-ph/0608064}{{\ttfamily
  astro-ph/0608064}}].

\bibitem{ForemanMackey:2012ig}
D.~Foreman-Mackey, D.~W. Hogg, D.~Lang and J.~Goodman, \emph{{emcee: The MCMC
  Hammer}}, \href{https://doi.org/10.1086/670067}{\emph{Publ. Astron. Soc.
  Pac.} {\bfseries 125} (2013) 306}
  [\href{https://arxiv.org/abs/1202.3665}{{\ttfamily 1202.3665}}].

\bibitem{Lewis:2019xzd}
A.~Lewis, \emph{{GetDist: a Python package for analysing Monte Carlo samples}},
   \href{https://arxiv.org/abs/1910.13970}{{\ttfamily 1910.13970}}.

\bibitem{Oddo:inprep}
A.~Oddo et~al., \emph{{in preparation}}.

\bibitem{Jeong2010}
D.~Jeong, \emph{Cosmology with high (z>1) redshift galaxy surveys}, Ph.D.
  thesis, The University of Texas at Austin, 2010.

\bibitem{Watkinson:2017zbs}
C.~A. Watkinson, S.~Majumdar, J.~R. Pritchard and R.~Mondal, \emph{{A fast
  estimator for the bispectrum and beyond -- a practical method for measuring
  non-Gaussianity in 21-cm maps}},
  \href{https://doi.org/10.1093/mnras/stx2130}{\emph{Mon. Not. Roy. Astron.
  Soc.} {\bfseries 472} (2017) 2436}
  [\href{https://arxiv.org/abs/1705.06284}{{\ttfamily 1705.06284}}].

\bibitem{Tegmark1997}
M.~{Tegmark}, A.~N. {Taylor} and A.~F. {Heavens}, \emph{{Karhunen-Lo{\`e}ve
  Eigenvalue Problems in Cosmology: How Should We Tackle Large Data Sets?}},
  \href{https://doi.org/10.1086/303939}{\emph{Astrophys. J.} {\bfseries 480}
  (1997) 22} [\href{https://arxiv.org/abs/astro-ph/9603021}{{\ttfamily
  astro-ph/9603021}}].

\bibitem{Knox1998}
L.~Knox, R.~Scoccimarro and S.~Dodelson, \emph{{The Impact of inhomogeneous
  reionization on cosmic microwave background anisotropy}},
  \href{https://doi.org/10.1103/PhysRevLett.81.2004}{\emph{Phys. Rev. Lett.}
  {\bfseries 81} (1998) 2004}
  [\href{https://arxiv.org/abs/astro-ph/9805012}{{\ttfamily
  astro-ph/9805012}}].

\bibitem{Amara2008}
A.~{Amara} and A.~{R{\'e}fr{\'e}gier}, \emph{{Systematic bias in cosmic shear:
  extending the Fisher matrix}},
  \href{https://doi.org/10.1111/j.1365-2966.2008.13880.x}{\emph{Mon. Not. Roy.
  Astron. Soc.} {\bfseries 391} (2008) 228}
  [\href{https://arxiv.org/abs/0710.5171}{{\ttfamily 0710.5171}}].

\bibitem{Blot:2015cvj}
L.~Blot, P.~S. Corasaniti, L.~Amendola and T.~D. Kitching, \emph{{Non-Linear
  Matter Power Spectrum Covariance Matrix Errors and Cosmological Parameter
  Uncertainties}}, \href{https://doi.org/10.1093/mnras/stw604}{\emph{Mon. Not.
  Roy. Astron. Soc.} {\bfseries 458} (2016) 4462}
  [\href{https://arxiv.org/abs/1512.05383}{{\ttfamily 1512.05383}}].

\end{thebibliography}\endgroup

\appendix

\section{1-dimensional basis functions}
\label{app:1d qn}

The separable basis functions, $Q_n(k_1,k_2,k_3)$, that we use in this work are constructed out of a product of three 1-dimensional basis functions, 
\begin{equation}
	Q_n(k_1,k_2,k_3) = q_{\{p}(k_1) q_r(k_2)q_{s\}}(k_3).
	\label{eq:Qn equals qpqrqs}
\end{equation}
The $p$, $r$, and $s$ subscripts on the right side index the different 1-dimensional functions that we have chosen, and the curly brackets require that the $Q_n$ functions are invariant to permutations of $k_1$, $k_2$, and $k_3$.

In principle, the $q_n$ basis can be any function, but in this work we implement and compare two choices: normal polynomials and shifted Legendre polynomials.
The normal $q_n$ basis we use is constructed as described in \cite{Fergusson:2009nv}, while the choice of shifted Legendre polynomials is 
\begin{equation}
	q_n(x) \equiv \tilde{P}_n(x) = P_n(2x-1),
\end{equation}
where $P_n$ are the usual non-shifted Legendre polynomials.

Given a definition of $q_n$, we still need to specify how we map between $\{ prs \} \leftrightarrow n$ in eq.~\eqref{eq:Qn equals qpqrqs}.
Our convention is that we group the $Q_n$ according to the maximum power of any term, $p+r+s$, and then within each group, we order the $Q_n$ by sorting by increasing $s$, $r$, and finally $p$.
Explicitly, our first 11 $Q_n$ have $prs$ as listed in Table \ref{tab:n prs}.

In choosing the number of modes in a basis, we will always choose a number of modes such that all modes up to a given $p+r+s$ are included.
For example, in determining how many modes are necessary for parameter constraints to converge when $k_{\rm max}=24.5 \, k_f$ and custom modes are not included, we compared results with $p+r+s \leq 3$ (7 modes), $p+r+s \leq 4$ (11 modes), $p+r+s \leq 5$ (16 modes), and so on, up to $p+r+s \leq 12$ (102 modes).
This is a somewhat arbitrary, but simple, way of grouping modes together to simplify the analysis whenever we check how our results depend on how many modes are used.

We note that this choice of ordering is mostly arbitrary, though larger values of $p+r+s$ generally correspond to $Q_n$ with smaller scale variations, such that the larger $n$ modes are expected to have smaller amplitudes in the weighted bispectrum. 
This would also be true for the alternative option of ordering the modes according to their `distance', $\sqrt{p^2+r^2+s^2}$, which was also described in \cite{Fergusson:2009nv}.

\begin{table}[h]
\centering
\begin{tabular}{c | c c c}
$n$ & $p$ & $r$ & $s$ \\
\hline
0 & 0 & 0 & 0 \\
\hline
1 & 0 & 0 & 1 \\
\hline
2 & 0 & 1 & 1 \\
3 & 0 & 0 & 2 \\
\hline
4 & 1 & 1 & 1 \\
5 & 0 & 1 & 2 \\
6 & 0 & 0 & 3 \\
\hline
7 & 1 & 1 & 2 \\
8 & 0 & 2 & 2 \\
9 & 0 & 1 & 3 \\
10 & 0 & 0 & 4 \\
\hline
etc. & ... & ... & ...
\end{tabular}
\caption[Modal basis ordering]{We show the $prs$ corresponding to the first 11 $Q_n$ basis functions, according to our convention of grouping functions first by $p+r+s$ (as shown by the horizontal lines) and within each group sorting by increasing $s$, $r$, and then $p$.}
\label{tab:n prs}
\end{table}

\section{Calculating the inner product using 1-dimensional FFT}
\label{app:1dfft}

In this appendix, we detail the calculation of eqs.~\eqref{eq:1dfft QQ} and \eqref{eq:1dfft Fx}, copied here for convenience, 
\begin{eqnarray}
	\llangle Q_n|Q_m \rrangle = \frac{1}{2\pi^5} \int {\rm d}x \,\frac{1}{x} 
	&& \{ F_{pa}(x) [ F_{rb}(x) F_{sc}(x) + F_{rc}(x) F_{sb}(x)] \nonumber \\
	&&+ F_{pb}(x) \left[ F_{rc}(x) F_{sa}(x) + F_{ra}(x) F_{sb}(x) \right] \nonumber \\
	&&+ F_{pc}(x) \left[ F_{ra}(x) F_{sb}(x) + F_{rb}(x) F_{sa}(x) \right] \},
	\label{eq:1dfft QQ app}
\end{eqnarray}
where
\begin{equation}
	F_{pa}(x) \equiv \int {\rm d}k \,q_p(k)\,q_a(k)\,\sin(kx).
	\label{eq:1dfft Fx app}
\end{equation}
Specifically, we use 1-dimensional FFTs to compute eq.~\eqref{eq:1dfft Fx app}, and we require it for a range and resolution of $x$ such that the outer integral over $x$ in eq.~\eqref{eq:1dfft QQ app} can be numerically calculated to sufficient accuracy.
Doing this requires some care, and we follow the steps described in Numerical Recipes (Chapter 13.9, ``Computing Fourier Integrals Using the FFT'', hereafter NR). 
Here we summarize the key equations that we use in our case, and we refer the reader to NR for the step-by-step derivation of the method and more general expressions.

The aim is to numerically  evaluate
\begin{equation}
	\int_a^b h(t) \, e^{iwt} \,{\rm d}t,
	\label{eq:dt integral}
\end{equation}
and we begin by approximating $h(t)$ using an interpolation
\begin{equation}
	h(t) \approx \sum_{j=0}^M h_j \, \psi\left(\frac{t-t_j}{\Delta}\right) 
	+ \sum_{j={\rm endpts}} h_j \, \phi_j\left(\frac{t-t_j}{\Delta}\right),
	\label{eq:ht interp}
\end{equation}
where $\Delta\equiv(b-a)/M$, $h_j \equiv h(t_j)$, and $t_j \equiv a+j \Delta$ for $j=0,...,M$.
$\psi$ and $\phi_j$ are kernel functions that depend on the interpolation scheme, which also determines which points at the boundary count as endpoints in the second sum in eq.~\eqref{eq:ht interp}.

We substitute this interpolating function into eq.~\eqref{eq:dt integral}, interchange the sum and integral in each of the two terms, and then make the change of variable $s\equiv (t-t_j)/\Delta$ in the first term and $s\equiv (t-a)/\Delta$ in the second term.
This leads to
\begin{equation}
	\int_a^b h(t) \, e^{iwt} \, {\rm d}t \approx \Delta \, e^{iwa} \left[ W(\theta) \sum_{j=0}^M h_j \, e^{i\theta j}
	+ \sum_{j={\rm endpts}} h_j \, \alpha_j(\theta) \right],
	\label{eq:1dfft result}
\end{equation}
where $\theta \equiv w\Delta$ and the two functions are
\begin{eqnarray}
	W(\theta) &\equiv& \int_{-\infty}^{\infty} {\rm d}s \, e^{i\theta s} \, \psi(s) \\
	\alpha_j(\theta) &\equiv& \int_{-\infty}^{\infty}{\rm d}s \, e^{i\theta s} \, \phi_j(s-j).
\end{eqnarray}
If the endpoint kernel is symmetric, then
\begin{align}
	\phi_{M-j}(s) &= \phi_j(-s) \\
	\alpha_{M-j}(\theta) &= e^{i w (b-a)} \, \alpha_j^*(\theta),
\end{align}
so the second sum in eq.~\eqref{eq:1dfft result} can be more explicitly written as
\begin{align}
	\sum_{j={\rm endpts}} h_j \, \alpha_j(\theta) &= 
	\alpha_0(\theta) \, h_0 + \alpha_1(\theta) \, h_1 + \alpha_2(\theta) \, h_2 + \alpha_3(\theta) \, h_3 + ... \nonumber \\ 
	&+ e^{i w (b-a)} \left[ \alpha_0^*(\theta) \, h_M + \alpha_1^*(\theta)\,h_{M-1} + \alpha_2^*(\theta)\,h_{M-2} + \alpha_3^*(\theta)\,h_{M-3} + ... \right],
\end{align}
where the ellipses represent terms that are dropped when the interpolation kernels are cubic splines (or lower order).

We perform the first sum in eq.~\eqref{eq:1dfft result} using an FFT. The FFT grid size $N$ must be $N \geq M+1$, and it determines the values of $w$ (and $\theta$) that are sampled,
\begin{equation}
	w_n \, \Delta \equiv \frac{2\pi n}{N},
\end{equation}
for $n=0,...,\frac{N}{2}-1$.
Therefore, we see that $M$ and $N$ are both free parameters of the calculation. 
Larger $M$ allows the integration to be sensitive to higher frequency oscillations in $t$, and $M$ must be higher if $h(t)$ is varying rapidly with $t$. 
$N$ on the other hand is the grid resolution used for the FFT that we use to evaluate the sum in the equation above, so for a fixed $M$, a larger $N$ produces finer sampling in $w$-space, which is especially important if the result of the integral is then going to be interpolated for different values of $w$, as we do when this is used to calculate the inner product.

In this work, we use the cubic order kernel functions in NR,
which are implemented in the subroutine \texttt{dftcor} provided there, and use a modification of the code in \texttt{dftint} also provided to calculate the integral for only the $\int_a^b \sin(wt) \, h(t) \, {\rm d}t$ part.

\end{document}